\documentclass{ar-1col-ah}

\usepackage[comma]{natbib}
\usepackage{url}
\usepackage{amsmath}
\usepackage{amssymb,graphicx}
\usepackage{enumitem}

\jname{Xxxx}
\jvol{AA}
\jyear{2020}
\newcommand{\sm}{\, {\rm M}_{\odot}}

\newcommand{\ndeg}{^{\rm o}}

\begin{document}

% Page header
\markboth{Amina Helmi}{Streams and substructures}

% Title
\title{Streams, substructures and the early history of the Milky Way}

%Authors, affiliations address.
\author{Amina Helmi
\affil{Kapteyn Astronomical Institute, University of Groningen, Groningen, 9700 AV, The Netherlands; email: ahelmi@astro.rug.nl}}

%Abstract
\begin{abstract}
 The advent of the 2nd release of the {\it Gaia} mission in combination with 
data from large spectroscopic surveys are revolutionizing our understanding of the Galaxy. Thanks to these transformational datasets and the knowledge accumulated thus far,  a new, more mature picture of the evolution of the early Milky Way is currently emerging. 
\begin{itemize}[rightmargin=25pt]
\item Two of the traditional Galactic components, namely the stellar halo and the thick disk, appear to be intimately linked: stars with halo-like kinematics originate in similar proportions, from a ``heated" (thick) disk and from debris from a system named Gaia-Enceladus.  Gaia-Enceladus was the last big merger event experienced by the Milky Way and probably completed around 10 Gyr ago. The puffed-up stars now present in the halo as a consequence of the merger have thus exposed the existence of a disk component at $z\sim 1.8$. This is likely related to the previously known metal-weak thick disk and may be traceable to metallicities [Fe/H]~$\lesssim -4$. As importantly, there is evidence that the merger with 
Gaia-Enceladus triggered star formation in the early Milky Way plausibly leading to the appearance of the thick disk as we know it. 

\item Other merger events have been characterized better and new ones have been uncovered. These include for example the Helmi streams, Sequoia, and Thamnos, which add to the list of those discovered in wide-field photometric surveys, such as the Sagittarius streams. Current knowledge of their progenitor's properties, star formation and chemical evolutionary histories is still incomplete. 

\item Debris' from different objects show different degrees of overlap in phase-space. This sometimes confusing situation can be improved by determining membership probabilities via quantitative statistical methods. A task for the next years will be to use ongoing and planned spectroscopic surveys for chemical labelling and to disentangle events from one another using dimensions other than only phase-space, metallicity or [$\alpha$/Fe]. 

\item These large surveys will also provide line-of-sight velocities missing for faint stars in {\it Gaia} releases and more accurate distance determinations for distant objects, which in combination with other surveys could also lead to more accurate age dating. The resulting samples of stars will cover a much wider volume of the Galaxy allowing, for example, linking kinematic substructures found in the inner halo to spatial overdensities in the outer halo. 

\item All the results obtained so far are in-line with the expectations of current cosmological models. Nonetheless, tailored hydrodynamical simulations to reproduce in detail the properties of the merger debris, as well as ``constrained" cosmological simulations of the Milky Way are needed. Such simulations will undoubtedly unravel more connections between the different Galactic components and their substructures, and aid in pushing our knowledge of the assembly of the Milky Way to the earliest times. 
\end{itemize}
\end{abstract}

%Keywords, etc.
\begin{keywords}
Galaxy: formation, evolution, kinematics and dynamics, thick disk, halo
\end{keywords}
\maketitle

%Table of Contents
\tableofcontents

\section{INTRODUCTION}

The current are very exciting times for research on streams and substructures, and their use to shed light onto the early history of our own galaxy, the Milky Way.  Although the field now known as Galactic Archaeology has a long history, it is hard to overstate the impact of 
the second data release from the {\it Gaia} Mission \citep[DR2,][]{2018A&A...616A...1G}, which took place on April 25th 2018. The combination with data already available from many large spectroscopic surveys such as APOGEE\footnote{\tt www.sdss.org/dr12/irspec/} \citep{2017AJ....154...94M}, GALAH\footnote{\tt galah-survey.org/} \citep{2015MNRAS.449.2604D}, RAVE\footnote{\tt www.rave-survey.org/} \citep{2017AJ....153...75K}, and LAMOST\footnote{\tt www.lamost.org/} \citep{2012RAA....12..735D}, has helped to obtain a much clearer picture of how the Milky Way, and in particular its older components, have evolved since $z \sim 2$, or equivalently 10 Gyr ago. 

These new datasets are allowing putting together and in a broader context, the many pieces of the puzzle previously reported in the literature to give a much more complete view of the Galaxy's past. The current generation is quite fortunate to be part of this chapter in the history of Galactic astronomy. It is very exciting that we might actually know how and when the Milky Way experienced its last big merger and that it seems likely that this event gave rise to most of the halo near the Sun, which would be predominantly be composed of debris from a single object that was accreted about 10~Gyr ago, and of heated disk present at the time. 
This is what {\it Gaia} has unraveled in conjunction with high resolution spectroscopic surveys, particularly with APOGEE. 
The rapid progress made in the field since {\it Gaia} DR2 has been possible thanks to the work of many scientists also before DR2, as their work 
allowed deriving relatively quickly a rather clear, although not yet fully settled, picture of the sequence of events. This is in fact an example of one of the pillars of the scientific enterprise: that we build on previous knowledge. It would have taken much longer to pin down Galactic history to the extent reached thus far had these earlier works not been carried out. The 2nd data release of 
the {\it Gaia} mission, even if only based on data taken during less than half of the mission's nominal lifetime (22 months out of 60), has really helped us to move from a fragmented view to seeing Galactic history in full glory.

Many excellent reviews have been written over the past 20 years on Galactic archaeology and near-field cosmology starting with \citet{2002ARA&A..40..487F}; and include also  \citet{2015ARA&A..53..631F} on first stars and their use for (near-field) cosmology; on the structure and dynamics of the  Galaxy by \citet{2016ARA&A..54..529B}; on substructure and tidal debris by \citet{2013NewAR..57..100B} and \citet{2016ASSL..420..141J}, as well as the introduction to the Galactic halo by \citet{2008A&ARv..15..145H}. An interesting exercise is to read the reviews using the information that we have recently acquired about our Galaxy. The reader is encouraged to put on the new {\it Gaia} glasses when going through the findings reported in those studies. Hopefully s/he will note that there is much consistency in the results obtained so far, and hopefully also these reviews will aid the readers in constructing their own narrative on the basis of the information and hints that we had but we did not fully understand at the time. 

The first objective of this review is thus to present the state-of-the-art in the context of what was previously known about our Galaxy. It should be noted that because we are still in the process of digesting the most recent results from the many ongoing surveys focused on the Milky Way, and because there is much more data to come in the next 5 to 10 years, it is particularly challenging to give an overview that is complete and that will stand the test of time. The emphasis and sometimes the interpretation of the recent discoveries reflects this author's own perspective and understanding, while still aiming for an objective and solid account of the facts.

Another objective of this review is also to bring out new venues for research now that we have a much better, albeit also sketchy and as just acknowledged, still in a state of flux understanding of the assembly of the Milky Way. As described especially in the second half of this review, there are still many small and not so small details missing. Solving these will require substantial effort. We will need more detailed modeling and better hydrodynamical and cosmological simulations. We will have to assemble large high resolution spectroscopic datasets with the chemical abundances of millions of stars to be able to pin down their site of formation,  to be able to label as it were, the stars' origin. It should be possible to go back in time even further than 10 Gyr ago,  perhaps out to redshift 6 -- 10  by studying stars in the different structures of the Milky Way.

This review starts in Sec.~\ref{sec:MW} with a brief description of the different Galactic components following a traditional approach. In Sec.~\ref{sec:GA} we move on to Galactic archaeology, and discuss the fossils and tools that are available to do this type of work. Then we dive in Sec.~\ref{sec:halo} into one of the components that holds clues to the evolution of the Galaxy at early times, namely the Galactic stellar halo. We describe the most recent discoveries and how they link to the formation of another ancient component, the thick disk. We focus on this latter component in Sec.~\ref{sec:thick-disc}. In this journey, we describe not just the data but we also discuss predictions from simulations and models. In Sec.~\ref{sec:next} we describe the next steps, those that would seem to be necessary to really fully unravel how the Milky Way was put together. These as well as the most important conclusions are summarized in Sec.~\ref{sec:concl}. 

\section{The Milky Way and its traditional components}
\label{sec:MW}

\subsection{Brief description}

The Milky Way is, in general terms, a fairly typical disk galaxy \citep{2016ARA&A..54..529B}. Its estimated stellar mass is $\sim 5 \times 10^{10} \sm$, which implies a luminosity close to the characteristic value $L_*$ of the galaxy luminosity function. Given its circular velocity of $V_{\rm max} \sim 240$ km/s \citep[see e.g.][]{2019A&A...625L..10G},
it may be slightly subluminous as it lies a bit below --but within 1$\sigma$, the Tully-Fisher relation. 

\begin{figure}[h]
\includegraphics[width=0.9\textwidth]{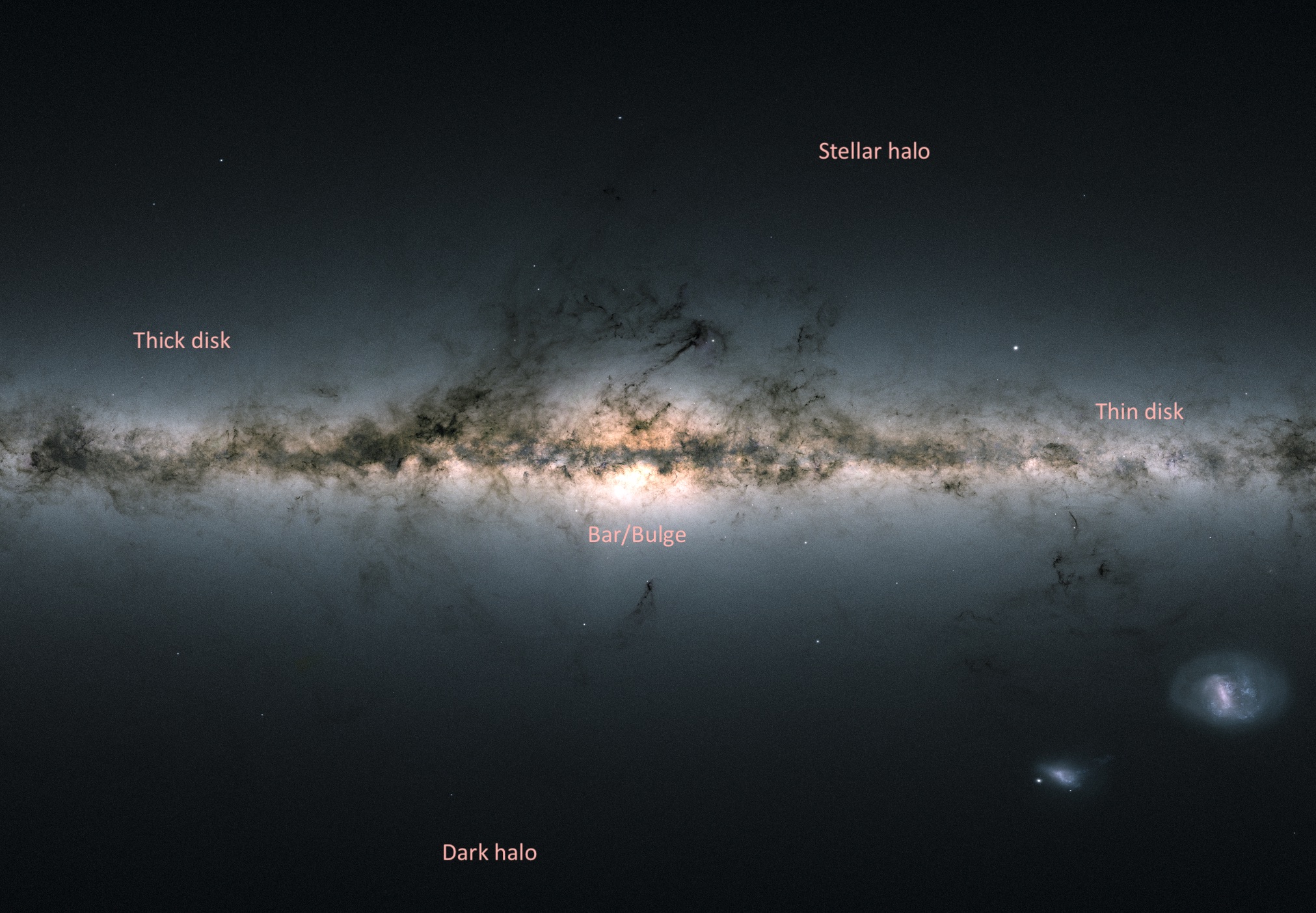}
\caption{The Milky Way and its various components. This image was obtained using data from the 2nd data release of the {\it Gaia} mission \citep{2018A&A...616A...1G}. {\it Credits: ESA/Gaia/DPAC, CC BY-SA 3.0 IGO}.}
\label{fig:mw}
\end{figure}

\begin{figure}[h]
\includegraphics[width=\textwidth]{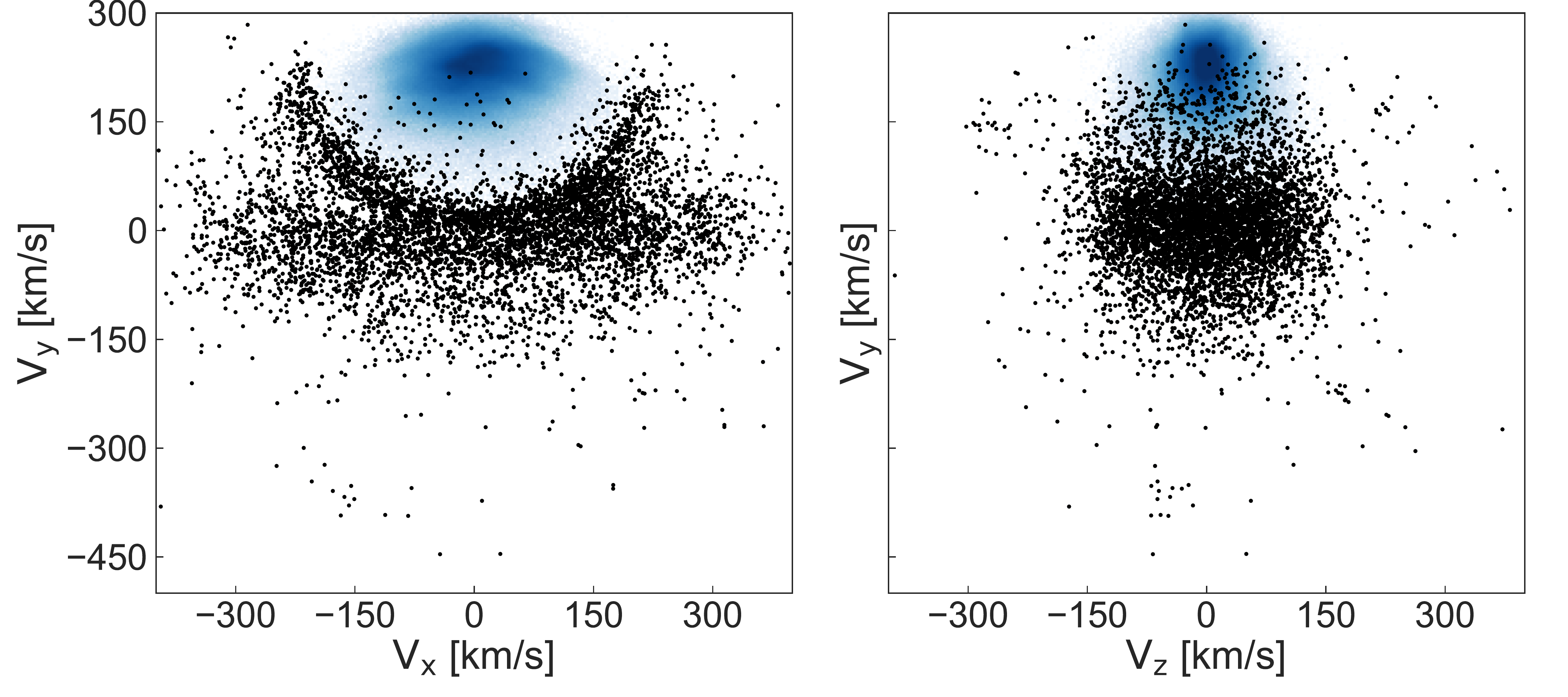}
\caption{Velocity distribution of stars in the solar neighborhood as determined by {\it Gaia}. In this figure, all stars 
from {\it Gaia} DR2 with full phase-space information, located within 1~kpc from the Sun, and with relatively accurate parallaxes, i.e. with $\varpi/\sigma_{\varpi} \ge 5$ have been considered. The 
nearby halo stars are plotted with black dots and defined as those that satisfy $|{\bf V} -  {\bf V}_{LSR}| > 210$ km/s,
for $V_{LSR}$ = 232 km/s. The blue density maps reveal the contribution of the thin and thick disks. The ``banana''-shaped structure seen in the left panel reveals an important contribution of ``hot'' thick disk-like stars to the halo.  {\it Credits:~H.H.~Koppelman \citep[see~also~Fig.~2~in][]{2018ApJ...860L..11K}}.}
\label{fig:vels}
\end{figure}

The Milky Way has several visible components: a thin disk, thick disk, bulge/bar and a stellar halo as shown in \textbf{Figure \ref{fig:mw}}. Each of these components has individual characteristics. Not only do their stars differ in their spatial distribution but of course, also kinematically as shown in \textbf{Figure \ref{fig:vels}}. Furthermore their age and chemical distributions are also different. This implies that the components are truly physically distinct. Their constituent stars inform us about the various processes that are important in the build up of a galaxy throughout its life. 

\subsubsection{Short summary of the main characteristics of the Galactic components} 
\begin{itemize}

\item The {\it thin disk} is the site of ongoing star formation and it is the most characteristic component of the Galaxy, giving it its name. Its current star formation rate is estimated to be $\sim 1.6 \sm$/yr \citep{2015ApJ...806...96L}, and it seems to have been forming stars at least for 8 or 9 Gyr \citep{2019A&A...628A..52T}. It is rotationally supported, and most stars move on fairly circular orbits. 

\item The {\it thick disk} is a thicker, more diffuse and hotter component than the thin disk. Its stars are older than the oldest stars in the thin disk, with estimates using white dwarfs in the Solar vicinity suggesting by at least $\sim 1.6$ Gyr \citep{2017ApJ...837..162K}. Its metallicity distribution function peaks at a lower metallicity value of [Fe/H] $\sim -0.5$, and its stars define a separate chemical sequence in e.g. [$\alpha$/Fe] vs [Fe/H] space, from that defined by the thin disk \citep{2003A&A...410..527B,2011MNRAS.414.2893F}, which can be attributed to a different (shorter and more intense) star formation history  \citep[see e.g.][]{1997ApJ...477..765C,2015A&A...579A...5H}. We discuss it in more detail in Sec.~\ref{sec:thick-disc}.

\item The {\it bar/bulge} is the most centrally concentrated component, and because it is heavily obscured, our current understanding is somewhat limited, although significant progress has recently been made thanks to new surveys, as described in e.g. the reviews by \citet{2018ARA&A..56..223B} and \citet{2019BAAA...61..137Z}. The presence of a 
``classical" bulge (i.e. of spherical shape, formed quickly, dispersion supported) is still debated but its contribution has been constrained by the
observed kinematics to be small \citep[$< 8$\% of the mass of the disk,][]{2010ApJ...720L..72S}. Most of the bulge is in a rotating triaxial structure, 
the Galactic bar, whose orientation, pattern speed and in particular, exact extent have undergone revision lately, where recent work
suggests a rather long bar \citep{2015MNRAS.448..713P,2015MNRAS.450.4050W}. Spectroscopic studies show a mix of populations present in the central regions \citep{2013MNRAS.430..836N}, some of which are very old (more than 13 Gyr), and metal-rich with [Fe/H] values up to +0.5 dex, and some that resemble other Galactic 
components, such as the thick disk and stellar halo, all of which of course, peak in terms of their spatial density in the inner Galaxy.
\item The {\it stellar halo} is the most extended component but at the same time it is rather centrally concentrated: the half-light radius traced by the metal-poor globular clusters is $\sim 0.5$~kpc \citep{2006A&A...450..105B}. It is oblate in the inner regions with $q \sim 0.6$, and its density is well-modelled by a broken power-law  \citep{2011MNRAS.416.2903D,2015ApJ...809..144X}. The most recent estimates of its total mass yield $\sim 1.3 \times 10^9 \sm$ \citep{Deason_2019,2019arXiv191003590M}. The stellar halo contains very metal-poor and old stars. It will be discussed in detail in Sec.~\ref{sec:halo} of this review. 
\item The above items refer to the stellar components of the Galaxy, but there is also warm-ionized gas 
\citep[in a halo or circum-galactic medium,][]{2017ASSL..430...15R,2019ApJ...871...35Z}, and cold gas mostly in the disk. 
\end{itemize}

If our understanding of Gravity is correct, the Galaxy is embedded in a dark matter halo, where most of the mass of the system is located. The characteristics of this halo are not very well constrained. Current estimates of its mass based on {\it Gaia} DR2  by \citet{2019A&A...621A..56P,2019ApJ...873..118W} give $\sim 1.3 \times 10^{12} \sm$  \citep[consistent with the range of values quoted in][]{2016ARA&A..54..529B}.  Its shape is uncertain and has been the subject of significant debate \citep{2001ApJ...551..294I,2004ApJ...610L..97H,2005ApJ...619..800J,2013ApJ...765L..15I}.
It is likely slightly oblate in the central regions \citep[][although \citet{2019MNRAS.485.3296W} argues for spherical]{2010ApJ...712..260K} and changes to a triaxial shape at large distances \citep{2010ApJ...718.1128L, 2013ApJ...773L...4V}, with the longest axis in the direction perpendicular to the disk \citep{2011ApJ...732L...8B,2013ApJ...773L...4V,2016MNRAS.460..329B,2019A&A...621A..56P}. 
The density profile of the dark halo has received less attention thus far \citep[but see][]{2016MNRAS.461.3483T,2019MNRAS.483.4724F,2019ApJ...875..159E,2019arXiv190802336Y}. An interesting question is the degree of lumpiness of the mass distribution 
and whether it is consistent with expectations from cold dark matter simulations, which predict myriads of (dark) satellites \citep{1999ApJ...524L..19M,1999ApJ...522...82K,2008MNRAS.391.1685S}. Recent work on streams is beginning to reveal a complexity that may require the consideration of perturbations by e.g. the Large Magellanic Cloud \citep{2013ApJ...773L...4V,2019MNRAS.485.4726K,2019MNRAS.487.2685E} as well as 
a certain amount of smaller scale lumpiness, as suggested by the ground-breaking analyses of \citet{2017MNRAS.466..628B,2018ApJ...863L..20P,2018MNRAS.477.1893D,2019ApJ...880...38B,2019arXiv190308141M}.

\subsection{Link between the components and physical processes in galaxy evolution}

The differing characteristics of the various Galactic components suggest each had its own formation path. Nonetheless it is likely that these paths were interlinked. It should largely be possible to unravel these formation paths using stars, since these retain memory of their origin. This idea constitutes the pillar of Galactic archaeology, as we discuss in greater detail in Sec.~\ref{sec:GA}.

The $\Lambda$-cold dark matter ($\Lambda$CDM) model provides a framework to understand how galaxies form and evolve from first principles \citep[see e.g. the review by][]{2012AnP...524..507F}. In this model, galaxies form inside dark matter halos \citep{1978MNRAS.183..341W}. Most of the dark halos properties (such as their mass function, abundance number, et cetera) depend on characteristics of the cosmological model, including for example the power spectrum of density fluctuations, the type of dark matter, and the values of the cosmological parameters \citep[as discussed extensively in the book by][]{2010gfe..book.....M}. 
Because in the concordance model there is $\sim 6 \times$ more mass in dark matter than in baryons \citep[this is supported by measurements of e.g. the fluctuations in the cosmic microwave background,][]{2016A&A...594A..11P}, 
many of the properties of galaxies such as how they cluster or their dynamics, are largely dictated by their dark halos. A direct example of this is the process of halo collapse and formation, during which dark halos attract baryonic material from which the (visible components of) galaxies can form. For the baryons in a gaseous configuration to be able to cool and form stars, several conditions need to satisfied \citep[dictated by e.g.~cooling and heating processes, and dynamical timescales, see][]{2010gfe..book.....M}. 
If these conditions are satisfied, the gas will cool and collapse to the center of their halos while conserving some amount of angular momentum.  This results in a gaseous disk that is rotationally supported \citep{1998MNRAS.295..319M}, with some amount of random motion depending on the state of the gas \citep{2009ApJ...707L...1B}.
Note that particularly in the early universe, gas can also be directly accreted as a cold flow and feed the forming galaxy \citep{2009Natur.457..451D}.
In the cold gas disk, stars will start to form. In fact, most star formation in the Universe takes place quiescently and is not associated to large starbursts \citep[see][]{2004MNRAS.351.1151B, 2011A&A...533A.119E}. 

In the $\Lambda$CDM model structure formation proceeds hierarchically, via mergers. At early times mergers were more frequent because of the higher density of the Universe. This means that galaxies were more prone to merge with other galaxies, and hence their disks were more vulnerable. Depending on the mass ratio, such an event could lead to the formation of a bulge \citep{1992ApJ...393..484B},
or merely to the thickening of the disk  \citep{1993ApJ...403...74Q}, and possibly also to the formation of a halo of stars from the original disk  and from the destroyed satellite \citep{2009ApJ...702.1058Z,2010MNRAS.404.1711P}, as seems to have happened for the Milky Way (see Sec.~\ref{sec:halo-new} for details). 
Depending on the characteristics of the merger, such an event could have also triggered the formation of a bar \citep{1990A&A...230...37G}. 
It is in fact likely that the Galactic bar has originated from a disk  instability. However it is not clear whether the bar had its origin in the thin disk \citep{2013ApJ...766L...3M}, 
or whether the metallicity gradient seen in the bar implies that some of the stars have their origin in the thick disk \citep[see][and references therein]{2015A&A...577A...1D,2018A&A...616A.180F}, as suggested also by their similar chemical abundance patterns \citep{2010A&A...513A..35A}.

These examples show that there may be strong links between different components of the Milky Way, and that some could even have gotten their current configuration during or triggered by the same event. Furthermore these components may share a fraction of their stellar populations, such as the bar with the (primordial) thick disk, or the halo with the primordial thick disk. On the other hand galaxies at earlier times had higher gas fractions, which could also imply that mergers may have indirectly led to the formation of a significant stellar population in a Galactic component via the triggering of a starburst, as perhaps was the case for the thick disk \citep[see][and Sec.~\ref{sec:disc-formation} for more details]{2019NatAs.tmp..407G}. 

These considerations highlight why we should probably not think of our Galaxy in terms of separate and independent components that have no connection to each other. Rather we should aim to establish if and how they may be related  given our ultimate goal to unravel the sequence of events that took place in the history of the Milky Way.

\section{Galactic Archaeology} 
\label{sec:GA}

\subsection{Introduction}

Today's commonly used phrase {\it Galactic archaeology} is often applied to describe research on the formation and history of the Milky Way and its stellar populations. The work of \citet[][and several of her subsequent papers]{1950ApJ...112..554R} showing that stars with different chemistry also have different kinematics, has been recognized to have been very influential\footnote{The reader may wish to consult the prefatory chapter of the 2019 volume of the Annual Reviews \citep{doi:10.1146/annurev-astro-091918-104446}, or listen to the associated podcast of J.~Bland-Hawthorn interviewing N.~G.~Roman a few months before her passing away in 2018.}. The papers by \citet{1962ApJ...136..748E} and \citet{1978ApJ...225..357S}, as well as \citet{1980FCPh....5..287T} more generally for galaxies, can arguably be considered as pioneering the field. 

In its modern form the idea behind ``Galactic archaeology" is to use the properties of long-lived stars to reconstruct the history much in the same way archaeologists use artifacts or ``rubble", to learn about the past. Possibly one of the first printed records of the use of the word ``archaeology" in an astronomical context is an article in The ESO Messenger 
by \citet{1979Msngr..16....7S}, where there is a reference to ``astro-archaeology". In this paper the authors aim to use old stars to understand the build-up of metals in the universe. The term ``Galactic archaeology" in a more dynamical context is used in the Report of IAU Commission 33: Structure and dynamics of the Galactic system \citep{1988IAUTA..20..377B}, in Section~13 (in charge of J. Binney):
\begin{extract}
``Perhaps it is not too fanciful to imagine a field of galactic archaeology opening up, in which painstaking sifting of the contents of each element of phase-space will enable us to piece together a fairly complete picture of how our Galaxy grew to its present grandeur and prosperity".
\end{extract}
The turn of the century is approximately the time that the phrase ``Galactic archaeology" was adopted widely by the community, as it begins to appear more frequently both in talks as in the printed literature, also in part because of the very influential reviews by \citet{2000Sci...287...79B,2002ARA&A..40..487F}
\citep[see also][who introduced the term ``near-field cosmology"]{1999Natur.400..220B}. 
Impetus to the field was undoubtedly given by the discovery by \citet{1994Natur.370..194I} of the Sagittarius dwarf as direct evidence of an ongoing merger, and to some extent  subsequently by discovery of debris streams near the Sun from a past merger in the Hipparcos\footnote{\tt sci.esa.int/web/hipparcos/} data \citep{1997A&A...323L..49P} by \citet{1999Natur.402...53H}. 

This time also coincides with the maturing of galaxy formation models \citep{1993MNRAS.264..201K,1998ApJ...498..504B,1999MNRAS.310.1087S},
and the establishment of the $\Lambda$CDM model as the concordance cosmological model. 
This allowed significant progress in the theoretical predictions concerning what a galaxy like the Milky Way should have experienced in its lifetime. 
Thus ``Galactic archaeology" could also be guided by theory and some aspects of the cosmological models could now tested directly from the perspective of the Milky Way. This spirit is particularly evident in the 3rd Stromlo symposium on the Galactic halo edited by \citet{1999ASPC..165.....G} 
which took place in Canberra in 1998. 
For example, the article describing the conference highlights \citep{1999PASP..111..653D}, as well as a quick inspection of the index of the proceedings will reveal that the theme of accretion and mergers and the use of the fossil record to reconstruct Galactic history was present  in many of the participants' contributions to the meeting. 

What does the ``Galactic archaeology" actually mean? As already mentioned the idea behind it is that stars have memory of their origin. Low-mass stars live longer than the age of the universe, and hence some will have formed at very early times and have survived until the present day.  They will have retained in their atmospheres a fossil record of the environment in which they were born. This is because the chemical composition of a star's atmosphere, particularly if it has not yet evolved off the main sequence, reflects the chemical composition of the interstellar medium (the molecular cloud) in which it formed. This means access to the physical conditions present at the time of formation of the star. For very old stars, the conditions might have been very different than today (leading for example to different initial mass functions, etc), and therefore such stars provide us with a window into the early Universe \citep{2015ARA&A..53..631F}. On the other hand, stars with similar chemical abundance patterns  likely have a common origin. The search for common ``DNA" would then allow the identification of stars  with a similar history, and is known as chemical ``tagging". 
 The foundations of this approach were put forward by \citet{2002ARA&A..40..487F} and are briefly discussed in Sec.~\ref{sec:chemistry}.

Another particularly useful way to track Galactic history is through precise measurements of stellar ages. Knowledge of the ages of stars would permit dating the sequence of events that led to the formation of the different components of the Galaxy. 
However obtaining precise ages for very old stars is very difficult. Even 10\% errors at 10~Gyr imply going from redshift 1.8 to 2.3, and a difference of only 2 Gyr exists for a star born at redshift 2 or at redshift 6. Nonetheless the combination of ages and chemical abundances of stars is very powerful and can be used to establish a timeline (i.e. in a closed system, stars born later will be more metal-rich). 

Stars also retain memory of their origin in the way they move. For example as a galaxy gets torn apart by the tidal forces of a larger system like the Milky Way, the stripped stars continue to follow similar trajectories as their progenitor system \citep{1996ApJ...465..278J,1998ApJ...495..297J}. 
This implies that if the Milky Way halo is the result of the mergers of many different objects, their stars should define streams that crisscross the whole Galaxy \citep{1999MNRAS.307..495H}. 
As will become clear in Sec.~\ref{sec:dynamics}, access to full phase-space information is critical to reconstruct the past history of the Galaxy using dynamics.

\subsection{Astrophysical properties of stars: Chemical abundances and ages as a tool}
\label{sec:chemistry}

\subsubsection{Chemical abundances}
The discovery that stars with different metallicity (or iron abundance) have different chemical abundance patterns was first hinted at the end of the 1960s \citep{1967ApJ...148..105C}, and one of the first systematic studies of metal-poor stars is the work of \citet{1979ApJ...234..964S}, and interpreted in the context of supernovae type I and II and Galactic nucleosynthesis models.  

The reason for the variety  of chemical elemental abundance patterns is that different elements are produced in different environments and on a range of timescales \citep{1997ARA&A..35..503M}. For  example, $\alpha$-elements such as O, Mg, Si, Ca, S, and Ti are released in large amounts during the explosion of a massive star as a Supernova, an event that occurs only a few million years after the star's birth. 
On the other hand, iron-peak elements are also produced in supernovae (the so-called Type Ia) that are the result of a thermonuclear explosion of a white dwarf in a binary system, although the details of the burning and the number of white dwarfs involved or their masses are under debate \citep[see e.g. the review][]{2014ARA&A..52..107M}. Because both stars in the binary are of lower mass, these SN explosions take place typically on a longer timescale than for SNII, of the order of 0.1 to a few Gyr \citep[e.g.][]{2001ApJ...558..351M}.
 In terms of the chemical evolution of a (closed) system, we thus expect that [$\alpha$/Fe] will eventually decrease as time goes by as the ISM of the system becomes polluted by SNIa. 
 When a significant number of such explosions have occurred, the initial nearly constant [$\alpha$/Fe] trend with [Fe/H] bends over and this leads to the appearance of a ``knee" after which [$\alpha$/Fe] can only decrease further (unless there is some fresh gas infall). 

Heavier elements beyond the iron-peak are created by neutron capture processes, through the so-called slow (s) and rapid (r) processes. When the neutron flux is relatively low, i.e. the timescale between neutron captures is large compared to that of the $\beta$-decay, the s-process can occur.  
This can take place for example in the envelopes of Asymptotic Giant Branch (AGB) stars \citep{1999ARA&A..37..239B}, 
and the contribution of low mass AGB stars (of 1 to $3 \sm$) appears to be particularly important in the chemical history of the Galaxy \citep[see e.g.][and references therein, and also \citealt{2016A&A...586A..49B}]{2010MNRAS.404.1529B}. For low metallicities (at early times) however, stars with such masses will not have had enough time to reach the AGB phase to be significant contributors of these elements \citep[see][and for a comprehensive review on s-process elements, see also \citealt{2011RvMP...83..157K}]{2004ApJ...601..864T}. A prime example of an element for which the s-process is dominant at [Fe/H]~$\gtrsim -1.5$ is Ba  \citep{1999ApJ...525..886A}. 

The r-process, on the other hand,  occurs when the neutron flux is sufficiently high to allow for rapid neutron captures.
This could occur in SN II environments for example, but also in the mergers of two neutron stars and of neutron stars with black holes, or in magneto-rotational supernovae \citep[as explored for example in the galactic simulations of][]{2019MNRAS.483.5123H}. The recent discovery of strontium in the spectra of the kilonova following the gravitational-wave event GW170817 \citep{2019Natur.574..497W}, clearly demonstrates that the r-process does occur in neutron star mergers. Nonetheless, the exact sites and conditions under which the various neutron-capture elements are produced, particularly at very low metallicities, have not yet been fully settled and there may be different channels for producing them \citep[see 
the excellent review of][and the more recent extensive review by \citealt{2019arXiv190101410C}]{2008ARA&A..46..241S}. 
Besides strontium, a very typical r-process element is Eu, while for example Nd is produced almost equally by r- and s-processes at the solar metallicity \citep{1999ApJ...525..886A}.

The information that can be obtained from detailed chemical abundances analysis is what underpins the principle of chemical tagging, as put forward by \citet{2002ARA&A..40..487F}. The chemical DNA of stars born in a variety of environments, will be different \citep{2015MNRAS.449.2604D}.  
Although in principle each molecular cloud will have its own chemical composition, and this is likely to differ from cloud to cloud in a galaxy, in practice the differences for clouds collapsing at the present day may be small, making it very difficult to disentangle (relatively young) groups of stars of common origin on the basis of their chemistry only unless extremely accurate measurements of many different elements are available \citep[although not impossible, see e.g.][]{2006AJ....131..455D}. 
It would be very interesting to associate each star in a galaxy to its parent molecular cloud because this would potentially reveal the physical processes acting on 1 -- 100 pc scales, i.e. the regime of the interplay between dynamics, star formation and stellar feedback.
Yet this is very challenging, and thus a less demanding form of chemical tagging, known as weak tagging or chemical labelling\footnote{This last term was coined by Vanessa Hill possibly in the year 2010.} has been put forward. With ``chemical labelling" we study different (larger) regions (or components) in the Galaxy to unravel for example migration mechanisms in the disk(s) \citep{2017ApJ...834...27M,2019BAAS...51c.238N}.
This allows establishing whether a star now part of the thick disk  actually formed in the thin disk  in the inner Galaxy and migrated to the Solar vicinity \citep{2009MNRAS.399.1145S}. 

\begin{figure}%[h]
\centering
 \begin{minipage}[b]{0.48\textwidth}
    \includegraphics[width=\textwidth]{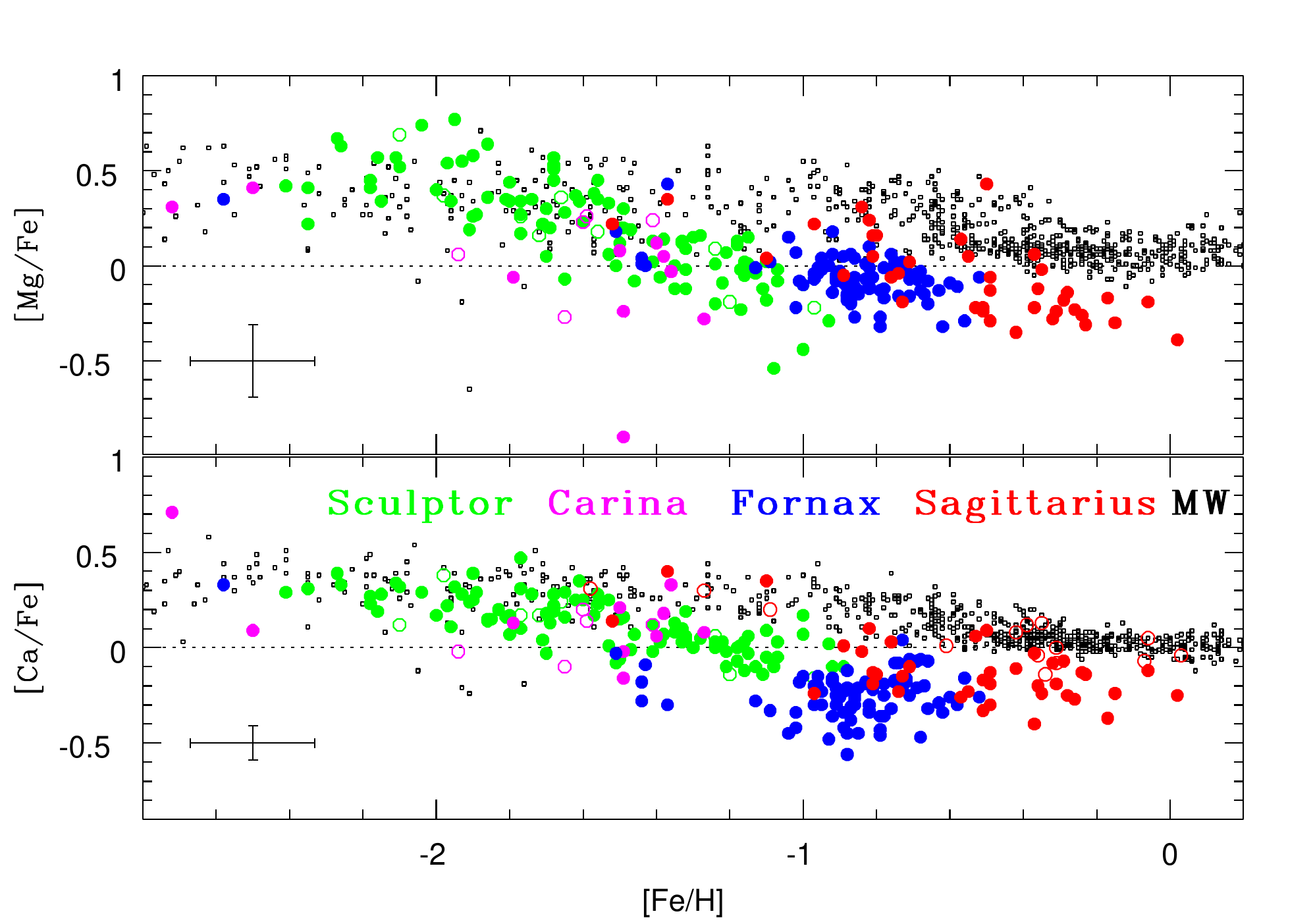}
 \end{minipage}
  \hfill
  \begin{minipage}[b]{0.48\textwidth}
    \includegraphics[width=\textwidth]{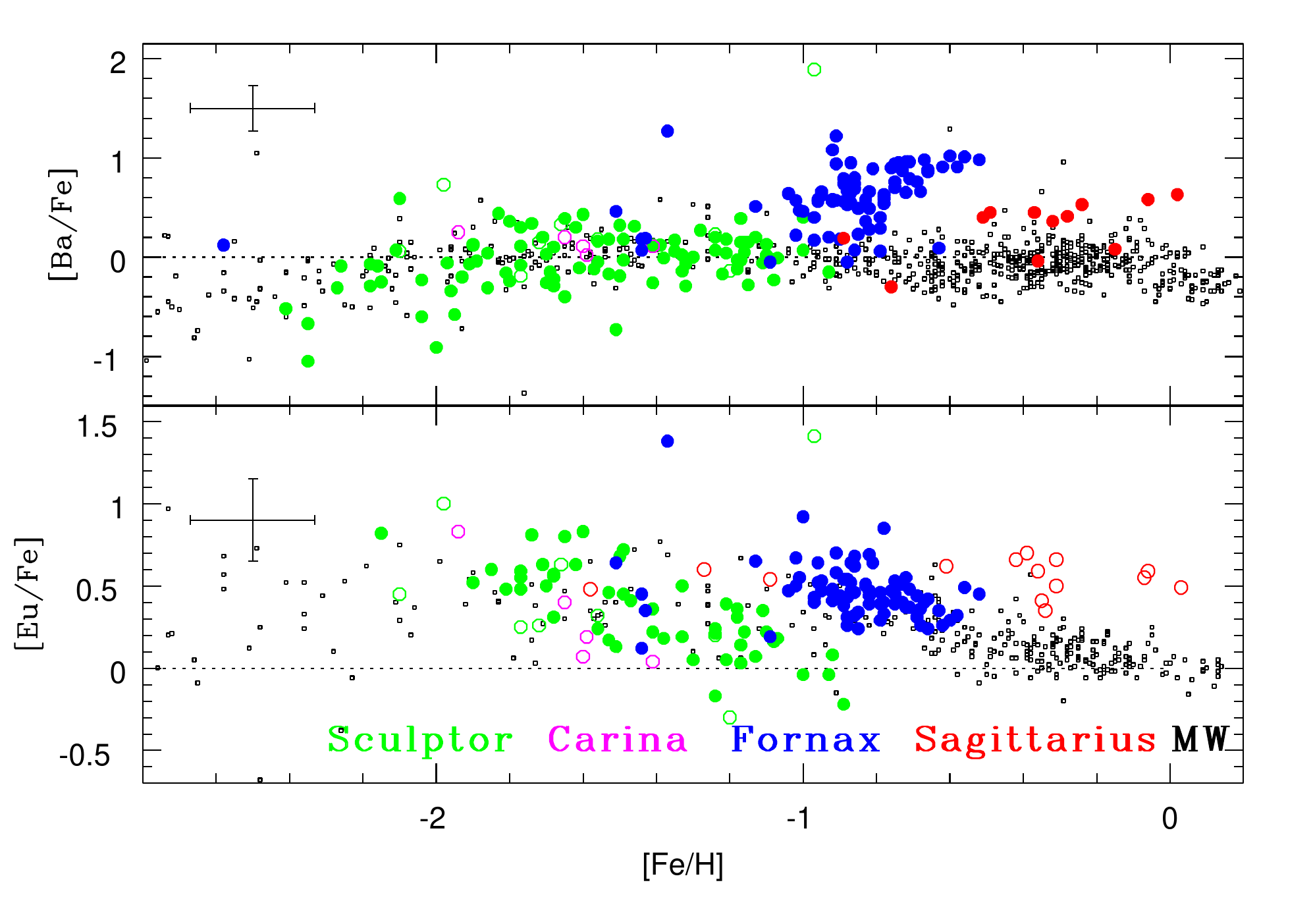} 
  \end{minipage}
\caption{Chemical abundances of stars in four dwarf spheroidal galaxies (in color) and in the Milky Way (in black). The left panels show the behavior with [Fe/H] of two $\alpha$-elements, namely Ca and Mg, while the panels on the right correspond to the trends followed for 
Ba (an s-process element at [Fe/H] $\gtrsim -1.5$ or thereabouts) and Eu (r-process) with metallicity. {\it Credits: Figure reproduced with permission from \cite{2009ARA&A..47..371T}}.}. 
\label{fig:chem-dwarfs}
\end{figure}

In the context of the halo, the underlying thought behind chemical labelling is that stars born in different systems (accreted galaxies or in the proto-Milky Way) will follow their own distinct chemical sequences, because each system had its own particular star formation and chemical enrichment history. This is in fact what  we see for stars associated to the different dwarf galaxies in the Local Group, as shown in \textbf{Figure~\ref{fig:chem-dwarfs}} from \cite{2009ARA&A..47..371T}.  Notice also how the sequences followed by the stars in the different galaxies appear to be sorted according to the mass of the system. In particular, the trend of [$\alpha$/Fe] with [Fe/H] could be an interesting discriminator of stars born in accreted dwarf galaxies. Low mass galaxies that have only formed one generation of stars will likely only have high [$\alpha$/Fe] at low [Fe/H], while those galaxies that have managed to sustain star formation longer, might have very low $[\alpha$/Fe] even at 
low [Fe/H] because of inherently inefficient star formation, and hence their debris may be more easily identifiable. 

Other potentially promising chemical labels for the identification of stars born in accreted dwarf galaxies appear to be r-process element abundances \citep[see e.g.][]{2019NatAs...3..631X}. In the Galactic halo, there is a large scatter in [r-process/Fe] at low metallicity (as seen to some extent in the bottom right panel of \textbf{Figure~\ref{fig:chem-dwarfs}}), which could be indicating a range of birth places. While most ultra-faint dwarf galaxies appear to be deficient in r-process
elements, the Reticulum II galaxy, contains in proportion many r-process enhanced stars \citep[$\sim 78$\% compared to less than 5\% in the Galactic halo,][]{2016Natur.531..610J}. The way to understand this is that the events leading to the formation of r-process elements are so rare, that they have not occurred in most ultra-faint dwarfs (given their low mass). But if one event does happen, it immediately enriches the entire galaxy.  Thus stars with extreme r-process abundances could have their origin in such galaxies \citep{2019ApJ...871..247B}.
For more massive galaxies, clustering in r-process elemental abundances might be expected \citep{2017ApJ...850L..12T}, which combined with the behaviour of [$\alpha$/Fe] or [Fe/H], could enhance their utility for chemical labelling \citep{2019arXiv190810729S}. 

Chemical labelling has also been used to identify field halo stars that may have originated in disrupted globular clusters \citep{2016ApJ...825..146M,2019ApJ...886L...8F}. Searches for these stars make use of peculiarities in the abundance partterns such as for example, anti-correlations in [Na/O] \citep{2009A&A...505..117C}, or more generally depletions in e.g. C, O, Mg, and enhancements in N, Na, Al, Si \citep[see][and references therein]{2019A&ARv..27....8G}. %% HERE

\subsubsection{Ages}
In comparison to chemical abundance estimation, the determination of precise ages, particularly for old stars, is much more difficult. Age determination has traditionally been done via isochrone fitting.
Recently Bayesian inference tools have been employed to derive ages for large numbers of stars by using not only 
multi-color photometry, but also astrometric data from {\it Gaia} and chemical abundance information provided by large spectroscopic surveys \citep[see for example][and also \citealt{2018A&A...618A..54M}]{2018MNRAS.481.4093S,2018MNRAS.476.2556Q}. Such ages tend to be more reliable, particularly in comparison to those based only on photometry.  

Recently a new way of estimating ages using information about the internal structure of a star (other than the Sun) has become possible via asteroseismology. This field is growing at a fast rate and providing new insights and understanding on stellar evolution, and as a consequence also on age determination \citep{2008Sci...322..558M,2014ApJS..210....1C}. Asteroseismology uses time series of photometry of outstanding quality with  campaigns that may take several months or years depending on the type of star (main sequence, RGB or AGB). The photometric variations are due to internal oscillations, and their frequencies depend on the star's mass, radius and effective temperature. Because the frequencies relate in different ways to each of these parameters, the mass of a star can in principle be derived with knowledge of the basic frequencies as well as of its temperature from, for example, broad-band photometry. The knowledge of the star's mass can then be used to determine its age using stellar evolution models \citep{2013ARA&A..51..353C,2013MNRAS.429..423M}. 

 In the recent past COROT\footnote{\tt http://sci.esa.int/corot/} \citep{2009A&A...506..411A} and Kepler\footnote{\tt https://www.nasa.gov/mission\_pages/kepler/overview/index.html} \citep{2010PASP..122..131G} have been providing new ``gold standards" that allow for better age determination from the frequencies of oscillations of the stars. By calibrating on these, it is possible to obtain independent constraints on e.g. the gravity of a star (log g), which can then be used as a prior for the analysis of spectroscopic surveys. This then results in a larger sample of stars that have been (indirectly) calibrated, and translates into more accurate stellar parameters determinations, which in combination with isochrone fitting can then yield ages for large samples of stars \citep[see e.g.][]{2017A&A...600A..66V}. 
The recently launched TESS\footnote{\tt https://tess.mit.edu/} \citep{2015JATIS...1a4003R} and the upcoming PLATO\footnote{\tt http://sci.esa.int/plato/}
mission \citep{2016AN....337..961R} will monitor and characterize large samples of stars for which ages will then be readily available, plausibly much more accurately than has ever been possible until now as argued by \citet{2019BAAS...51c.503K}.

\subsection{Kinematical properties of stars: Dynamics as a tool}
\label{sec:dynamics}

As mentioned earlier, when a galaxy is disrupted by tidal forces, its stars continue to follow closely the trajectory of the system they used to belong to. A regular orbit (a trajectory), may be characterized by the integrals of motion, such as energy $E$, total angular momentum (for a spherical system) or one of its components (in the case of an axisymmetric galaxy, $L_z$), or by the associated actions, such as $J_R$, $J_\phi$ and $J_z$ for an axisymmetric system \citep{2008gady.book.....B}. 
Since a small galaxy may be seen as an ensemble of stars with similar positions and velocities, this implies that also their integrals of motion (or their orbits) are similar. Hence if these are conserved through time (as expected to hold to first order for a collisionless system such as a galaxy), this implies that the tidally stripped stars will follow very similar orbits as their progenitor. This then results in the formation of a stream \citep{1999MNRAS.307..495H}. 
A stream may thus be seen  as a portion of an orbit populated by stars \citep[to first order, see][for the caveats]{2013MNRAS.433.1813S}. This explains why streams are long and narrow if originating in a small system or if formed recently \cite[see the excellent review by][for more information]{2016ASSL..420..141J}.

\begin{figure}[!h]
\centering
\includegraphics[width=0.7\textwidth]{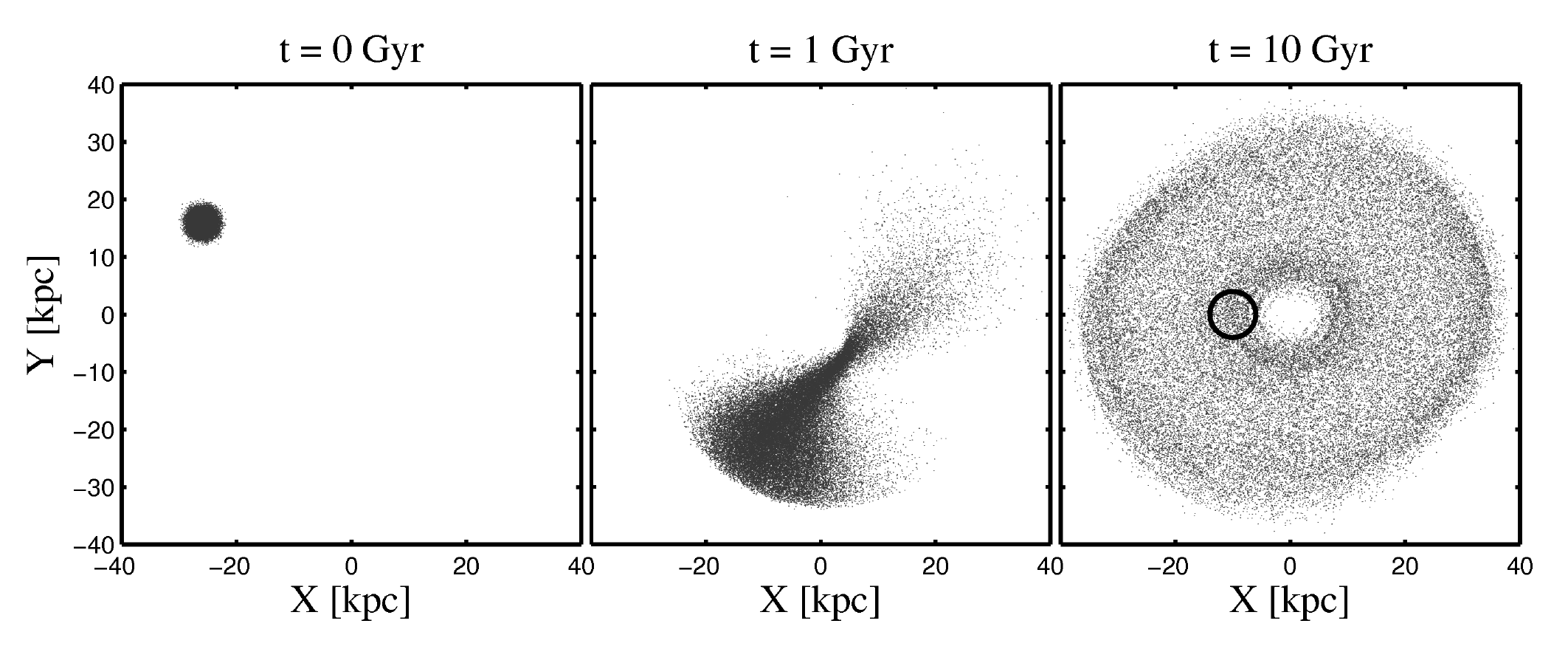}
\includegraphics[width=7cm]{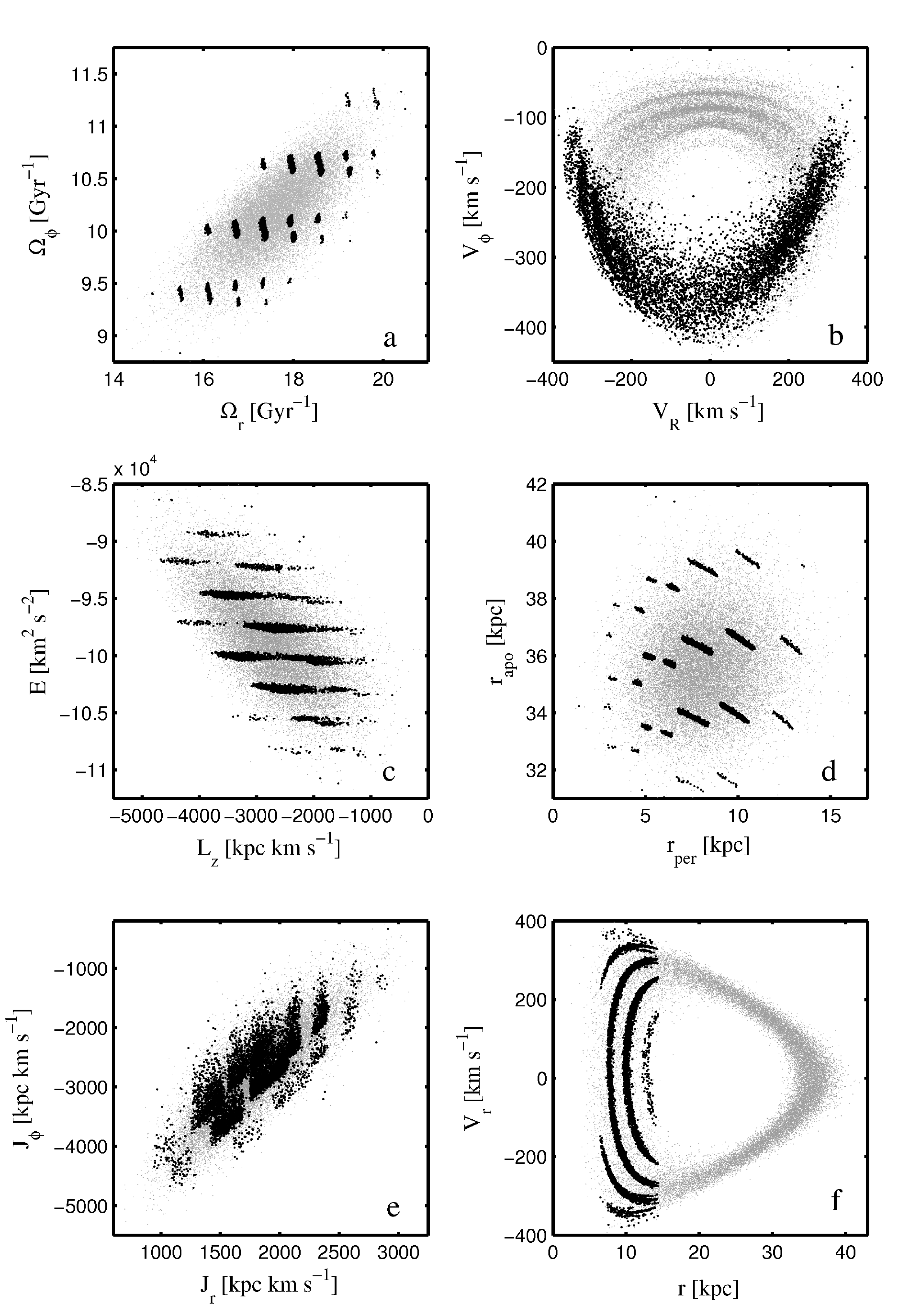}
\caption{Comparison of various spaces commonly used to identify merger debris, namely ({\bf a}) frequency space; ({\bf b}) velocity space; ({\bf c}) energy and $L_z$; ({\bf d}) orbital pericenter vs apocenter; ({\bf e}) actions space and ({\bf f}) phase-space slice of $r$ vs $v_r$. The accreted satellite depicted here was evolved in a spherical Plummer potential (of mass $10^{12}\,{\rm M_{\odot}}$ and $b \sim 22$~kpc) for 10 Gyr. It was non-self-gravitating, spherical and represented with a 6D-Gaussian with $\sigma_{x} \sim 1$~kpc and $\sigma_{v} = 22$ km/s. These characteristics make it comparable to the dwarf elliptical galaxy NGC185 and whose luminosity is only a factor of a few lower than that of the whole Galactic stellar halo.  As a result of this large initial extent, the debris occupies a large volume in phase-space. 
The top panels show X-Y distribution at three different times. The black circle indicates the location of a ``Solar neighborhood'' sphere of $4$ kpc radius. In the bottom panels grey dots show all satellite particles whereas the black dots represent those inside the
  sphere, and reveal the presence of multiple streams in the system. {\it Credits: Adapted from \citet{2010MNRAS.401.2285G}, their Figs.~4 and 5}.}
\label{fig:freq} 
\end{figure}

In the case of a more massive object, tides act in  the same way, but the stars that are stripped at any given point in time have a larger range of values of the integrals (e.g. of energies), which results in a broader population of orbits, and hence in broader streams (and sometimes even hard to distinguish spatially). The process is not different, but the end-product has a different visual appearance and higher complexity, particularly if the parent object is disky, in which case sharper and asymmetric tails may arise depending on the details of the configuration of the merger \citep[][see also \citealt{1972ApJ...178..623T,1973A&A....22...41E}]{1984ApJ...279..596Q}. Furthermore, the morphological features of the debris depend also on the type of orbit of the system \citep{2015MNRAS.454.2472H,2017MNRAS.464.2882A}. For example, if the orbit of the progenitor is fairly radial, then shells are very pronounced. These correspond to the turning points of the orbits of the stars \citep{1999MNRAS.307..495H,1999MNRAS.307..877T}. If the orbit is circular, then there are no turning points, and hence no shells. An example of the spatial evolution of debris on a somewhat radial orbit (with apocenter/pericenter $\approx 4.5$) is shown in the top panels of \textbf{Figure~\ref{fig:freq}}.

The properties of a stream depend on the extent of the parent object, the time since it formed (i.e. since a star became unbound) $t$ and the characteristic orbital timescales, which we denote as $t_{orb}$. For a dispersion-supported progenitor, the 
density of a stream at a given point in space may roughly be expressed as $\rho \propto  (t_{orb}/t)^3 \times 1/(R \sigma^2)$ \citep[see][for details and the full derivation]{1999MNRAS.307..495H}. 
Here $R$ and $\sigma$ are the characteristic size and velocity dispersion of the progenitor system. The dependence on time $t$ is related to the form of the potential and the number of independent orbital frequencies \citep[see][]{2008MNRAS.385..236V}. 
Here it is assumed to be axisymmetric and the orbit to be quasi-regular (and non-resonant), hence the dependence on $t^{-3}$. The expression shows that in the first stages of the dispersal ($t \sim t_{orb}$), the debris will have a high density and therefore remain spatially coherent, leading to easily detectable overdensities on the sky, such as those discovered in the Sloan Digital Sky Survey (SDSS\footnote{\tt https://www.sdss.org/}) by \citet{2006ApJ...642L.137B}. 
This is typically the regime of streams orbiting in the outer halo, since there $t_{orb}$ is large and the tidal forces are less strong, implying also that $t$ is small. For the inner halo, however, the orbital timescales are short, and therefore the density will decrease quickly, also for streams originating in small objects. 

The behavior of stars in a stream is different if their orbits are irregular or chaotic. In that case the rate of divergence will no longer be a power-law but exponential, and phase-mixing is therefore much faster, see \citet{2016MNRAS.455.1079P}. On the other hand, if the orbit is resonant stars take longer to spread out and the debris can remain spatially coherent over more extended timescales.

As time goes by, debris streams mix spatially, i.e. they become long enough that they may cross each other, and therefore one single system can be responsible for multiple streams in a given location in the Galaxy. What characterizes each of the streams is that locally the stars have very similar velocities (in this sense, the stars are truly streaming through the host galaxy). Furthermore, because of the conservation of phase-space density (or volume), as a stream becomes longer and longer, its velocity dispersion will have to decrease, i.e. $\Delta^6 w \sim \Delta^3 x \Delta^3 v$ and since $\Delta^3 x$ (the spatial extent covered by the debris or $1/\rho$) grows in time, this means that $\Delta^3 v$ decreases with time locally as shown by \citet[][and see \citet{2019arXiv190700987B} for a slightly different and interesting application of these concepts]{1999MNRAS.307..495H}. 
This implies that in a given location in the Galaxy, one may have many different moving groups sharing a common origin, as clearly apparent in  \textbf{Figure~\ref{fig:freq}f} which depicts a phase-space slice of stars in a simulation of a relatively massive accreted satellite. 

From these considerations, it transpires 
that to detect each of the predicted moving groups large samples of stars are needed with accurate kinematics.  
\citet{1999MNRAS.307..495H} estimated analytically \citep[this was later confirmed using cosmological simulations by][]{2003MNRAS.339..834H,2013MNRAS.436.3602G} that if the whole stellar halo had been built via mergers, approximately 500 streams would be expected in the halo near the Sun (independently of whether 10 or 100 galaxies had been accreted). Given that the velocity dispersion of the halo is $\sim 100$~km/s, the velocity resolution required would be $100/(500)^{1/3} \sim$~13~km/s, and the sample size needed would have to contain at least as many as 5000 halo stars to yield on average 10 stars per stream. These estimates have nearly been met by {\it Gaia} DR2. Of course, higher precision and larger numbers of tracers would be necessary to go beyond the simple detection of granularity \citep{2003ApJ...592L..63G}
to the full characterization of the streams and their parent objects.

As described above,  debris originating in a single galaxy is thus expected to have similar integrals of motion (which includes of course, the adiabatic invariants). This has led to the search for ancient accretion events by looking for lumpiness in a space of ``Integrals of Motion" (IoM). The first application of this method was by \citet{1999Natur.402...53H} which led to the discovery of the Helmi streams. 
Then a proof of concept of what would be possible with a mission like {\it Gaia} was given in \citet{2000MNRAS.319..657H}. Diagrams of $E$ vs. $L_z$ or $L_z$ vs. $L_\perp$ (where $L^2_\perp = L^2 - L_z^2$, acts as a proxy for a third integral) are now being widely used to establish Galactic accretion history. The advantage of using IoM is that all the individual streams (or wraps) of a single object fold as it were into defining a single clump (compare for example, the ensemble of grey points in panels {\bf c} and {\bf e} to panel {\bf b} in \textbf{Figure~\ref{fig:freq}}). Therefore, the precision required on the measurements is less demanding and the signal of a clump in IoM is higher, because the number of stars in a clump is the total number of stars from each of the streams from a given object added together. 

There are two caveats however. In an ideal case, energy or other integrals of motion would be conserved. However, the gravitational potential in which the streams have evolved must have changed with time, implying that this is not exactly true. Nonetheless, for example \citet{2013MNRAS.436.3602G,2019arXiv190509842S} have shown that substructure is still present in these spaces, even in simulations of the full hierarchical assembly of the halo. Actions being adiabatic invariants, are better conserved although more difficult to compute \citep[but see e.g.][]{2016MNRAS.457.2107S}. 
Thus far however, there has not been a real need to resort to them for the identification of merger debris. 
The likely reason is that the volumes probed so far by data with full phase-space information (6D) are sufficiently small that in the expression $E = 1/2 v^2 + \Phi({\bf x})$, the potential term is approximately constant, i.e. $\Phi({\bf x}_{sun}) = \Phi_0$, and so time variation, or even limited knowledge of the exact form of the potential has not been a limiting factor. The situation will change as we begin to explore beyond the solar neighborhood, especially with {\it Gaia} DR3 and subsequent {\it Gaia} data releases. 

On the other hand, only if all the stars from a given accreted system would be mapped, the defined clump would be fully smooth (in the absence of dynamical friction). As discussed above, when we observe locally we typically only probe portions of debris streams. This implies that we expect substructure to be present within a clump associated to a given object in IoM space when using spatially localized samples of stars. This is clearly seen in \textbf{Figure~\ref{fig:freq}c}, where the grey particles denote all the stars from the system (independent of their final location within the host) and those in black only those inside a small volume (indicated by the circle in the top right panel of \textbf{Figure~\ref{fig:freq}}). Substructure in IoM may also appear if the system is very massive, and thus suffered dynamical friction. In that case, the orbit will have changed with time, and material lost early can be on significantly different orbits than that lost later. 

Individual (portions of) streams are particularly apparent in frequency space,  as can be seen in \textbf{Figure~\ref{fig:freq}a}. This is because  the individual streams each have their own characteristic frequency \citep[which defines their phase along the orbit, see][]{2008MNRAS.390..429M,2010MNRAS.401.2285G}. The regular pattern seen in  \textbf{Figure~\ref{fig:freq}a} depends on the time of accretion of the system since 
$\Delta \Theta = \Delta \Omega t$, where $\Delta \Theta$ represents the difference in phase, and $\Delta \Omega$ would be the separation between neighboring clumps, i.e. a characteristic scale in frequency space. Therefore, since the stars plotted in this figure have all roughly the same location but differ in phase by  $\Delta \Theta \sim 2\pi n$ (with $n$ an integer), this implies that $t$ could be inferred by applying a Fourier analysis, provided enough stars are found in each stream \citep{2010MNRAS.401.2285G}. It turns out frequency space is also useful for constraining the mass growth or time variation of the gravitational potential as the characteristic regular pattern becomes distorted depending on how the system has evolved \citep{2015A&A...584A.120B,2017A&A...601A..37B}. It may be possible to measure these effects using samples of
nearby main sequence halo stars, as these stars are numerous and their velocities and distance estimates may be more accurate because of their relative proximity. 

\section{The Galactic halo }
\label{sec:halo}

\subsection{Generalities}

Mergers play a key role in the hierarchical cosmological paradigm.  This is, after all, by and large the way that galaxies build up their dynamical mass, i.e. their dark halos \citep{2011MNRAS.413.1373W}. 
Therefore, tracking mergers becomes of great importance for the goal of unraveling the build up of galactic systems. The only way we have to track past mergers over long timescales is by resorting to stars. 

This is why the stellar halo of the Galaxy could be considered {\it the} component  to disentangle the merger history of the Galaxy. It is here where disrupted galaxies, cannibalized by the Milky Way, will most likely have deposited their debris. Some debris may be deposited in the thick disk by satellites on low inclination orbits \citep{2003ApJ...597...21A}. It is also a place where we may find heated stars from the disk, i.e. from those present at the time of the mergers and which were perturbed on to hotter orbits \citep{2009ApJ...702.1058Z,2013MNRAS.432.3391T}.
Most of the mass in the inner regions of Milky Way-like dark halos are predicted to originate in a few massive progenitors \citep{2002PhRvD..66f3502H,2011MNRAS.413.1373W}, 
implying that these will have hosted sizable luminous galaxies \citep{2010MNRAS.406..744C}. Therefore most of the information regarding these mergers will be traceable in the stellar halo.

Not only is the stellar halo interesting from the point of view of the merger history, but as mentioned earlier also because it contains (some of) the oldest stars and the most metal-poor ones (possibly together with the bulge). This is not necessarily a coincidence. The existence of a mass-metallicity relation for galaxies, implies that the proto-Milky Way will generally have been the more massive object in its cosmic neighborhood. 
This implies that accreted galaxies will have been less massive than the proto-Milky Way and hence on average, more metal-poor than the disk. 
Since these objects deposit debris in the stellar halo, it is natural for it to have a lower metallicity. 
\citep[Of course, this shifts the question to a different one, which is understanding why and how such a mass-metallicity relation arises, see e.g.][]{2004ApJ...613..898T}. Since there is also a correlation between mass and star formation rate (SFR hereafter), and even though the first stars to form in the Galaxy might have been very metal-poor (or Pop III), the ISM of the proto-galaxy likely was enriched more quickly because of its high SFR, reaching quite fast a higher overall metallicity, as observed for example in the Galactic bulge/bar region \citep[see e.g.][]{2018gbx..confE..25M}.

Understanding the age distribution is trickier, no less because of the fewer precise constraints. However, there is a simple explanation for why the halo should generally be older than the thin disk. Since mergers were much more frequent in the past, it is only after the major epoch of merger activity that a thin disk could grow to its full current extent. The concordance model predicts that the first stars will form in the highest density peaks, which will collapse first and which are typically associated to the more massive objects at later times \citep[e.g.][]{2005MNRAS.364..367D}. This would mean that the first stars in our cosmic environments ought to have formed in the proto-Milky Way. 
Cosmological simulations suggest these first stars are likely part of the bulge or inner spheroid \citep{2000fist.conf..327W,2010ApJ...708.1398T,2017MNRAS.465.2212S,2018MNRAS.480..652E}, 
where the outer halo would be slightly younger. Thus a slight age gradient (remember we are still discussing the epoch before the thin disk as we know it was in place) could arise from the fact that lower mass objects typically form their first stars a bit later. Later accreted objects would also have continued forming stars longer, and so contributed to the trend \citep{2018ApJ...859L...7C}. An age gradient was what \citet[][SZ, hereafter]{1978ApJ...225..357S} 
discovered when studying the age distribution of halo globular clusters, and what led to the SZ-fragments model of the formation of the halo. Not only outer globular clusters are younger but this is also apparent in recent studies of blue horizontal branch stars \citep{2015ApJ...813L..16S}. 
Note however, that  the age distribution as well as the metallicity, particularly of the outer halo could be rather patchy, and could depend on the specifics of the merger history (e.g orbits, time of infall) and mass spectrum of accreted objects \citep[e.g.][]{2006ApJ...646..886F}.

In summary, because the stellar halo contains in proportion more pristine stars, it gives us a window into the physical conditions present in the early universe \citep[e.g.][]{2015ARA&A..53..631F}, and also on the early phases of the assembly of the Milky Way. Hence its relevance in a cosmological context.

\subsection{State of the art / Most recent discoveries}
\label{sec:halo-new}

Our knowledge of the Galactic halo has increased greatly in the last 20 years. Relatively deep wide-field photometric surveys such as SDSS \citep{2000AJ....120.1579Y} and PanSTARRS\footnote{\tt https://panstarrs.stsci.edu/} \citep{2016arXiv161205560C}, and DES\footnote{\tt https://www.darkenergysurvey.org/}   \citep{2018ApJS..239...18A} more recently, have revealed large overdensities on the sky and many narrow streams \citep{2016MNRAS.463.1759B,2018ApJ...862..114S}. These are direct testimony of accretion events that have built-up the (outer) halo, as discussed in e.g. the reviews by \citet{2013NewAR..57..100B} and \citet{2016ASSL..420...87G}, as well as other articles in the book edited by \citet{2016ASSL..420.....N}. 

The second data release of the {\it Gaia} mission is, on the other hand, currently driving a true revolution in our understanding of the (inner) Galactic halo. This might have been expected because of the need for full phase-space coordinates for large samples of stars to pin down formation history discussed in Sec.~\ref{sec:dynamics}. What was unexpected perhaps was the discovery that a large fraction of the halo near the Sun appears to be constituted by the debris from a single object, named Gaia-Enceladus \citep{2018Natur.563...85H,2018ApJ...863..113H}. This object is sometimes referred to as {\it Gaia} sausage because of its kinematic signature \citep{2018MNRAS.478..611B,2018ApJ...862L...1D}. The other very important contributor in the vicinity of the Sun to stars on halo-like orbits is the (tail of the) Milky Way thick disk \citep{2018A&A...616A..10G,2018ApJ...860L..11K,2018ApJ...863..113H}, as can be seen in the left panel of \textbf{Figure \ref{fig:vels}}. These (proto-)thick disk stars have likely been dynamically heated during the merger with Gaia-Enceladus \citep{2018Natur.563...85H, 2018arXiv181208232D}. We elaborate on these points below.

\subsubsection{Gaia-Enceladus}
\label{sec:GE}

\begin{figure}[!h]
\centering
 \begin{minipage}[b]{0.46\textwidth}
    \includegraphics[width=\textwidth,trim={0 9.5cm 0.5cm 9cm},clip]{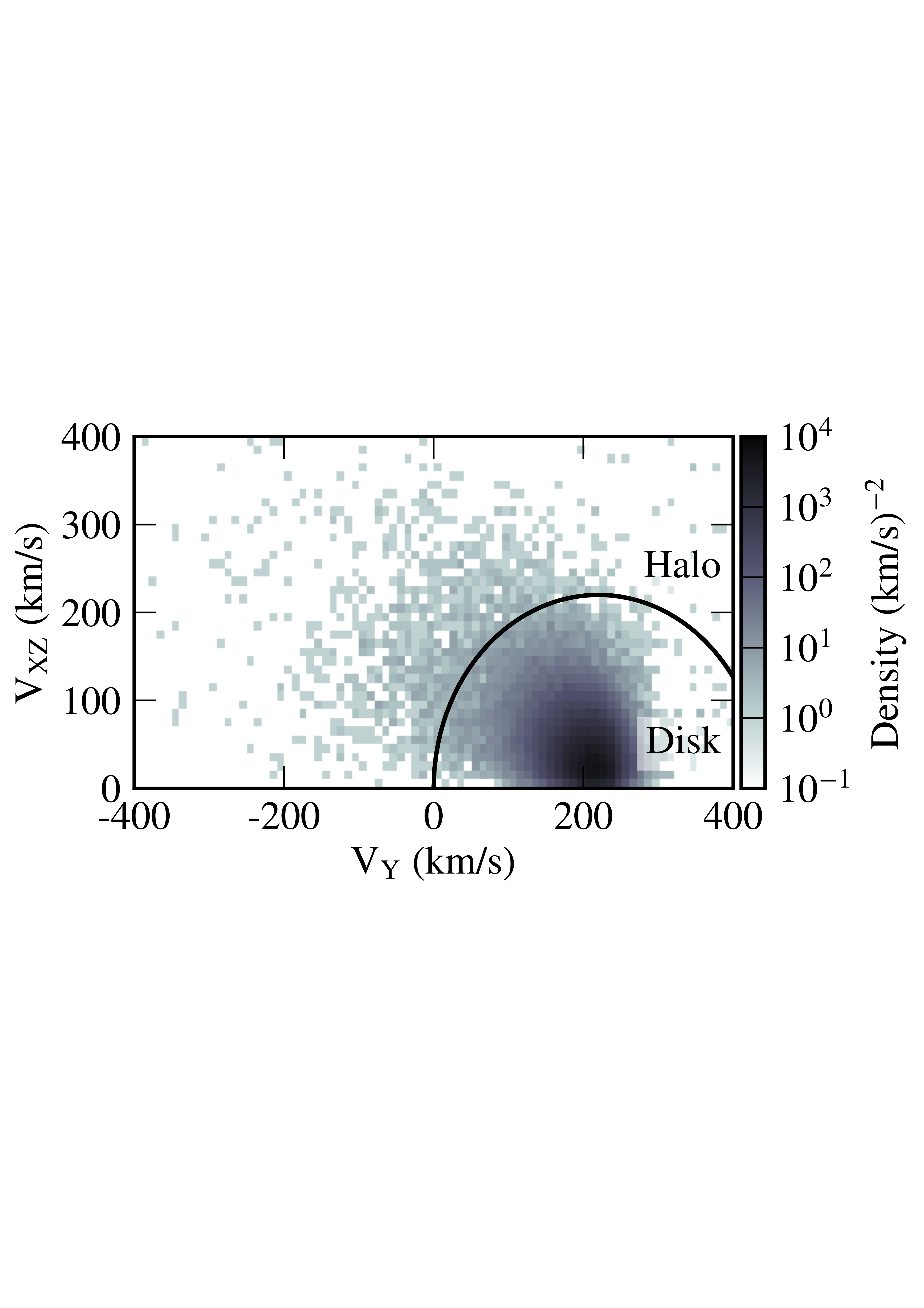}
 \end{minipage}
  \hfill
  \begin{minipage}[b]{0.42\textwidth}
    \includegraphics[width=\textwidth]{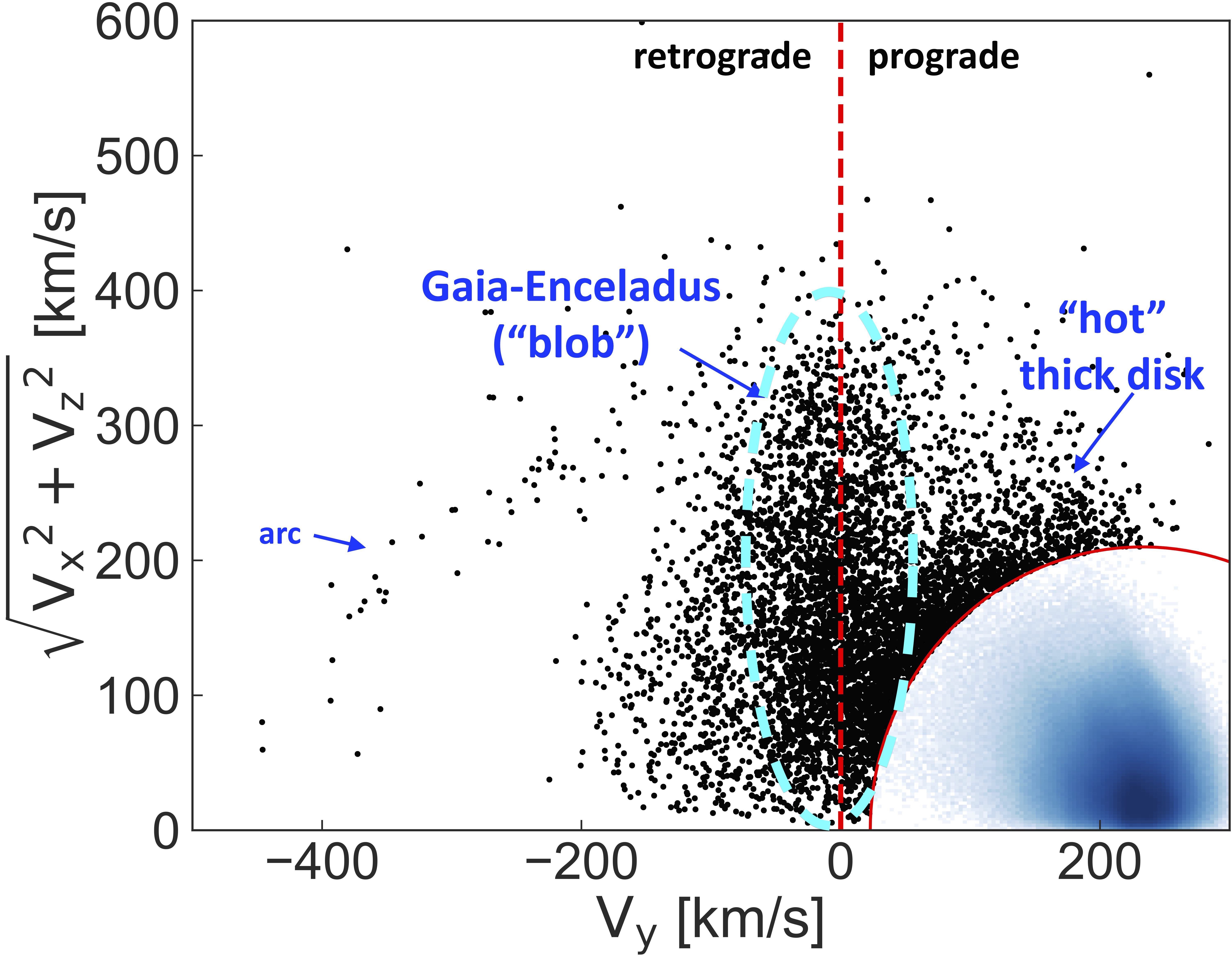}
  \end{minipage}
\caption{Toomre diagram for nearby stars before (left) and after (right) {\it Gaia} DR2. The left panel shows the state-of-the-art before DR2  constructed using proper motions from TGAS \citep[released with {\it Gaia} DR1, \citealt{2016A&A...595A...2G},][]{2016A&A...595A...4L} and line of sight velocities and distances from RAVE \citep{2017AJ....153...75K,2017ApJ...840...59C}. {\it Credits: Figure reproduced with permission from \cite{2017ApJ...845..101B} and from the AAS.} The  panel on the right shows the diagram obtained using DR2 data. The disjoint kinematic nature of the ``blob" and the ``hot" thick disk are unmistakably apparent in this figure. {\it Credits: Figure adapted from \citet{2018ApJ...860L..11K}, and reproduced with permission from the AAS.}}
\label{fig:comparison}
\end{figure}

\begin{figure}[!h]
\centering
\includegraphics[width=\textwidth]{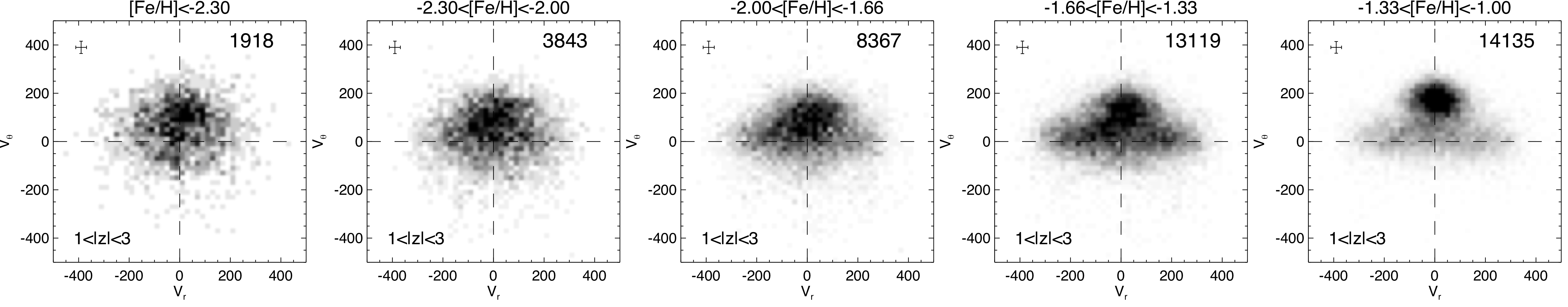}
\includegraphics[width=\textwidth]{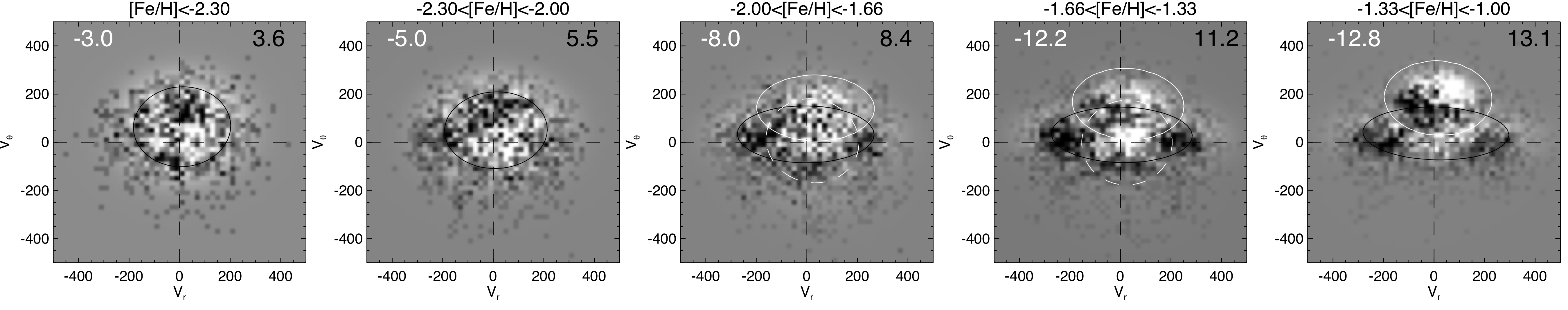}
\caption{Map of the kinematics of halo stars in the sample used by \citet{2018MNRAS.478..611B}. This was obtained from the cross-match of the positions in {\it Gaia} DR1 and SDSS and use of the long time baseline to derive proper motions. The top panels show the distribution for stars in different metallicity bins, while in the bottom panels the residuals resulting from a Gaussian mixture model are plotted, with the contribution of 2 and sometimes 3 subcomponents as indicated by the ellipses. Although a complex kinematic distribution can be retrieved statistically, comparison to the left panel of \textbf{Figure~\ref{fig:vels}} using {\it Gaia} DR2 data reveals how the striking increase in quality of this dataset leads to a true distinction of the various components,  as was also noted in \textbf{Figure~\ref{fig:comparison}}. The $V_R$ asymmetry seen in the rightmost bottom row panel for the faster moving component is also apparent in  \textbf{Figure~\ref{fig:vels}}, and likely the result of the impact of the Galactic bar on the kinematics of these stars, as also reported by \citet{2015ApJ...800L..32A} for the canonical thick disk stars. {\it Credits: Reproduced from \citet{2018MNRAS.478..611B}, top panels of their Figs.~2 and 3}.}
\label{fig:bel}
\end{figure}

Although the presence of stars with metallicities typical of the thick disk but with halo-like kinematics had been reported before {\it Gaia} DR2  \citep[most recently in e.g.][]{2017ApJ...845..101B}, the distinction in the kinematics had not been so clearly seen until DR2, as can be appreciated from the comparison between the left and right panels of \textbf{Figure~\ref{fig:comparison}}, and by inspection of \textbf{Figure~\ref{fig:bel}} compared to the left panel of \textbf{Figure~\ref{fig:vels}}. For stars within 2.5 kpc from the Sun and with $|{\bf V} -  {\bf V}_{LSR}| > 200$ km/s, i.e. traditionally the regime of the halo, approximately 44\% of the stars are in the ``hot" thick disk region ($200 < |{\bf V} -  {\bf V}_{LSR}| < 250$ km/s), while a large fraction of those remaining (between 60\% and 80\% depending on the exact definition) are in the elongated structure that is due to Gaia-Enceladus and indicated in the right panel of \textbf{Figure~\ref{fig:comparison}} \citep[see][]{2018ApJ...860L..11K}. Similar percentages have been reported in e.g. \citet{2017ApJ...845..101B,2018arXiv181208232D,2019arXiv190904679B}. 
%within 2.5 kpc:  total: 20958
%thick disc = 9142, total halo = 11816, GE = 9777. where GE is defined as all the stars with -1500<Lz<150, the thick disc as 200<vtoomre<250 and the total halo as vtoomre>250. These numbers are a bit inflated for GE because I used no energy cut, so in reality, slightly fewer stars are part of it. And for the thick disc, the number of stars increases quickly when the lower velocity limit is changed.

\begin{figure}[!h]
\centering
\includegraphics[width=6.5cm,trim={0.2cm 0.2cm 0 0.2cm},clip]{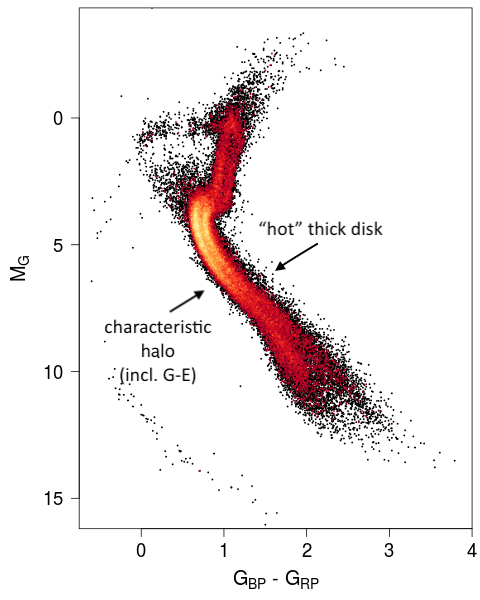}
\includegraphics[width=6.5cm,trim={0 2cm 1cm 14cm},clip]{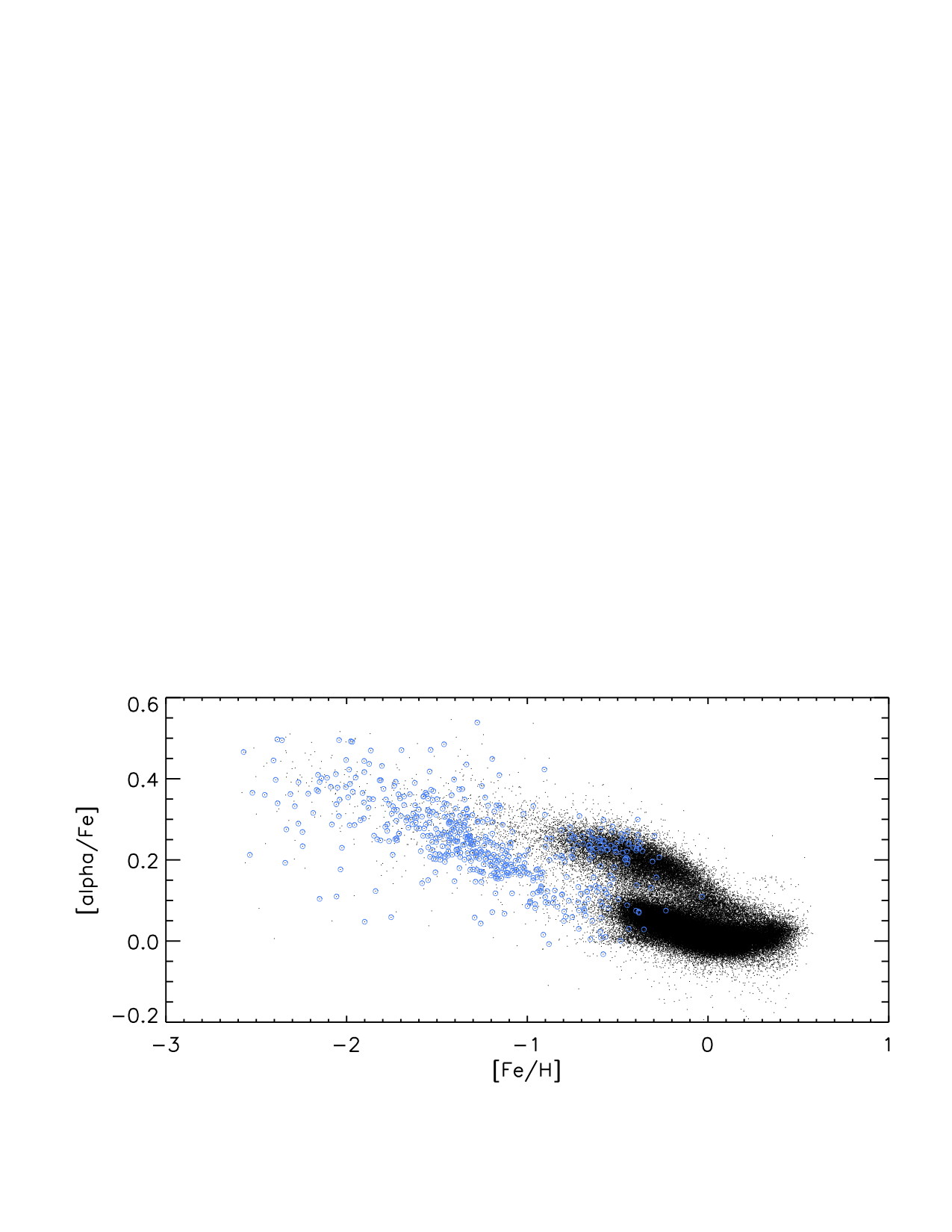}
\caption{The top panel shows the CMD from \citet{2018A&A...616A..10G}, which revealed two sequences in a sample of stars kinematically selected to be part of the halo (with tangential velocities, $V_T = 4.74 \sqrt{\mu_{\alpha*}^2 + \mu_\delta^2}/\varpi > 200$~km/s). 
The bottom panel shows that, when performing a selection in energy and angular momenta \citep[similar to that marked by the ellipse in the right panel of \textbf{Figure~\ref{fig:comparison}}, see for details][and their Extended Data Fig.~1]{2018Natur.563...85H}, most of the stars in the ``blob" region (in blue) define a distinct sequence [$\alpha$/Fe] vs [Fe/H]. 
The stars in blue on the [$\alpha$/Fe]-rich sequence would correspond to thick disk stars on hot halo-like orbits, as discussed in Sec.~\ref{sec:thick-disc}. {\it Credits: Top figure reproduced from Gaia Collaboration, Babusiaux et al., A\&A, 616, A10, 2018, reproduced with permission © ESO; bottom figure adapted from \citet{2018Natur.563...85H}.}}
\label{fig:gaia-halo}
\end{figure}

These findings link to what was arguably one of the first stunning surprises on the halo in {\it Gaia} DR2: namely the color-(absolute) magnitude diagram of stars with ``halo"-like kinematics (i.e. selected to have tangential velocities $V_T \gtrsim 200$~km/s) and 
presented in \citet{2018A&A...616A..10G}  revealed the presence of two clearly distinct sequences, as shown in the top panel of 
\textbf{Figure~\ref{fig:gaia-halo}}. These well-defined sequences point to the presence of distinct stellar populations (i.e. with different ages and metallicities), and are evocative  of a ``dual" halo \citep[see][and discussed in some detail in Sec.~\ref{sec:dual-halo}]{2007Natur.450.1020C}. 
\citet{2018A&A...616A..10G} tentatively suggested that  the older and more metal-poor sequence corresponded to low $\alpha$-abundance stars on retrograde orbits first reported by \citet{2010A&A...511L..10N,2011A&A...530A..15N}. Then, \citet{2018ApJ...860L..11K}
demonstrated that this sequence was dominated by the large kinematic structure (or ``blob" as it was referred to by the authors), seen in the right panel of \textbf{Figure~\ref{fig:comparison}}. 

Driven by these findings, and by the fact that the mean motion of the stars in the kinematic structure was slightly retrograde (as appreciated from the location of the red vertical line in the right panel of \textbf{Figure~\ref{fig:comparison}}), \citet{2018Natur.563...85H} selected these stars 
and showed that they define a well-populated extended chemical sequence of at least 1 dex in [Fe/H] and which runs below that of the thick disk in [$\alpha$/Fe] vs [Fe/H], as seen in the bottom panel of \textbf{Figure~\ref{fig:gaia-halo}}. 
Because the 
stars in question have lower [$\alpha$/Fe] at the [Fe/H] where there is overlap with thick disk (which by and large must have formed in-situ), this immediately implies that the stars formed in a different system than the thick disk, as [$\alpha$/Fe] will generally decrease as [Fe/H] increases. This means that these stars must have been accreted. Furthermore because the majority of the stars in the nearby halo are part of the ``blob" (if they are not in the tail of the thick disk), this implies that a large fraction of the halo near the Sun has been accreted. The accreted system is what has been called Gaia-Enceladus. 

\begin{textbox}[!h]
\section{Gaia-Enceladus/Gaia-Sausage}
The prominence of  Gaia-Enceladus had in fact been noticed by \citet{2018MNRAS.478..611B} using an unpublished catalogue of proper motions obtained by combining SDSS and {\it Gaia} DR1 astrometric positions \citep[also used in][]{2017MNRAS.470.1259D}. Although the separation between the thick disk and the halo is less sharp because of the lower astrometric precision (as can be seen from \textbf{Figure~\ref{fig:bel}}),  the authors found (after performing a Gaussian mixture model), that a high fraction of the halo stars had very large radial motions. Through comparison to zoom-in cosmological simulations, this significant radial anisotropy was interpreted as implying that the halo stars originated in a significant merger the Galaxy experienced between redshift 1 and 3. In the spirit of what was known before {\it Gaia} DR2, this however was not the only possible interpretation,  particularly because the lower quality of the proper motions did not reveal a retrograde mean signal (and hence somewhat ``abnormal") in the multi-Gaussian component decomposition, and chemical abundance information (in particular the sequence of [$\alpha$/Fe]) was not used in the study. As the authors themselves acknowledge in their paper, perhaps this blob or ``sausage" structure as it was called \citep[see][and the available versions on the ArXiv]{2018ApJ...863L..28M}, was the result of a monolithic-like collapse of the kind proposed by \citet{1962ApJ...136..748E},
in the traditional model of formation of an in-situ halo. The result of  such a collapse would likely put the stars formed on radially biased orbits. 
Now with the knowledge provided by {\it Gaia} DR2 data in combination with that of the APOGEE survey, 
revealing respectively the retrograde mean motion of the halo and the distinct chemical sequence defined by the majority of its stars, 
there is absolutely no question that \citet{2018MNRAS.478..611B}  had seen Gaia-Enceladus' mark in the kinematics of halo stars, and interpreted with remarkable insight correctly its accreted origin.
 \end{textbox}

The presence of a {\em significant} $\alpha$-poor (low Mg) chemical sequence was first reported by \citet{2018ApJ...852...49H},
who used data from the APOGEE survey.  It was however hard for these authors to determine to which component the stars in this sequence belonged to because of the lack of proper motion information (their study was carried out just a few months before {\it Gaia} DR2) or the magnitude of this population. 
Nonetheless from the line-of-sight velocities, \citet{2018ApJ...852...49H} concluded that the stars had ``halo"-like kinematics and because of the characteristics of the sequence that they likely represented an accreted population. These considerations 
led 
\citet{2018ApJ...852...50F}  to fit a chemical evolution model. These authors showed that the sequence could be reproduced if the stars had formed in a system with an average SFR of 0.3~$\sm$/yr over a period of 2~Gyr or so. 
By integrating this SFR, \citet{2018Natur.563...85H} subsequently estimated a stellar mass of $\sim 6 \times 10^8\sm$ for Gaia-Enceladus. 
Since the existence of a sequence had been known for some years since \citet{2010A&A...511L..10N},
there were studies in literature that had compared the ages of the stars in the sequence to those in the thick disk sequence in the metallicity range $-1 \lesssim$ [Fe/H] $\lesssim -0.6$ \citep[][more recently \citealt{2019MNRAS.487L..47V}]{2012A&A...538A..21S,2014MNRAS.445.2575H}.
The low-$\alpha$ stars were found to be younger than those in the thick disk. 
Since these are the most metal-rich stars and likely formed in Gaia-Enceladus before it was fully disrupted, this dates also the time of the merger, and at the same time it demonstrates that a disk was already in place in the Milky Way at the time, roughly 10~Gyr ago 
\citep[see][and Sec.~\ref{sec:thick-disc}]{2019NatAs.tmp..407G}.  

\begin{figure}[!h]
\centering
 \begin{minipage}[b]{0.49\textwidth}
    \includegraphics[width=\textwidth,trim={1.5cm 9cm 0.5cm 8cm},clip]{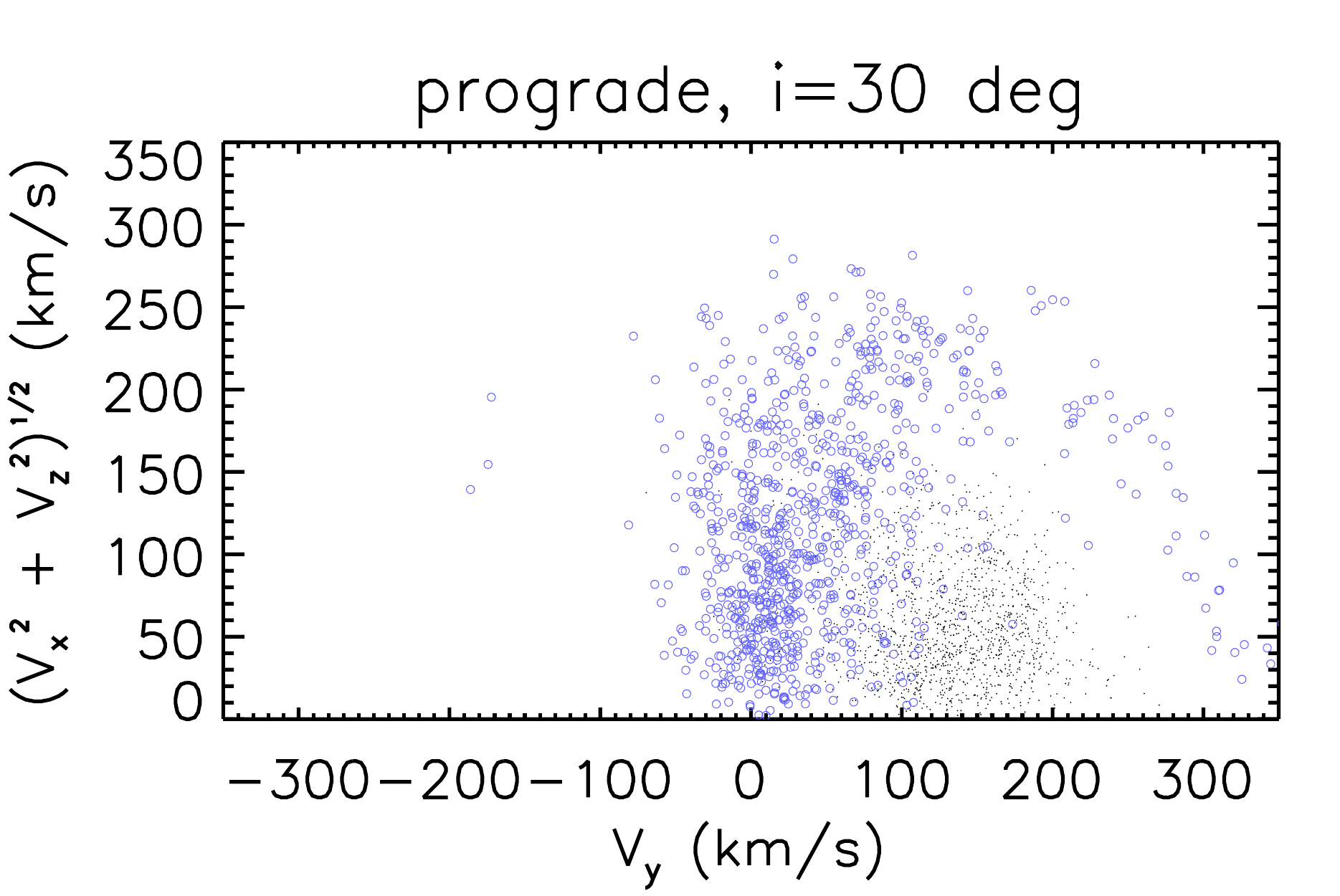}
 \end{minipage}
  \hfill
  \begin{minipage}[b]{0.49\textwidth}
	\includegraphics[width=\textwidth,trim={1.5cm 9cm 0.5cm 8cm},clip]{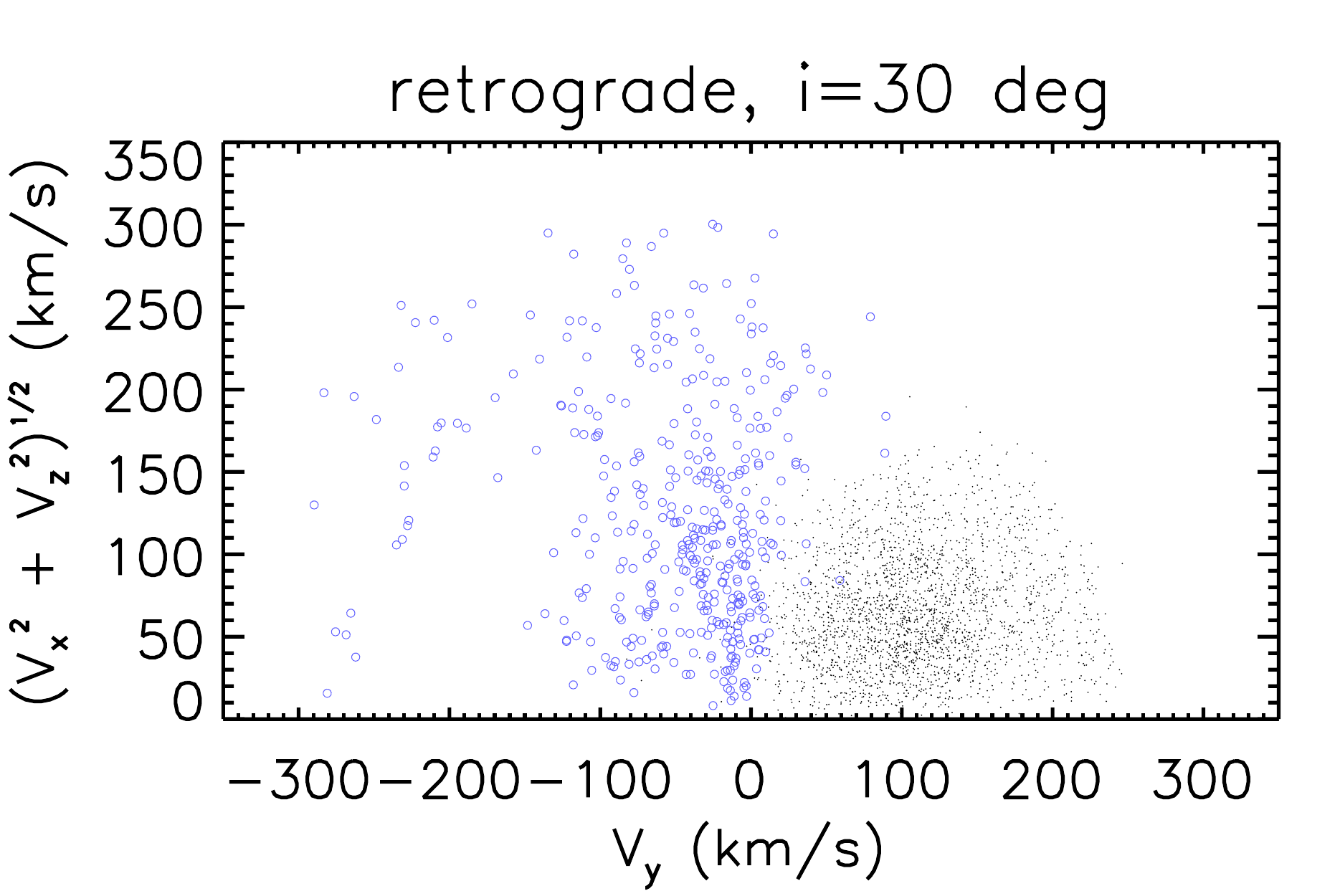}
 \end{minipage}
\caption{Toomre diagrams obtained from a set of simulations aimed to study the formation of the Galactic thick disk via a 20\% mass ratio merger  \citep{2008MNRAS.391.1806V}. In these simulations, a merger between a disky satellite and a host disk is modeled for different orbital configurations. In the left panel the orbit is prograde, while on the right panel it is retrograde. In both cases the initial orbital inclination is 30$^{\rm o}$. The star particles plotted are located inside a solar neighbourhood-like volume, 4 Gyr after infall, with black and blue corresponding respectively to those from the host and from the satellite. Some differences are apparent in the kinematic properties of the merger product, one of the most noticeable being the presence of an arc at high rotational velocities, positive for the prograde and negative for the retrograde case. This arc is composed by star particles lost early during the merger before the object had fully sunk in via dynamical friction \citep[see][]{2020arXiv200607620K}. The arc on the right hand side panel is very reminiscent of that seen in the {\it Gaia} DR2 data and shown in the right panel of \textbf{Figure~\ref{fig:comparison}}. Also \citet[][their Fig.~6]{2017ApJ...845..101B} report the presence of a similar structure in the LATTE cosmological simulation of a Milky Way-like galaxy, but its origin is not discussed in the paper.}
\label{fig:thick-disk-sims}
\end{figure}

The details of the merger have still to be pinned down. For example, a coarse comparison of the {\it Gaia} kinematical data to existing simulations of the merger of a disk galaxy with a massive disky satellite by \citet{2008MNRAS.391.1806V,2009MNRAS.399..166V}, as shown in \textbf{Figure~\ref{fig:thick-disk-sims}}, 
suggests that the merger was counter-rotating because the mean rotational motion of associated halo stars in the solar vicinity is (slightly) retrograde. The presence of specific features in velocity space, namely the arc seen in  the right panel of \textbf{Figure~\ref{fig:comparison}},
and their resemblance to those seen in the simulations, support the retrograde infalling direction and also suggest that the merger's inclination was initially approximately 30$\ndeg$. However other configurations might also be possible, as similar characteristics are found for a merger that is coplanar but where the accreted object is spinning in the opposite sense than the host as shown by \citet{2019arXiv190807080B} using the EAGLE cosmological simulations suite.

\citet{2018Natur.563...85H} suggest a mass ratio of $\sim 4:1$ for the merger \citep[which has been confirmed by][]{2019NatAs.tmp..407G} on the basis of the estimated stellar mass of Gaia-Enceladus, the expected stellar mass-to-halo mass ratio for objects of this size, and assuming a $10^{10}\sm$ thick disk' stellar mass present at the time. Other authors have suggested stellar mass of $10^9 \sm$ up to $5 \times 10^9\sm$ \citep{2019MNRAS.482.3426M,2019MNRAS.484.4471F}. 
Clearly these estimates are uncertain but all pin-point to a significant merger, with lower masses \citep[as proposed by][]{2019arXiv190310136D} being inconsistent with the chemical abundances of the stars (i.e. they would violate the mass-metallicity relation)\footnote{This high-metallicity as well as the tightness of the $\alpha$-poor sequence for [Fe/H]~$\gtrsim -1.3$ (whose width is consistent with being due to measurement errors only) are the reasons why this debris is unlikely to have originated in several (small mass) objects accreted on radial orbits.}. The exact configuration of the merger will have to be pinned down with more detailed modeling and by probing the 3D motions and properties of stars well beyond the solar neighborhood. 

\begin{textbox}[!h]
\section{Mean rotational velocity of Gaia-Enceladus debris}
Determining the mean motion of the debris from Gaia-Enceladus is important as it reveals properties regarding the initial configuration of merger. Support for a retrograde encounter is apparent from the comparison of the simulations shown in \textbf{Figure~\ref{fig:thick-disk-sims}} to the data shown the right panel of \textbf{Figure~\ref{fig:comparison}}, as well as from the location of the vertical line (denoting null rotation) in the latter figure. To quantify this further, we select stars on the [$\alpha$/Fe]-poor sequence, for [Fe/H]~$\ge -1.3$, i.e. the region dominated by Gaia-Enceladus debris (see \textbf{Figure~\ref{fig:gaia-halo}}). By imposing a 20\% relative parallax (or distance) errors quality cut, using distances determined either via $1/\varpi$ or from the Bayesian method by \citet{2018RNAAS...2...51M}, a selection is made that satisfies $-500 \le L_z \le 500$~kpc~km/s. The resulting sample is small but within 1~kpc from the Sun, $\langle v_\phi \rangle = -21.1 \pm 1.8$~km/s (6 stars) and within 2~kpc $\langle v_\phi \rangle = -16.1 \pm 2.8$~km/s (23 stars), where the uncertainty has been estimated from re-sampling the individual velocity errors. For the stars within 1~kpc, these errors are all smaller than 7.2~km/s (the mean is 3.3~km/s), while for those within 2 kpc, the mean error is 10.8~km/s. 
The signal is therefore weak but it is robust to measurement uncertainties (statistical and systematic). This is because within 1~kpc from the Sun, the parallax is $\gtrsim 1$ mas, the relative errors are small ($\lesssim$ 20\%) for the bright stars in the {\it Gaia} RVS sample  \citep{2018A&A...616A..11G}, 
and the systematic parallax bias \citep{2018A&A...616A...2L,2018A&A...616A..17A}
is negligible  within this parallax range (even if as large as 0.05~mas). A similar amplitude mean retrograde motion has been reported by \citet{2019MNRAS.482.3426M} using APOGEE distances. 
\end{textbox}
%distance bin       0       1           6
%mean, stddev, median of mean     -21.0612      1.80382     -21.0772
%distance bin       1       2          23
%mean, stddev, median of mean     -14.8263      3.39309     -14.8407
%distance bin       0       2          29
%mean, stddev, median of mean     -16.1114      2.75143     -16.0822

In summary, the halo near the Sun (if not in a ``hot" thick disk) is dominated by debris from Gaia-Enceladus, a very massive object that was accreted 10 Gyr ago. As such it most likely represents the last significant merger that the Milky Way experienced. Probably after this was completed, the (current) Milky Way thin disk could start a more quiescent growth phase, and this would be consistent with the ages of its oldest stars. Other large mergers the Milky Way has experienced since then include that with the Sagittarius dwarf, but because of the late time of infall ($\sim 8$ Gyr ago) and mass ratio of 1 - 5\% \citep[$M_\star \sim 5 \times 10^8 \sm$, see for example][]{2017ApJ...836...92D,2019MNRAS.483.4724F}, 
this has resulted in a less dramatic impact. Even the ongoing merger with the Large Magellanic Cloud is less important (with a mass ratio of probably $\sim$10\%), although in both cases we do see their effect on the disk of our Galaxy, in the form of phase-space spirals\footnote{The authors termed the structure discovered in {\it Gaia} DR2 near the Sun originally ``snail-shell" (``caracol" in Spanish). This structure is a result of phase-mixing, but because it is apparent in a slice of phase-space (the $z$-$V_z$ plane), ``phase-space
spiral" might be preferred over ``phase spiral" \citep[see also][]{Bland_Hawthorn_2019,2019MNRAS.489.4962K}.} \citep{2018Natur.561..360A}, and waves  \citep[vertical and radial oscillations, see e.g.][]{2018MNRAS.473.1218L,2019MNRAS.485.3134L,Bland_Hawthorn_2019}.

\paragraph*{Evidence with hindsight.}
Besides \citet{2018MNRAS.478..611B}, several other authors had in fact come across traces of Gaia-Enceladus without knowing. We had a rather fragmented view of the inner halo until {\it Gaia} DR2 because it was based on small samples and hence our knowledge was ``patchy" or ``incomplete". For example, the sample with the low-$\alpha$ sequence of \citet{2010A&A...511L..10N}, contained less than 100 stars \citep[which nonetheless was a significant increase in comparison to the original discovery paper, which had just 13 halo and 16 thick disk stars, see][]{1997A&A...326..751N}. Although these authors were able to put forward a scenario rather similar to what has now been revealed with {\it Gaia} DR2 and APOGEE, it is not until recently that the realization came that {\it most} of the halo follows the Nissen \& Schuster low-$\alpha$ sequence, thanks to the work of \citet{2018ApJ...852...49H} on the chemistry and \citet[][]{2018ApJ...860L..11K} on the kinematics \citep[see also][]{2018ApJ...863..113H}. Possibly the reason for this is that the halo near the Sun is known to peak at a metallicity of [Fe/H]~$\sim -1.6$ and have [$\alpha$/Fe]~$\sim 0.2 - 0.4$ around this iron abundance. It is only at higher metallicity that the sequence by Gaia-Enceladus becomes clearly separate and has lower [$\alpha$/Fe] than the thick disk as seen from the lower panel of \textbf{Figure~\ref{fig:gaia-halo}}. Other hints of this structure were present in the \citet{2000AJ....119.2843C}
study based on Hipparcos data. 
These authors identified a concentration of stars for [Fe/H]~$ \sim -1.7$ with very high eccentricities ($e \sim 0.9$), but they interpreted it as being stars formed during an ELS-like collapse in the early Milky Way.  \citet{2003ApJ...585L.125B} compared the \citet{2000AJ....119.2866B} sample 
to  a cosmological hydrodynamical simulation of the formation of a Milky Way-like galaxy. In this comparison, the authors identified a clump of stars with retrograde motions which they proposed could be satellite debris. The presence of clumps of stars with retrograde motion goes back even further, to the work of e.g. \citet{1996ApJ...459L..73M}
and \citet{1996AJ....112..668C} and more recently \citet{2007MNRAS.375.1381K}.
The globular cluster Omega Cen became the natural culprit, because it was known to have a retrograde orbit and because of its peculiar multiple populations, which led to the suggestion that it could have been the nucleus of a disrupted dwarf galaxy 
\citep{2002ASPC..265..365D,2003MNRAS.346L..11B,2005MNRAS.359...93M} .

\subsubsection{The dual halo}
\label{sec:dual-halo}

The two sequences revealed in the CMD of \citet{2018A&A...616A..10G} are a direct reminder of the suggestion of a ``dual halo" in the Milky Way. As just discussed it has been possible to attribute the first, more metal-poor sequence largely to Gaia-Enceladus. On the other hand, the second sequence is populated by stars which kinematically are clearly part of the tail of the thick disk seen from the right panel of \textbf{Figure~\ref{fig:comparison}} \citep{2018ApJ...863..113H,2018arXiv181208232D}. 

The idea of a dual halo was discussed already quite thoroughly by \citet{1994ApJ...431..645N} \citep[see also the nice historical introduction in][]{2010ApJ...712..692C}. \citet{1994ApJ...431..645N}, like several other authors around the same time, found that at lower metallicities the retrograde component becomes more and more prominent. In his interpretation of the data, this dual halo would consist of an accreted component \citep[a la][]{1978ApJ...225..357S}
 and a contracted halo \citep[a la ELS,][]{1962ApJ...136..748E}.

We now know from the {\it Gaia} DR2 data that what  \citet{1994ApJ...431..645N} called ``the contracted halo" is actually largely heated (proto)-thick disk. So it was indeed formed mostly in-situ, however the stars did not form during a collapse but  in a disk that was heavily dynamically perturbed during a merger, as in the simulations of \citet{2009ApJ...702.1058Z,2013MNRAS.432.3391T}. On the other hand, the accreted component is predominantly debris from Gaia-Enceladus. 
This explains why the local inner halo was for a long time considered to be slightly prograde (since it is a mix of prograde thick disk and slightly retrograde Gaia-Enceladus, and was concealed as a single component because of the large velocity errors as shown in the left panel of \textbf{Figure \ref{fig:comparison}}), and more metal-rich locally than in the outskirts. In comparison, the debris from Gaia-Enceladus dominates farther out (at several kpc away from the Sun and above the Galactic plane), implying that the outer halo consists of more metal-poor stars on retrograde orbits (at least in comparison to the proto-thick disk). This is also what \citet{2009ApJ...694..130M} had found in a carefully constructed sample of nearby metal-poor red giant branch stars with accurate distances, and what the more recent study by \citet{2017ApJ...845..101B} using {\it Gaia} DR1 data \citep{2016A&A...595A...2G} also revealed. 

\citet{2007Natur.450.1020C, 2010ApJ...712..692C} and \citet[][and references therein]{2012ApJ...746...34B}
based on SDSS and SEGUE data put the idea of a ``dual halo" on firmer ground, despite concerns that the retrograde mean motion signal was driven by biases in the distances as suggested by \citet{2011MNRAS.415.3807S}.
The {\it Gaia} DR2 data has put the debate fully to rest since the retrograde signal is very strong and is seen also in the nearby halo, where the distances have been measured very precisely and accurately with trigonometric parallaxes \citep[or with Bayesian methods,][]{2018RNAAS...2...51M}. 
There is however, still some tension with the work of \citet{2007Natur.450.1020C}, because what they called outer halo peaked at a metallicity of [Fe/H]~$\sim -2.2$, lower than what is typical for Gaia-Enceladus stars. A possible way out would be to consider that large galaxies like Gaia-Enceladus probably had a metallicity gradient. As the galaxy was destroyed, some of the material in the outer regions would be put on more unbound orbits. This material would have lower metallicity and form part of the outer portions of the inner halo (i.e. we are referring here to inner halo as that within 20 kpc from the Sun). Another possibility could be an issue with the metallicity scale calibration, or perhaps a small bias in the sample selection, which because it is based mostly on the colors for main sequence turn-off stars, might preferentially select more metal-poor stars \citep{2019ApJ...885..102L}. 

\subsubsection{More substructures in the inner halo}
\label{sec:more_subs}

\paragraph{Helmi streams} 
\label{sec:h99}
One of the first genuine accreted inner halo kinematic substructures discovered near the Sun were the Helmi streams\footnote{This naming has been used informally in workshops and conferences for years and it appears in written print in \citet{2010A&ARv..18..567K}.} \citep{1999Natur.402...53H}. Before their discovery, there had been reports of substructure \citep[e.g.][]{1994ApJ...427L..37M} and in particular, Eggen argued for the presence of streams in the halo but their reality was often questioned because of the methods used (because in some cases, for example, the very uncertain distances were enforced to match those expected for a kinematic group\footnote{It would nonetheless be an interesting exercise to revisit in a systematic fashion the Eggen moving groups with the data newly available, particularly from {\it Gaia} DR2. For example, \citet{2015ApJ...808..103N} have analysed the Kapteyn group and the chemical abundances of its stars and shown that they do not really constitute a physical unity.}). The Helmi streams were identified in a dataset largely compiled by \citet{1998AJ....115..168C}, who used the Hipparcos dataset supplemented by line of sight velocities and distances from the literature. The streams are also readily apparent in {\it Gaia} DR2 \citep{2018A&A...616A..12G,2018ApJ...860L..11K}.  A characteristic of the streams is that they have rather high $z$-velocities, which makes them easy to identify as they are well separated from the rest of the stars with the same rotational velocity ($v_\phi \sim 150$~km/s). One of the streams has positive $v_z$ while the second (and better populated) has $v_z < 0$.  \citet{2019A&A...625A...5K} have shown that the streams likely stem from a progenitor of $M_* \sim 10^8 \sm$.
Seven globular clusters have been associated to this object, including e.g. NGC~5024 and NGC~5053 \citep{2019A&A...625A...5K,2019arXiv190608271M}. These clusters follow a well-defined age-metallicity relation and their number is consistent with that expected from the globular clusters' specific frequency relation given the progenitor's mass \citep[see for example][]{2016ApJ...818...99Z}. The time of infall estimated by \citet{2019A&A...625A...5K} is between 5 and 8 Gyr ago \citep[see also][]{2007AJ....134.1579K} and it is largely driven by the asymmetry seen in the number of stars with positive and negative $v_z$. It is somewhat puzzling that an object that is not so massive would have made it to the inner halo so recently. Since its
orbit seems to lie close to a resonance (J. Hagen, PhD thesis, Univ. of Groningen, 2020), the inferred timescale could be underestimated as streams will then have spread out more slowly. Alternatively the progenitor system may have fallen in together with a heavier galaxy as part of a group \citep[as is the case for the Magellanic Clouds, e.g.][]{2017MNRAS.465.1879S}.

\begin{figure}[h]
\begin{minipage}[b]{0.48\textwidth}
\includegraphics[width=\textwidth]{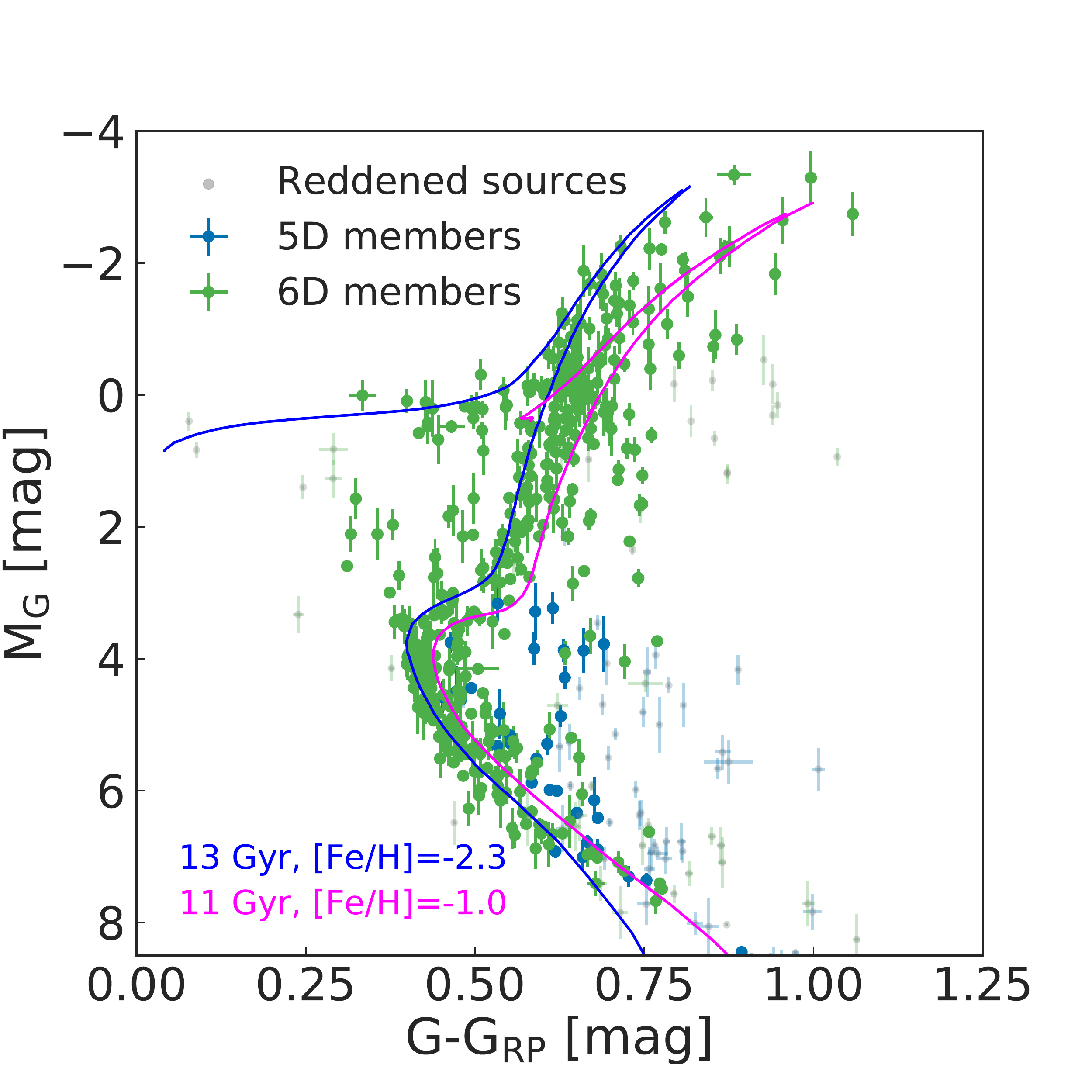}
\end{minipage}
\hfill
\begin{minipage}[b]{0.48\textwidth}
\includegraphics[width=\textwidth]{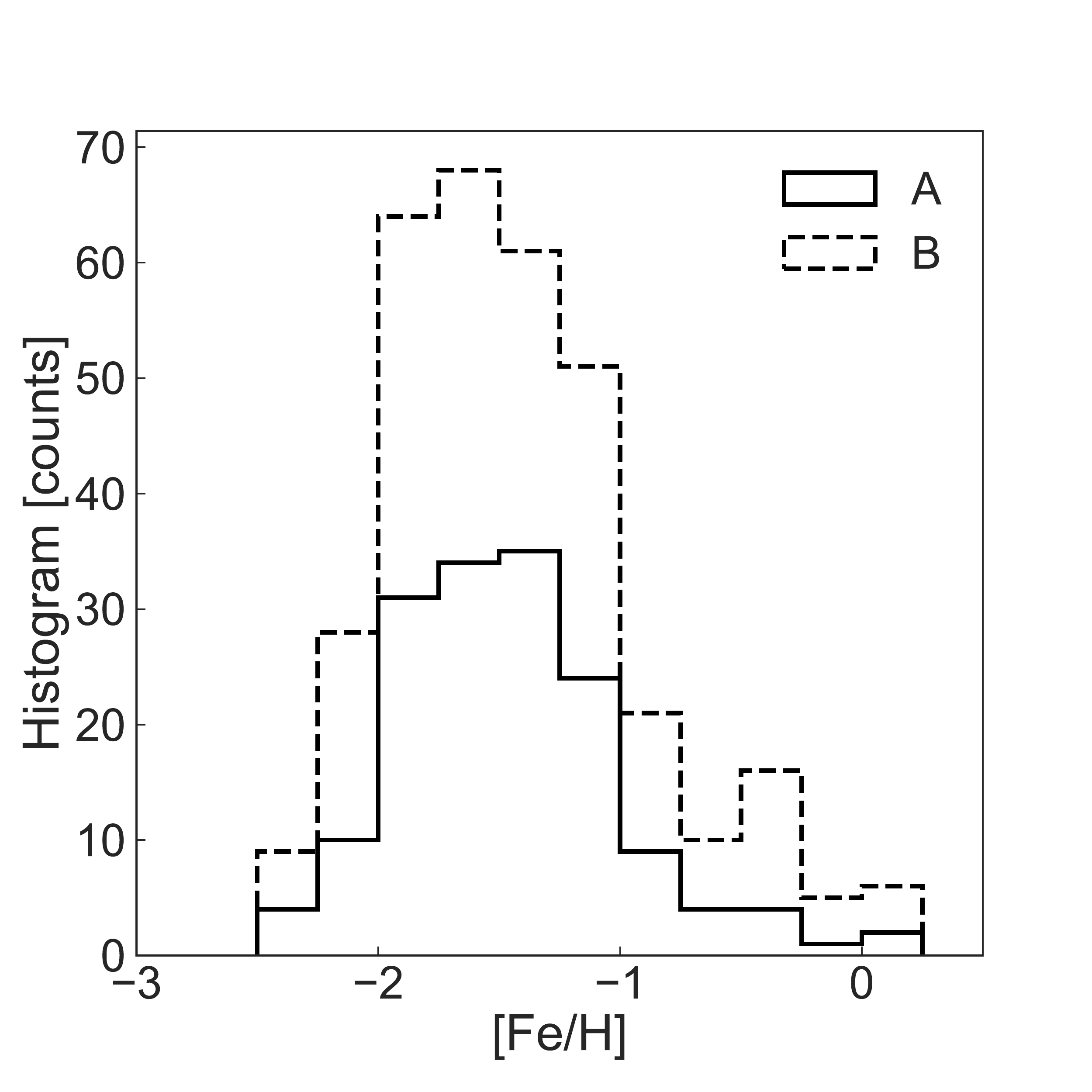}
\end{minipage}
\caption{Properties of the stars from the Helmi streams. The left panel displays the CMD of members identified in a sample with 6D coordinates (from {\it Gaia} DR2 and supplemented further by radial velocities from RAVE, APOGEE and LAMOST), in green. Members identified using the 5D sample from {\it Gaia} DR2 in fields within 15$\ndeg$ of the (anti)center are plotted in blue. Highly reddened members are indicated in light grey. The  isochrones serve to illustrate that the Helmi streams include a range of old, metal-poor stellar populations. This is confirmed by the panel on the right showing the metallicity distribution function using the APOGEE/RAVE/LAMOST datasets. The two distributions marked A/B correspond to two different selections in IoM space. Both peak at [Fe/H]~$\sim -1.5$ and reveal
a broad range of metallicities. {\it Credit: Figures reproduced from Koppelman et al., A\&A, 625, A15, 2019, with permission © ESO}.}
\label{fig:helmi-streams}
\end{figure}

There are now enough members of the Helmi streams known that is is possible to produce a well-populated CMD, which is very similar to those constructed for many decades for the dwarf spheroidal satellites of the Milky Way \citep[see e.g.][]{2009ARA&A..47..371T}  as shown in the left panel of \textbf{Figure~\ref{fig:helmi-streams}}. The information contained in this CMD should allow deriving a star formation history. In combination with the metallicity distribution function shown in the right panel of \textbf{Figure~\ref{fig:helmi-streams}}, this should enable  the reconstruction of the full chemical history of the system \citep[in a similar manner as done for several dSph in e.g.][]{2015ApJ...799..230H}. Detailed elemental chemical abundances \citep[such as those already measured by][]{2010ApJ...711..573R}, might also help in this respect \citep[see for example][]{2010A&A...512A..85L}. It is intriguing that such an exercise of reconstructing the star formation and chemical enrichment history is now possible using nearby stars and for objects long-gone. This truly lies at the heart of what it means to carry out Galactic archaeology.

\paragraph{Other reports of kinematic substructures}
Several other kinematic substructures have been reported in the literature prior to {\it Gaia} DR2. Typically though, these structures were less conspicuous as they were based on Hipparcos/Tycho data in combination with ground-based radial velocities \citep[see the reviews by e.g][]{2010A&ARv..18..567K, 2016ASSL..420..113S}. \citet{2007AJ....134.1579K} and \citet{2009ApJ...694..130M} report lumpiness in a sample of red giant branch stars with accurate distances, some of which can be attributed to Gaia-Enceladus and to the hot thick disk, but also ``outliers" on very prograde orbits. \citet{2015AJ....150..128R} report substructures among the fastest moving halo subdwarfs.
More recently other authors used LAMOST and TGAS \citep{2017ApJ...844..152L}, RAVE and TGAS \citep{2017A&A...598A..58H}, SDSS and {\it Gaia} DR1 \citep{2018MNRAS.475.1537M} and identified several small clumps, some of which were related to those previously known \citep[e.g. the S2 group in ][overlaps very substantially with the Helmi streams]{2018MNRAS.475.1537M}. Also in {\it Gaia} DR2 other small kinematic groups have been identified by \citet{2018ApJ...860L..11K,2019arXiv190702527B} and \citet{2019arXiv191007538Y}. 

\begin{figure}[h]
\includegraphics[width=\textwidth, trim={0cm 4cm 0 4cm},clip]{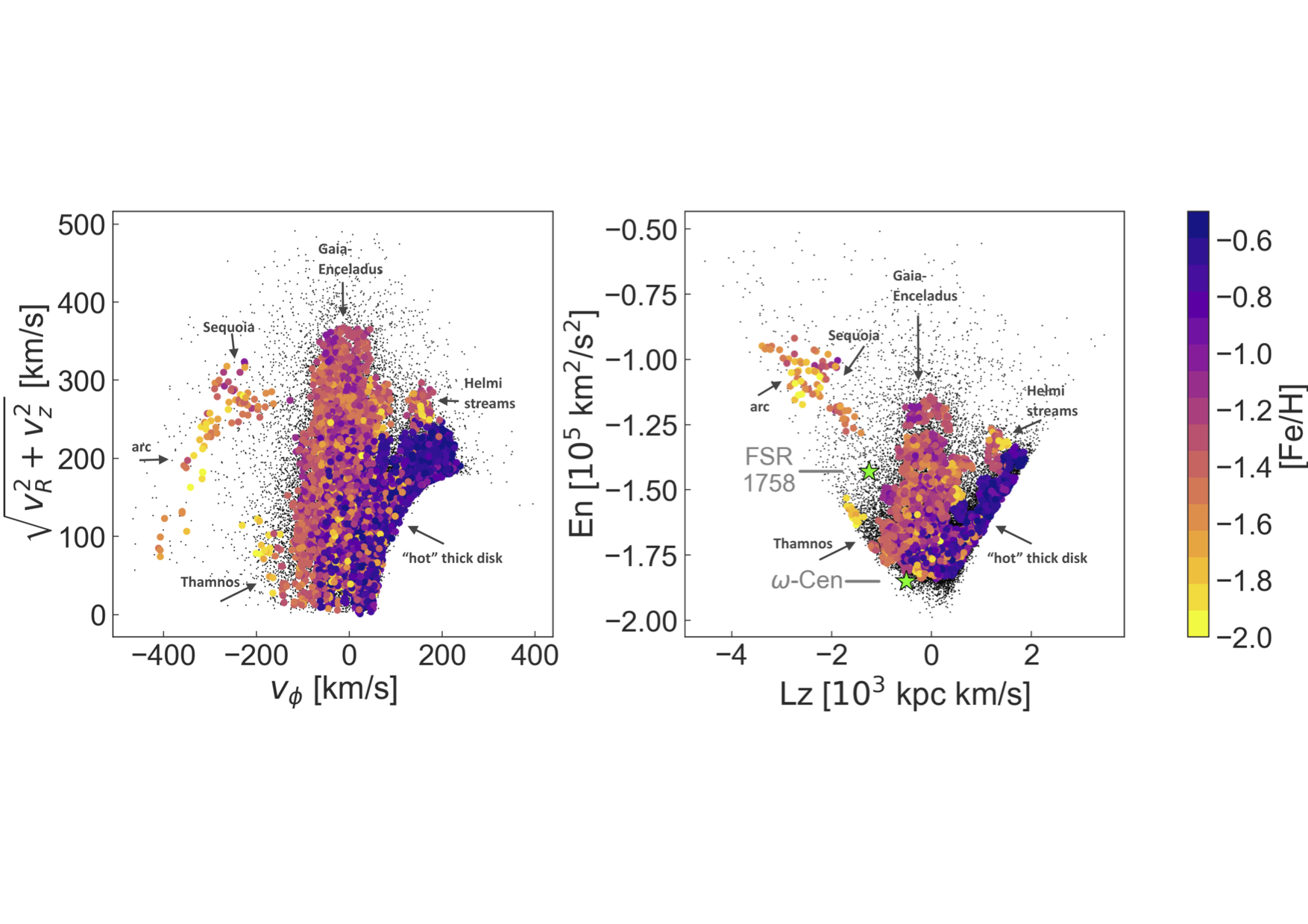}
\caption{Distribution of halo stars selected kinematically, i.e. $|{\bf V} - {\bf V}_{LSR}| > 180$ km/s, within 3 kpc and with $\varpi/\sigma_\varpi > 5$, from the {\it Gaia} RVS sample extended with radial velocities from APOGEE, RAVE, and LAMOST with metallicities from the latter.  The left panel shows their distribution in the Toomre diagram, while on the right energy $E$ and $z$-angular momentum are plotted. Stars identified by HDBSCAN to be in clumps in the space of $E$, $L_z$, eccentricity, and [Fe/H], are color-coded by [Fe/H], with the rest shown as black dots \citep[see][for details]{2019arXiv191007684C}. The arrows indicate the approximate location of the various groups discussed in the text. Comparison of their extent in $E$-$L_z$ to N-body simulations suggest that $M_\star \sim 6 \times 10^5 \sm$ for Thamnos while for Sequioa $ M_\star \sim 10^7 \sm $. {\it Credit: Figure adapted from Koppelman et al., A\&A, 631, L9, 2019, and reproduced with permission © ESO}.}
\label{fig:clumps}
\end{figure}

\paragraph{Sequoia and Thamnos}
\label{sec:mess}
Two of the clumps identified in \citet{2018ApJ...860L..11K}
have been suggested to be part of a larger substructure associated to the debris of a dwarf galaxy dubbed as ``Sequoia" \citep{2019MNRAS.488.1235M}. The associated stars have more retrograde motion and a lower metallicity than Gaia-Enceladus as can be seen from \textbf{Figure~\ref{fig:clumps}}, and they are on a less eccentric orbit. \citet{2019MNRAS.488.1235M}
suggest that Omega Cen could have been Sequoia's nuclear star cluster,  and that several other clusters, including the recently characterized large cluster FSR~1758 \citep{2018A&A...618A..93C,2019ApJ...870L..24B} would also be associated. On the other hand, \citet{2019arXiv190608271M} argues that it is more likely that Omega Cen is associated with Gaia-Enceladus, given its location in the age-metallicity diagram of the Galactic globular clusters.
As can be seen the right panel of \textbf{Figure~\ref{fig:clumps}}, the extent of Sequoia in the space of energy and $L_z$ if the two clusters would be associated to it, is as
large as that of Gaia-Enceladus, implying that it would need to have been as massive  \citep[see][for details]{2019arXiv191007684C}. This however would be in tension with its lower metallicity. 

The presence of additional (besides Gaia-Enceladus) debris with retrograde motions put forward by \citet{2019MNRAS.488.1235M} 
to a certain extent confirms the analysis by \citet{2019MNRAS.482.3426M} of a dataset that is the cross-match of APOGEE DR14 and {\it Gaia} DR2. Taking advantage of the chemical abundance information provided by APOGEE, these authors have applied a $k$-means algorithm to identify clustering using [Fe/H], [Mg/Fe], [Al/Fe] and [Ni/Fe] as well as eccentricity. \citet{2019MNRAS.482.3426M} found evidence for the presence of a group of stars with low eccentricity and {\it high} [Mg/Fe], of which roughly 50\% have (very) retrograde motions (with the rest being likely associated to the thick disk).

On the other hand, \citet{2019ApJ...874L..35M}, using chemical abundances from the SAGA database\footnote{\tt http://sagadatabase.jp/} \citep[see][and subsequent papers]{2008PASJ...60.1159S}, showed that the high-energy more retrograde stars have {\it lower} [Mg/Fe] at [Fe/H] $\lesssim -1.6$ than Gaia-Enceladus stars and especially, different [Na/Fe]. The authors further argue that the other elemental abundances are too different from those measured for Omega Cen for the cluster (which has a higher binding energy) to be related. 

Finally, \citet{2019arXiv191007684C}  report evidence that the lower energy region occupied by stars tentatively associated to 
Sequoia, is likely part of a different separate structure, which the authors name ``Thamnos" and indicated in \textbf{Figure \ref{fig:clumps}}. Thamnos stars define the separate clump to the lower left of the region occupied by Gaia-Enceladus ($v_y \sim -150$~km/s and $v_\perp < 150$~km/s). Their progenitor would be a small galaxy accreted on a low inclination orbit, and this is consistent with the lower mean metallicity of its stars. Thamnos, which in ancient greek means ``shrubs", overlaps also in part with substructures that have been reported earlier in \citet{2017A&A...598A..58H,2018ApJ...860L..11K}. The majority of the low eccentricity stars with high [Mg/Fe], low metallicity and retrograde motions identified by \citet{2019MNRAS.482.3426M} overlap with those from ``Thamnos". 

It should also be noted that \textbf{Figure \ref{fig:clumps}} reveals that some of the stars from Sequoia overlap directly with the arc-like structure assigned to Gaia-Enceladus on the basis of the resemblance to numerical simulations, as discussed before. If Gaia-Enceladus had a metallicity gradient, which might be expected given that it was of comparable size to the Large Magellanic Cloud, then one might expect the outskirts to be more metal poor and part of the material lost first, to be on a more a retrograde orbit, as found for some Sequoia stars. 

\begin{textbox}[!h]
\section{The very retrograde halo}
Put together, a possible scenario to explain the findings reported in this Section %Sec.~\ref{sec:mess} 
would be the existence of (a bonsai) Sequoia as embodied by a clump of high-energy retrograde stars that is truly distinct from Gaia-Enceladus \citep[because of the differing Na/Fe abundances reported in][]{2019ApJ...874L..35M}, but whose extent is smaller than that originally proposed by  \citet{2019MNRAS.488.1235M}. It is not inconceivable that the region of the arc-like structure contains a mix of debris from two different objects, stars from the outskirts of Gaia-Enceladus and from the Sequoia galaxy. An alternative possibility would be that we are seeing only stars from the outskirts of Gaia-Enceladus and that these followed a somewhat different chemical enrichment path. On the other hand, the lower energy region, with stars of low eccentricity and high [Mg/Fe] identified by \citet{2019MNRAS.482.3426M} and \citet{2019arXiv191007684C}, would be associated to ``Thamnos". 
\end{textbox}

The above discussion serves to demonstrate the high-state of flux of the field and that it is currently in some turmoil. It evidences that it is very difficult, given the available data and methods, to pin down the origin of the different clumps and overdensities being discovered in the halo near the Sun. This is particularly manifested in the analysis of \citet{2019arXiv191007538Y} who identified more than 50 small clumps using neural network based clustering on the very metal-poor sample of stars from LAMOST. Nonetheless, it is possible that very detailed modeling and chemical analysis of a large sample of stars will help shed light on their true nature. At the moment, the modeling is sketchy, carried out in idealized (time-independent or non-cosmological) configurations, often hydrodynamics and star formation (as well as chemical enrichment) are not followed, and the numerical resolution (especially in terms of particle number) is typically too low to do a proper comparison to the features seen in the data. Furthermore, the samples with detailed and precise chemical abundances of halo stars, even for those that are bright and nearby, are relatively small. Finally, more robust statistical, or probabilistic tools are needed to assign stars to overdensities, as these very likely have some degree of overlap in phase-space.

\subsubsection{Substructures in the outer halo}
 
At larger distances from the Sun, i.e. in the outer halo, many overdensities have been uncovered, especially with photometric wide-field surveys such as SDSS. Examples are the Sagittarius streams \citep{2000AJ....120..963I,2000ApJ...540..825Y}, Hercules Aquila Cloud \citep{2007ApJ...657L..89B} and the Virgo overdensity \citep{2004AJ....127.1158V,2008ApJ...673..864J}. At this point the nature of these high latitude structures (beyond the Sagittarius streams) and the link to those in the solar vicinity discussed in previous sections remains unknown, although \citet{2019MNRAS.482..921S}
have proposed that both Hercules-Aquila and Virgo are related to Gaia-Enceladus. To pin this down requires large samples of stars with accurate full phase-space information, some of which will become available in future {\it Gaia} data releases combined with data from large spectroscopic surveys such as WEAVE\footnote{\tt https://ingconfluence.ing.iac.es:8444/confluence//display/WEAV/The+WEAVE+Project} \citep{2016SPIE.9908E..1GD}, 4MOST\footnote{\tt https://www.4most.eu/} \citep{2019Msngr.175....3D}, DESI\footnote{\tt https://www.desi.lbl.gov/the-desi-survey/} \citep{2019BAAS...51g..57L} and smaller efforts like the H$^3$ survey of \citet{2019arXiv190707684C}. Again, orbital integrations and extensive modeling are needed to understand what is what. 

Other known overdensities located closer to the Galactic plane, are the Monoceros ring \citep{2002ApJ...569..245N}, and the feathers at slightly higher latitudes \citep[seen in SDSS and in PanSTARRS, respectively]{2011ApJ...738...98G,2016MNRAS.463.1759B}. These  low latitude structures are likely the result of the response of the disk to a massive perturber, such as Sagittarius, and perhaps also contain some satellite debris \citep{2018ApJ...854...47S,2018MNRAS.481..286L}. 

\begin{figure}[h]
\includegraphics[width=\textwidth]{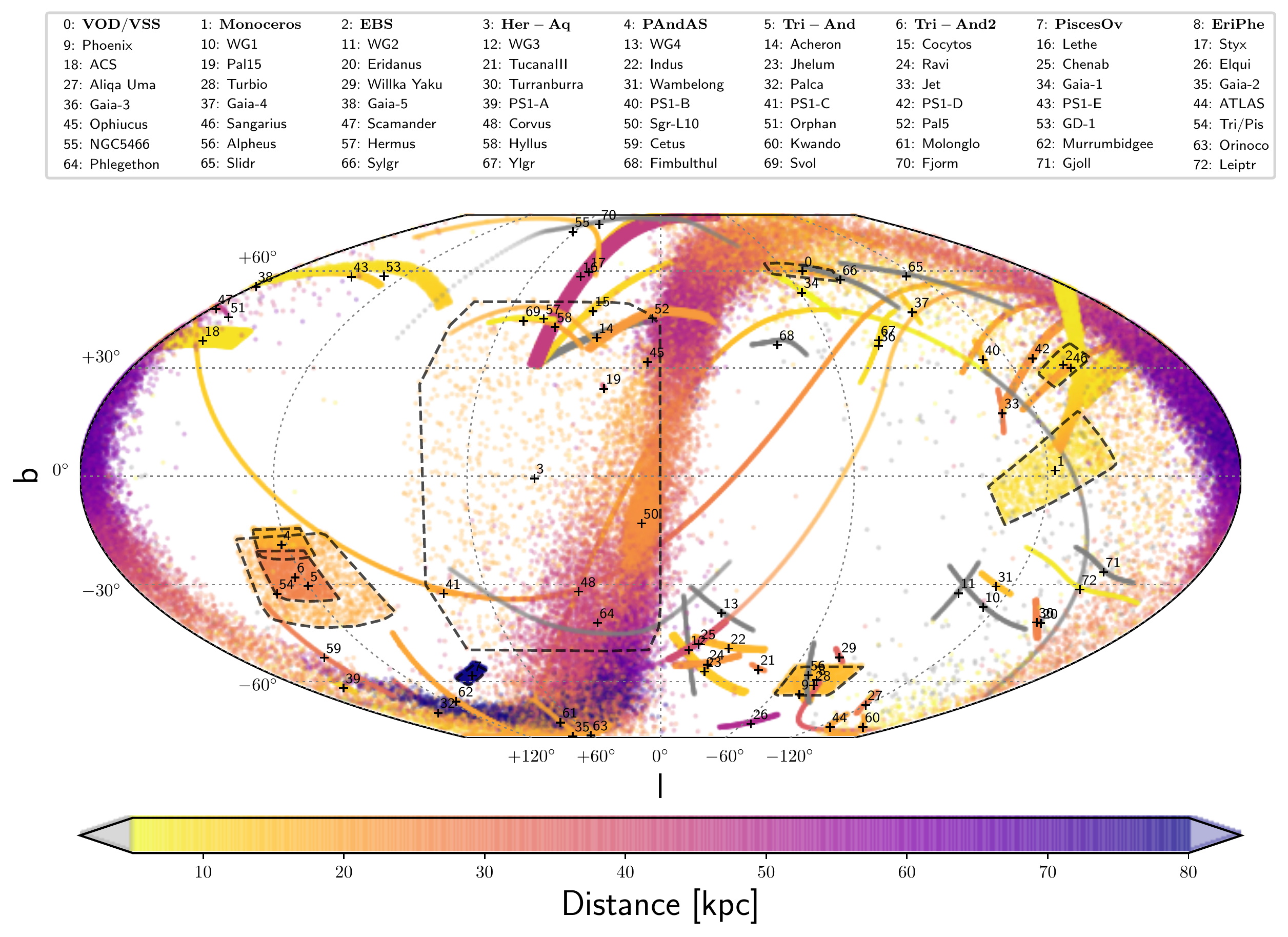}
\caption{Sky distribution of currently known spatially coherent streams and overdensities (indicated as regions delimited by dashed lines, and in boldface in the inset) produced using the {\sc GalStreams} package by \citet{2018MNRAS.474.4112M}. {\it Credits: C. Mateu and E. Balbinot}.}
\label{fig:mateu}
\end{figure}

Besides large overdensities, many thin streams crisscrossing the halo at larger distances have been uncovered \citep[see][]{2016ASSL..420...87G, 2018ApJ...862..114S}. These include for example, GD-1, Atlas, Orphan, Pal 5 and many more. \citet{2018MNRAS.474.4112M} has also identified 14 candidate streams using RR Lyrae stars and produced a very interesting and useful interface, the {\sc GalStreams} Footprint Library and Toolkit for Python\footnote{\tt https://github.com/cmateu/galstreams}, to visualize and keep track of all known structures. An example of the distribution on the sky of currently known streams and spatial substructures is shown in \textbf{Figure~\ref{fig:mateu}}. Thin streams tell us the story of the destruction of less massive objects, i.e. smaller dwarf galaxies and globular clusters, and are particularly useful for constraining the Galactic potential and the distribution of mass at large radii. An exciting development is the detection of substructure, gaps and overdensities in these streams, potentially revealing the presence of (dark) satellites orbiting the halo of the Milky Way \citep{2019ApJ...880...38B}.

The {\sc Streamfinder} algorithm developed by \citet{2018MNRAS.477.4063M} has also allowed the discovery of less distant thin streams. These include Gaia-1 and Gaia-2 \citep{2018MNRAS.481.3442M}, and the stream from Omega Cen itself \citep{2019NatAs...3..667I}. 
{\sc Streamfinder} works by randomly sampling radial velocities (which have not been measured for the majority of the stars in {\it Gaia} DR2), 
while making use of the proper motion and photometric information of stars \citep[i.e. for a given 
color there are at most 3 possible absolute magnitudes, see for details][]{2019ApJ...872..152I}. From these tentative phase-space coordinates, orbits are integrated in a Galactic potential and if stars can be found along the orbit on a ``spaghetti" or stream-like structure, then a stream is identified (after proper statistical assessment and comparison to a suitable background).  

\subsubsection{Link between inner and outer halo (sub)structures}

Several authors including  \citet{2009MNRAS.398.1757W,2013ApJ...763..113D} have reported the presence of a break in the density profile of the Galactic stellar halo, at a distance of $\sim 20-25$ kpc from the Galactic center. \citet{2018ApJ...862L...1D} demonstrated that is likely 
marking the orbital turning-points (i.e. a shell), of the debris from Gaia-Enceladus. These authors  integrated the orbits of the stars belonging to the Sausage (Gaia-Enceladus), and showed that their apocenters  lie at roughly this distance. Several shells might be expected from the debris, possibly at different distances  because the object must have experienced significant dynamical friction. This break manifests itself also in a change in the shape of the velocity ellipsoid of halo stars around this distance, as reported in \citet{2014ApJ...794...59K}.

In an attempt to make the link between the outer and the inner stellar halo even stronger, we have followed a similar approach to that of \citet{2018ApJ...862L...1D}, and have integrated the orbits of halo stars located within 1~kpc from the Sun. Since we expect the inner halo to be well mixed, these stars' trajectories should give us a broader view of the 3D (spatial) structure of the halo for example. This is much in the same way as orbits can be seen as the building blocks of a galaxy in Schwarzschild's modeling \citep{1979ApJ...232..236S} and can be used to reproduce, for example, their light profile. 

\begin{figure}[h]
\includegraphics[width=0.82\textwidth,trim={-2cm 2cm 0cm 12cm},clip]{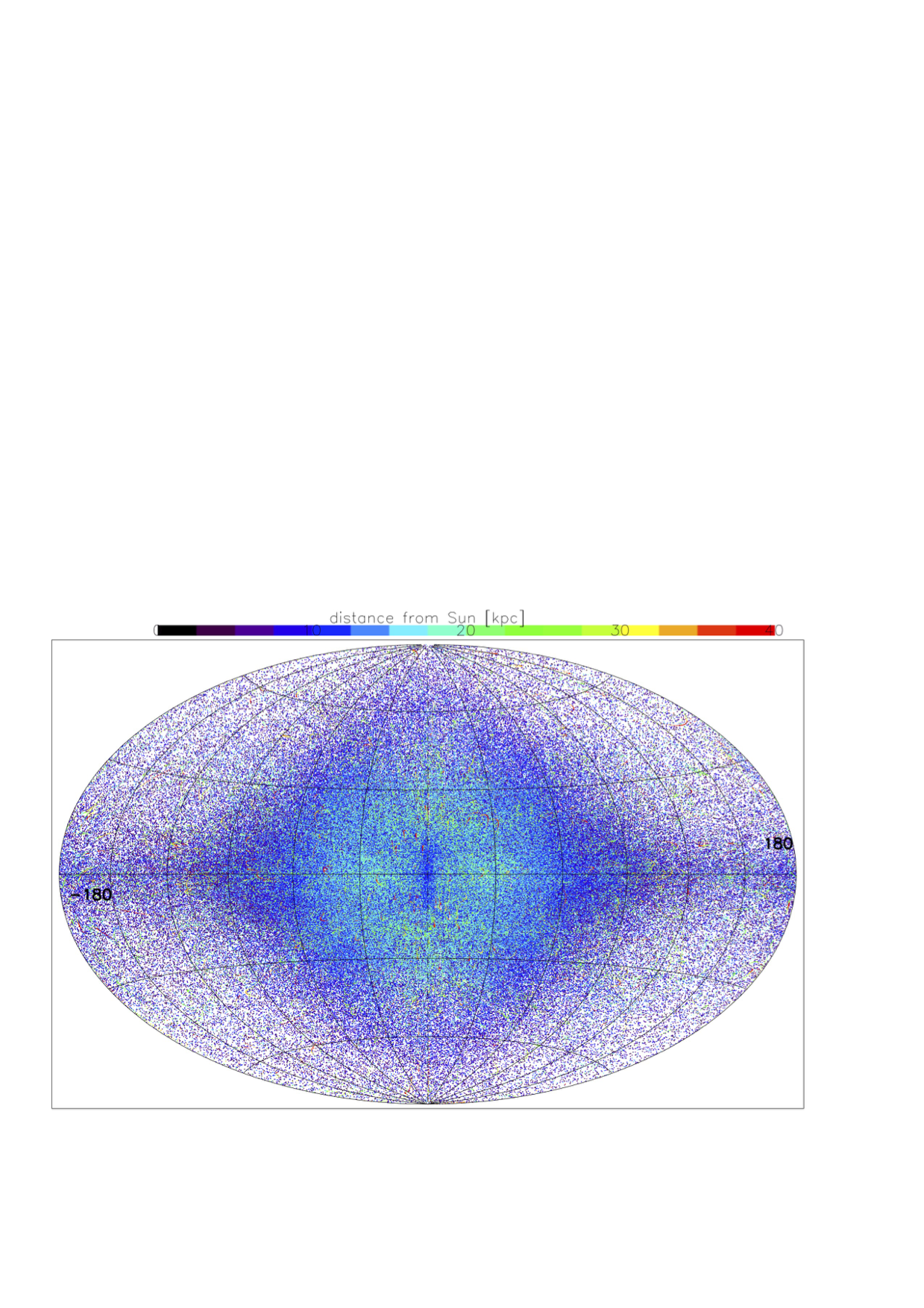}
\includegraphics[width=0.82\textwidth]{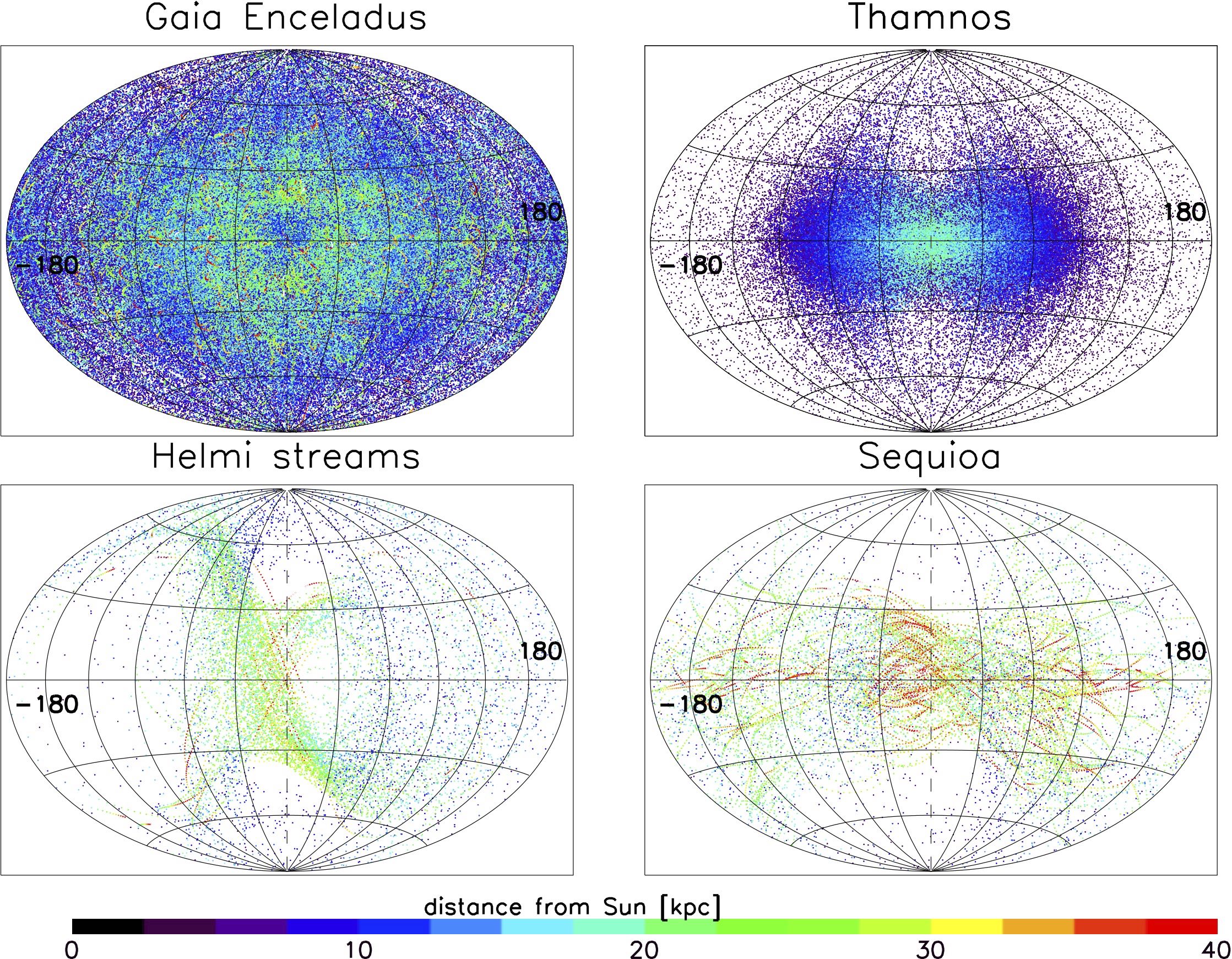}
\caption{Sky projections of the orbits of stars with halo-like kinematics currently located within 1 kpc from the Sun (and with $|L_z| \ge 10$~kpc~km/s). The orbits have been integrated in a Milky Way-like potential for 2 Gyr, and each point in the diagram corresponds to a point on a star's orbit color-coded by its distance. In the top panel we show all the stars together, while in the four lower subpanels we have separated the stars in the structures identified thus far following the assignment of \citet{2019arXiv191007684C}, which is similar to that schematically shown in \textbf{Figure~\ref{fig:clumps}}. The decrease in the density of points towards the Galactic center is not physical and is due to removal of stars ($\lesssim 1.3\%$) with very radial orbits of low $|L_z|$ angular momenta .}  %%5946 to start, 5870 kept.
\label{fig:orbits-sky}
\end{figure}

\begin{figure}[h]
\hspace*{0.35cm}\includegraphics[width=0.6\textwidth]{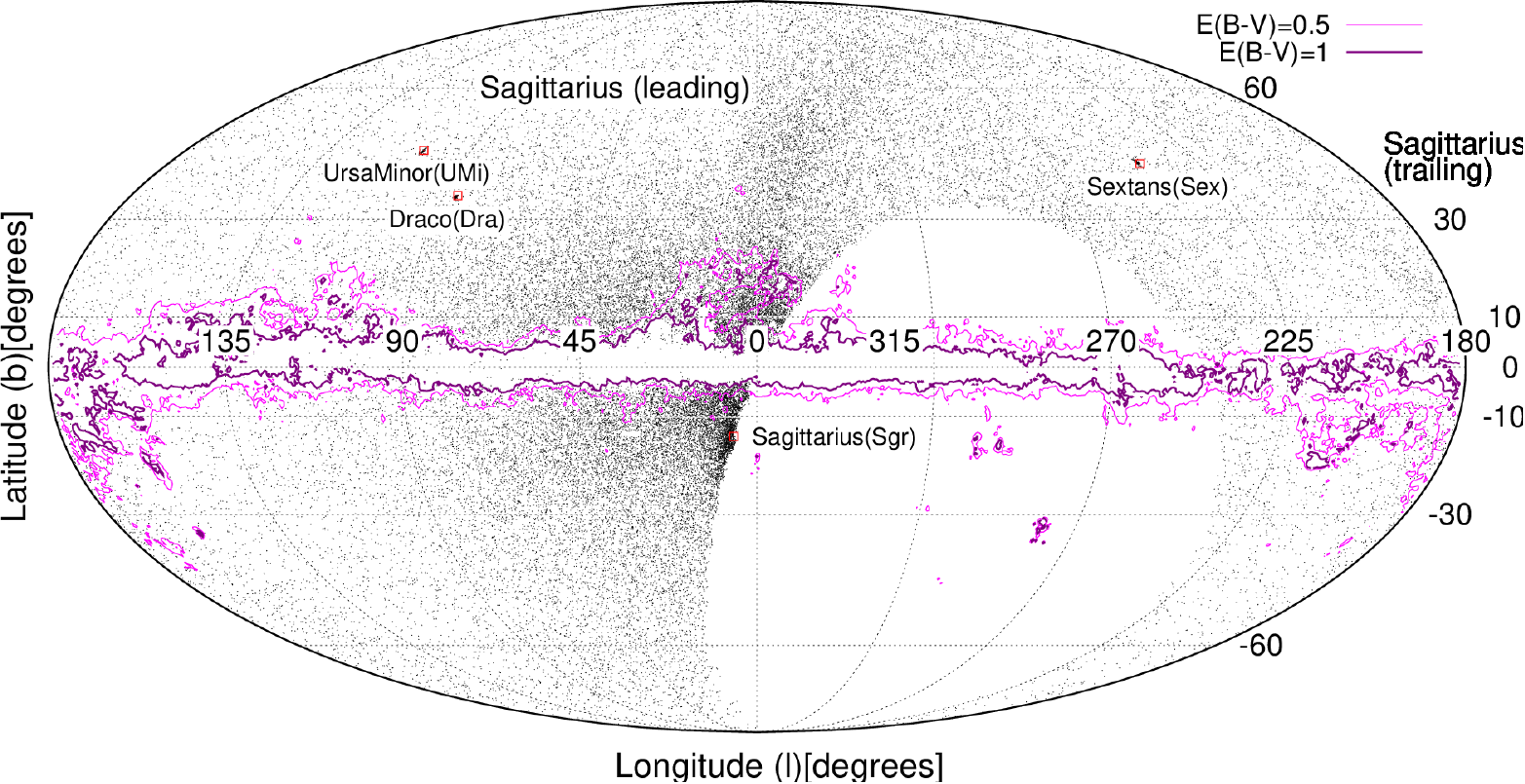}
\hspace*{0.75cm}\includegraphics[width=0.75\textwidth]{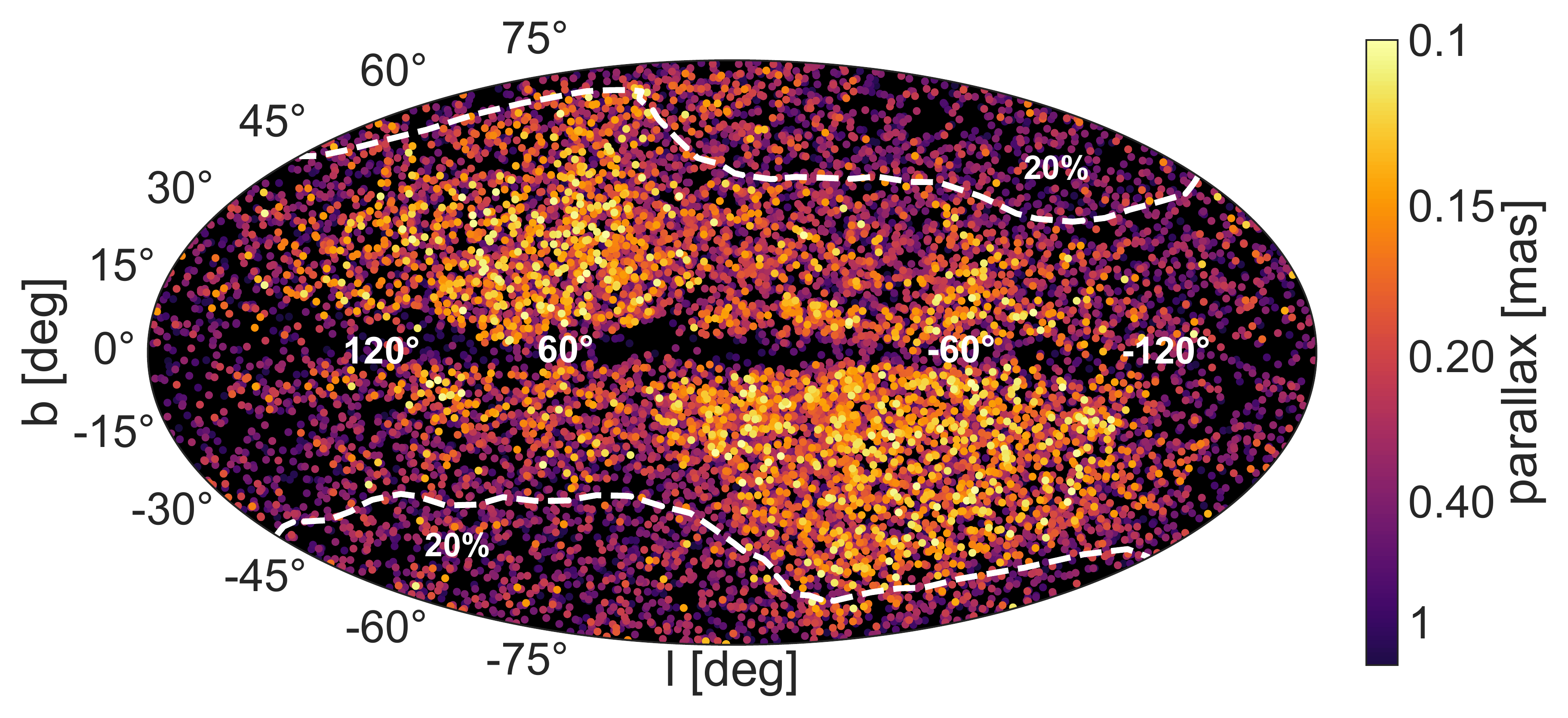}
\includegraphics[width=0.7\textwidth]{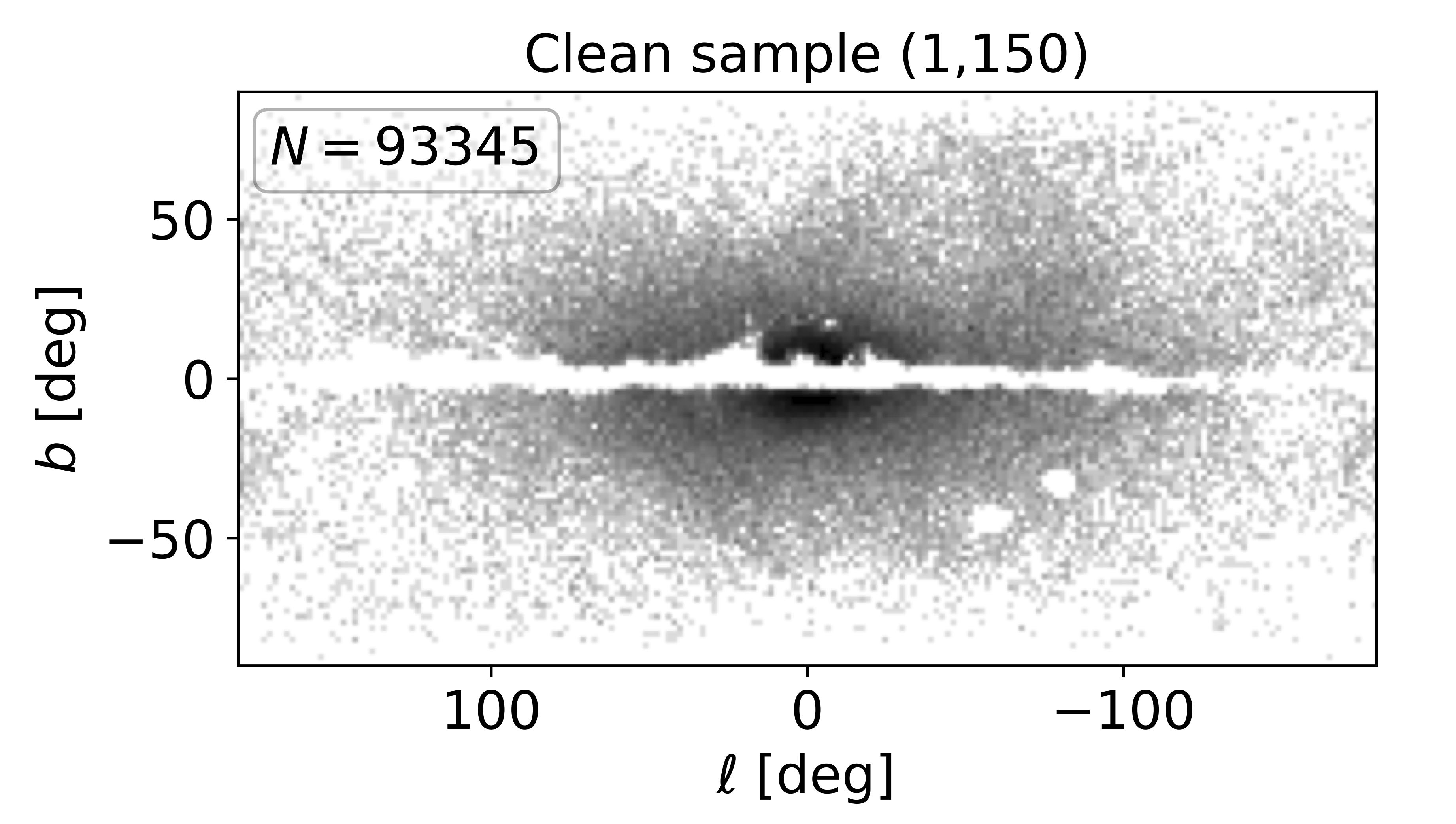}
%\hspace*{0.35cm}\includegraphics[width=0.625\textwidth]{figures/RRLyrae_mollweide_3pi_finalcandidates.pdf}
%\hspace*{0.75cm}\includegraphics[width=0.775\textwidth]{figures/Gaia-Enceladus-on-sky-parallax-20p-effect}
%\includegraphics[width=0.725\textwidth]{figures/iorio-allsky_square_final}
\caption{Top panel: Distribution of RR Lyrae stars from PanSTARRS. Middle panel: Distribution of 
potential members of Gaia-Enceladus from {\it Gaia} DR2 (with $-1500 < L_z < 150$ km/s kpc) and with parallaxes $> 0.1$~mas. The color coding indicates parallax, with more distant stars in yellow. The white contour encompasses 90\% of {\it all} stars in the 6D {\it Gaia} DR2 sample with $0.1 < \varpi < 0.2$~mas and 20\% relative parallax error. Tentative members of Gaia-Enceladus follow a similar distribution on the sky as all
the distant stars implying that the parallax quality cut introduces a selection bias: distant stars outside of these contours could be missed because these regions of the sky have been visited less frequently by {\it Gaia}. 
Bottom panel: Distribution of RR Lyrae from {\it Gaia} DR2, after removal of the Sagittarius streams and other satellites of the Milky Way. A similar selection effect is apparent in this figure. 
{\it Credits: Top panel: Reproduced with permission from \citet{2017AJ....153..204S} and from the AAS; Middle panel: Figure adapted by H.H. Koppelman and this author from 
\citet{2018Natur.563...85H}, their Figs.~3 and 4. Bottom panel:  Reproduced from \citet{2019MNRAS.482.3868I}, top right panel of their Fig.~1.} } 
\label{fig:rrlyr}
\end{figure}

The top panel of \textbf{Figure~\ref{fig:orbits-sky}} shows the orbits of all stars from {\it Gaia} DR2, located within 1 kpc from the Sun, with halo-like kinematics, defined as having $|{\bf V} - {\bf V}_{\rm LSR}| >$ 210 km/s. In this case, we have used the distances from \citet{2018RNAAS...2...51M}. We have used their current positions and velocities (corrected for the Solar motion and the Local Standard of Rest) as initial conditions for the orbital integrations. These were performed in the MilkyWay Gala potential \citep{gala,Price-Whelan:2017}, which contains an NFW halo, a nucleus, and a disk and bulge \citep{2015ApJS..216...29B}. The integration covers 2 Gyr. We have plotted each orbit sampled in 0.1 Gyr intervals, color-coded by their distance from the Sun. The top panel of \textbf{Figure~\ref{fig:orbits-sky}} reveals the presence of many overdensities, boxy shapes and some sharp edges, some of which bear some resemblance to those seen in the distribution of RR Lyrae in PanSTARRS by \citet{2017AJ....153..204S}, and shown in the top panel of \textbf{Figure~\ref{fig:rrlyr}}. 

%!The Sun is placed at (x,y,z)=(-8.3,0,0.014) kpc, where the Galactic Center is at (0,0,0) kpc.
%!The Sun is moving with (vx,vy,vz) = (11.1, 252.24, 7.25) km/s
%!The LSR moves with vy(x=-Rsun, y=0, z=0)=240 km/s [=-vphi(R=Rsun,phi,z=0)], i.e. V=12.24 km/s.
% vtoomre_cut = 210.
% within 2.5 kpc
 
 %% the parameters of the potential used for the integration are:
 % HQ bulge: M = 5 x 10^9, c = 1 kpc
 % HQ nucleus: M = 1.71 x 10^9, c = 0.07
 % MN disk (as in Bovy): M = 6.8 x 10^10, a = 3, b = 0.28
 % NFW: Ms = 1.043 x 10^{11}, rs = 15.62

The bottom four panels of  \textbf{Figure~\ref{fig:orbits-sky}} show the orbits of stars now separated by the progenitor they are presumably associated with according to \citet{2019arXiv191007684C} (i.e. roughly following what is shown in \textbf{Figure \ref{fig:clumps}}). 
Comparison of the panels reveals that these four objects do indeed have different orbital properties as they have deposited their debris in different spatial configurations. For example, the debris from Gaia-Enceladus follows a symmetric configuration with respect to $b = 0\ndeg$, with sharp boundaries at $l \sim \pm 120\ndeg$.  
The limited extent in longitude is similar to that reported in \citet{2018Natur.563...85H} and shown in the middle panel of \textbf{Figure~\ref{fig:rrlyr}}. This map shows the sky distribution of stars selected to be part of Gaia-Enceladus (with relative parallax errors $< 20$\% and with a rather generous cut on $L_z$ towards the retrograde halo, and a sharper more conservative one towards prograde moving stars to avoid contamination from the hot thick disk). The remarkable difference with the sky map obtained via the orbit integration is the degree of incompleteness and the effect of selection biases particularly for distant stars (with $\varpi \lesssim 0.2$~mas) that appear as a result of the imposed parallax quality cut.
Incompleteness also affects the RR Lyrae map of \citet{2019MNRAS.482.3868I} shown in the bottom panel of the same figure, although in this case it is due to poorer time-sampling and hence characterization of their light curves in {\it Gaia} DR2. This explains the significant asymmetries with respect to the Galactic plane seen in middle and bottom panels of  \textbf{Figure~\ref{fig:rrlyr}}, which would not be 
consistent with the estimated time of accretion and the size of the progenitor, as its debris is expected to have fully phase-mixed. This is indeed what the velocities of nearby stars from Gaia-Enceladus predict, as shown in \textbf{Figure~\ref{fig:orbits-sky}} (as pointed out by H-W. Rix, private communication). On the other hand the distribution in longitude revealed in the studies of \citet{2018Natur.563...85H}, \citet{2019MNRAS.482.3868I} and also in \citet{2017AJ....153..204S} are not too dissimilar to that predicted for Gaia-Enceladus as seen in \textbf{Figure~\ref{fig:orbits-sky}}. They all reveal a very centrally concentrated distribution of the debris.  

\textbf{Figure~\ref{fig:orbits-sky}} could also serve as guidance in searches for associated debris particularly at large radii. Furthermore, 
since the exact location of the features is sensitive to the gravitational potential, a comparison between the outcome of the orbit integrations and observational data (position in the sky, distance and kinematics of the various features) could be used to constrain better the distribution of mass in the Milky Way. Some of the features may even link to the overdensities discussed in the literature and already mapped at larger distances from the Sun, as a coarse comparison to \textbf{Figure \ref{fig:mateu}} suggests. 

\section{The thick/early disk}
\label{sec:thick-disc}

Not so many review articles have been written on the Galactic thick disk \citep[but a good starting point is the introduction of][]{2014A&A...569A..13R}. This is likely because its reality (independent of that of the thin disk) has been highly debated throughout the years \citep{1983MNRAS.202.1025G,1984ApJS...55...67B}
and also recently \citep{2011MNRAS.414.2893F,2012ApJ...751..131B}. 
Another likely reason is that sometimes conflicting answers regarding its properties have been obtained depending on the type of observational tool used to characterize its properties \citep[e.g abundances, kinematics, star counts, see for example,][]{2010ApJ...714..663D,2012ApJ...752...51C}. 
This is discussed in \citet{2016AN....337..976K} and an insightful explanation is given in \citet{2015ApJ...804L...9M}. 
We will not attempt to provide here a review but we will mostly focus on some observational facts and on recent discoveries, particularly in relation to the stellar halo, which help us understand at least in part the formation or evolutionary history of this component. 

\subsection{Overview of its properties}

The thick disk was discovered through star counts by \citet{1983MNRAS.202.1025G}. These authors found an excess of stars at large heights above the plane, beyond what would be expected from a single exponential fit corresponding to the thin disk. The excess could be fit by invoking a second component following also an exponential functional form, but with a larger scale height. Subsequent work revealed that the stars in the thick disk had different kinematics, which although mostly rotationally supported, 
have lower rotational speeds 
(by about 30-50 km/s) and  higher velocity dispersions than the thin disk. 
The first spectroscopic studies showed the thick disk to be more metal-poor than the thin disk and to be composed of stars which were older, see e.g.~Sec.~4 of the extensive review by \citet{1989ARA&A..27..555G}.

More detailed high-resolution chemical elemental abundance studies demonstrated that thick disk stars organize themselves in a segregated sequence from that of the thin disk stars in the solar neighborhood, for example in [$\alpha$/Fe] vs [Fe/H] \citep{1995AJ....109.1095G}. Several authors have recently provided definitive evidence that the sequences are truly separate, and hence that the two components are really physically distinct, as they are made up of stars that do not overlap in their properties \citep[e.g.][]{2011A&A...535L..11A,2014A&A...567A...5R,2015ApJ...808..132H}. 
\citet{2013A&A...560A.109H} also showed that the stars in the thin and thick disks follow very tight and well defined tracks in [$\alpha$/Fe] and [Fe/H] with age, with a break occurring at  $\sim 8$--9 Gyr, which marks the oldest stars present in the thin disk. 
These distributions display small scatter, a result that although based on a local sample can be extended beyond the solar vicinity since the orbits of the stars probe a relatively large radial range (from 2--10 kpc from the Galactic center). 
This small scatter (which implies no radial gradient) can be explained if the (majority of) thick disk stars formed rather quickly in a massive gaseous disk, possibly supported by turbulence \citep{2014ApJ...781L..31S}.

\begin{figure}[h]
\includegraphics[width=0.8\textwidth]{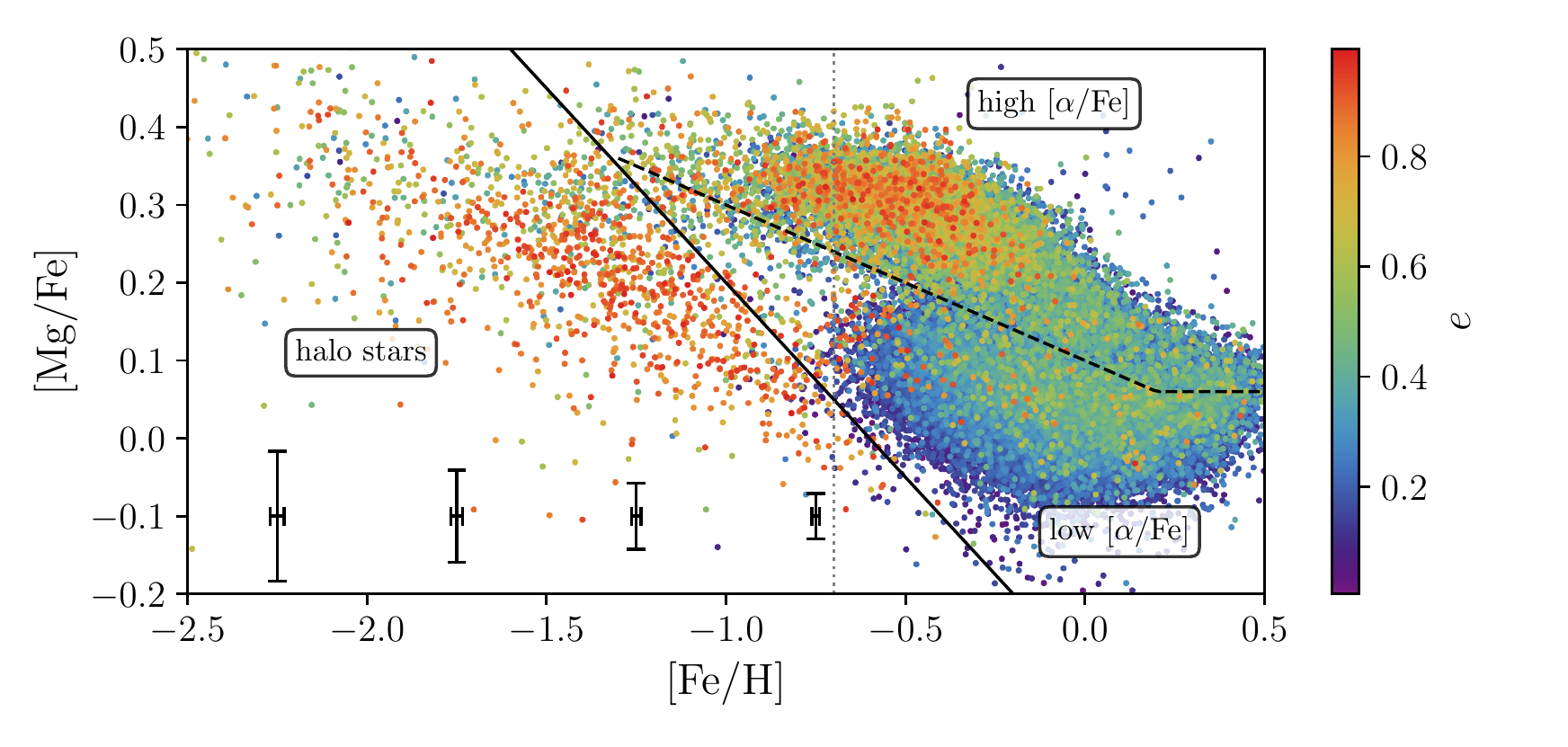}
\caption{Distribution of [Mg/Fe] vs [Fe/H] color-coded by eccentricity, using APOGEE and {\it Gaia} data.  Notice the presence of low eccentricity stars (in blue/green colors) for [Fe/H] $\lesssim -1$, many of which appear to be following a well-defined sequence that appears to be the extension of the traditional thick disk towards lower metallicity. Note as well the increase in scatter in [Mg/Fe] for [Fe/H]~$\lesssim -1.75$. The vertical line indicates the highest metallicity considered for the sample used to perform a $k$-means analysis that reveals the presence of a high and a low eccentricity populations amongst halo stars as discussed in Sec.~\ref{sec:mess}. {\it Credit: Reproduced from \citet{2019MNRAS.482.3426M}, their Fig.~1}.} 
\label{fig:thickd-ecc}
\end{figure}

The thick disk metallicity near the Sun peaks at [Fe/H]~$\sim -0.5$, and extends on the metal-rich side up to solar metallicity. It also has a very significant tail, which is often referred to as the metal-weak thick disk \citep{1985ApJS...58..463N,1990AJ....100.1191M,1993AJ....105..539M,2002AJ....124..931B}. 
The stars associated to this tail are clearly visible as the blue/green data points with high [Mg/Fe] and [Fe/H]~$\lesssim -1$
in \textbf{Figure~\ref{fig:thickd-ecc}}, which is based on APOGEE data \citep{2019MNRAS.482.3426M}. This metal-weak thick disk could potentially be related to the very first disk or the oldest disk that was ever formed in the proto-Milky Way. 

\subsection{Formation paths}
\label{sec:disc-formation}

Typically four different scenarios are discussed in the literature for the formation of the thick disk \citep{1989ARA&A..27..555G,2014A&A...569A..13R}. The traditional/oldest is {\it via a minor merger} onto a pre-existing disk, which leads to dynamical heating and the formation of a hotter but still rotation-supported component \citep{1993ApJ...403...74Q}. 
The {\it accretion} scenario is based on cosmological simulations which showed that if satellites are preferentially accreted from specific directions, this can lead to their debris being deposited in a planar configuration \citep{2003ApJ...597...21A}. On this preferred plane, gas would later cool down and form the thin disk. 
The {\it gas-rich} scenario is inspired in cosmological hydrodynamical simulations that show that disks were highly turbulent and hotter in the past, partly because they were more gas rich and also because of the ongoing merger activity that prevented full settling \citep{2004ApJ...612..894B,2013ApJ...773...43B}. 
This is also what observations of high redshift disks appear to suggest \citep[e.g.][]{2007ApJ...670..237B}. 
A last scenario is that of {\it migration}, that is stars from the inner (thin) disk have migrated with time to the outer regions of the disk. Because of inside-out formation and metallicity gradients, these stars would be older and have different chemical composition. \citet{2009MNRAS.399.1145S} were the first advocates of this model that have quantitatively explored its feasibility.

\citet{2009MNRAS.400L..61S} proposed that one way to determine the dominant formation mechanism of the thick disk would be from the eccentricity distribution of its stars. These authors showed that the different paths discussed above led to different distributions, with radial migration changing only slightly the low eccentricities of the stars. On the other hand, a dry large minor merger would leave behind a distribution of stars with intermediate eccentricity (the heated disk) and a high eccentricity bump formed mostly by accreted stars. A comparison to data from RAVE and SDSS carried out later by \citet{2011MNRAS.413.2235W} and by \citet{2010ApJ...725L.186D} showed that the most likely path was through gas rich mergers, i.e. turbulent disks in which stars were forming during mergers \citep{2004ApJ...612..894B}. This interpretation and idea has been largely confirmed by the latest analyses based on {\it Gaia} DR2. As \citet{2019NatAs.tmp..407G} 
have shown, the majority of the thick disk stars likely formed after/during the merger with Gaia-Enceladus, and not before. 
However, some fraction did form before, as in the dry manager scenario, 
although the predicted bump with high eccentricities associated to the accreted stars  \citep{2009MNRAS.400L..61S,2011A&A...525L...3D} is not seen in the thick disk. In fact, these stars exist but now we know they make up a large fraction of the Galactic halo (i.e. this is Gaia-Enceladus debris). It is interesting that the connection between thick disk and halo had not been fully made until recently \citep[although see][who using numerical simulations discussed this possibility]{2010MNRAS.404.1711P}. 

\begin{figure}[h]
\includegraphics[width=0.9\textwidth]{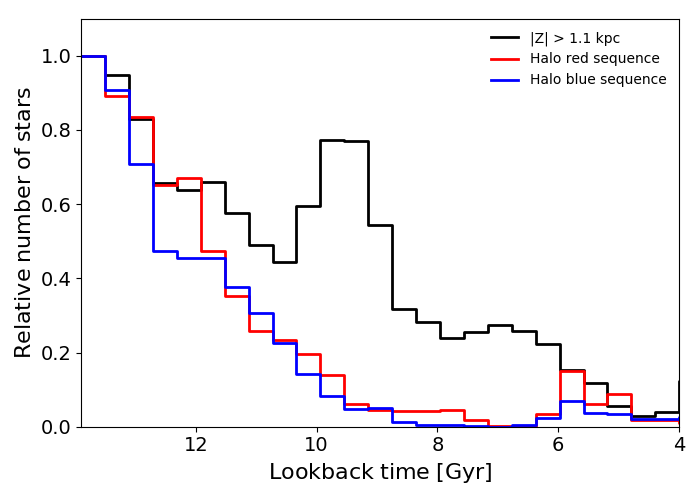}
\caption{Age distribution for stars with $|Z| > 1$ kpc and $V_T < 200$~km/s (i.e. the thick disk, in black) and for stars on the red and blue ``halo'' sequences (with $b > 30\deg$ and $V_T > 200$~km/s) as derived from the analysis of {\it Gaia} DR2 photometric data. The stars on the blue and red sequences, which correspond to Gaia-Enceladus and to the ``hot" thick disk respectively, are both old but have different colour because of their different metallicity. The thick disk ``proper" (in black) is younger and more metal-rich.  This figure reveals agreement with earlier work showing that the youngest stars in Gaia-Enceladus are 9--10 Gyr old (the tail for younger ages is likely contamination), and also shows that a starburst in the thick disk appears to have been triggered 10~Gyr ago. Probably this is the time of the closest encounter between the two interacting systems, after which Gaia-Enceladus was fully engulfed by the Milky Way. {\it Credits:  Figure adapted by T. Ruiz Lara and C. Gallart from \citet{2019NatAs.tmp..407G}, left panel of their Fig.~2}.}
\label{fig:thick-disc}
\end{figure}

Although it is probable that radial migration has played some role in the evolution of the thick disk, and that some fraction of the stars in the thick disk have an (inner) thin disk origin \citep[see e.g.][]{2011A&A...535L..11A}, it is now likely that the efficiency of this process was initially overestimated \citep[as argued by][]{2012A&A...548A.127M}. 
The evidence discussed above, and particularly the work of \citet{2019NatAs.tmp..407G} supports a scenario in which a gas-rich disk experienced a merger with Gaia-Enceladus, where the stars already present were dynamically heated, and star formation was triggered (possibly in a starburst) leading to the formation of the bulk of the stars in the thick disk. This interpretation is based on what is shown in \textbf{Figure~\ref{fig:thick-disc}}. This figure reveals that the distribution of ages of stars in the thick disk ``proper" (in black) peaks at $\sim 10$~Gyr, while the stars in the ``hot" thick disk \citep[the red sequence in the top panel of \textbf{Figure~\ref{fig:gaia-halo}} revealed by][]{2018A&A...616A..10G}, have similar older ages as those in Gaia-Enceladus (i.e. the blue sequence). 

This scenario is also consistent with the rather uniform distribution of [$\alpha$/Fe] with age and radius discussed in \citet{2015A&A...579A...5H}, as such a merger likely triggered a global response of the whole disk. Although there was probably a large amount of radial migration during the merger as the disk became dynamically hotter, this migration (only ``blurring", no ``churning") would not have been due to internal mechanisms as proposed in \citet{2009MNRAS.399.1145S}, but externally induced.

\subsection{Further insights on the early disk from chemistry and dynamics}
\label{sec:early-disc}

The evidence accumulated so far and discussed above suggests that we may have identified the reality of the proto-disk as the metal-weak (or ``hot") thick disk \citep[see also][]{2019ApJ...887...22C}, and that this was present more than 10 Gyr ago. It is interesting that relatively low eccentricity stars have survived as such the dynamical impact of a massive minor merger\footnote{Although the simulations of \citet{2008MNRAS.391.1806V} even show that a thin disk-like component with 15\%-25\% is present after the merger is completed.}, allowing this disk to 
be traced back to lower and lower metallicities as \textbf{Figure~\ref{fig:thickd-ecc}} shows. In fact in recent work, \citet{2019MNRAS.484.2166S} on ultra metal-poor stars (with [Fe/H]$ < -4$) have shown that a fair fraction of these (26\%) actually are rotationally supported and have thick disk-like kinematics. 
These stars would potentially be tracing the most pristine disk in the Milky Way. It will be interesting to bridge the gap in metallicity between the metal-weak thick disk and the regime probed by the most metal-poor stars to trace the history of the very first disk-like component in our Milky Way. This is particularly relevant in the context of linking the Milky Way to studies of high-redshift disks \citep[e.g.][]{2013ApJ...771L..35V,2014ApJ...789L..30L,2019MNRAS.490.3196P}. 

Just like we have done for the halo, we can now put previous work on the thick disk in the recently gained context. For example,  \citet{2002ApJ...574L..39G} 
discovered an excess of stars towards the rotation fields ($l \sim 90^{\rm o}, 270^{\rm o}$), with lags of approximately 100 km/s (as well as a minor contribution from a retrograde component). They argue this is due to ``shear" in the kinematics of the thick disk such that at higher latitudes ($b \sim 33\ndeg, 45\ndeg$), thick disk stars rotate more slowly than near the plane. They interpreted this as evidence for a shredded satellite, but instead (given the evidence we have discussed so far) it is likely they were seeing ``kicked-out" thick disk stars, and a bit of debris from Gaia-Enceladus. They were nonetheless ``Deciphering the Last Major Invasion of the Milky Way" as the title of their paper suggested. 
Further analysis of \citet{2006ApJ...639L..13W} and \citet{2013A&A...555A..12K}
towards other lines of sight confirmed their first results. Other evidence hinting at the dynamical consequences of a significant merger on the early disk, are the overdensities discovered by \citet{1996ApJ...468L..99L,2011AJ....141..130L} suggesting that the thick disk may be triaxial. Such a configuration is not an uncommon end-product of simulations of disks experiencing a massive minor merger \citep{2008MNRAS.391.1806V}. 
In these simulations this shape is delineated only by some of the stars already present at the time of the merger, which as \citet{2019NatAs.tmp..407G} argue, do not comprise the majority of the thick disk of the Milky Way. 

Other evidence of ``substructure" in the thick disk was put forward by \citet{2006A&A...445..939S}, where two different groups of stars in the thick disk with different mean metallicities and mean rotational velocity were identified. This is in fact one of the key papers preceding the \citet{2010A&A...511L..10N}
discovery of the two sequences, since what \citet{2006A&A...445..939S} were seeing was in fact stars from Gaia-Enceladus and from the thick disk. 
Analysis of the Geneva-Copenhagen survey \citep{2004A&A...418..989N} led \citet{2006MNRAS.365.1309H} 
to also propose the presence of substructure in the region kinematically dominated by thick disk stars. What these authors demonstrate in a follow up paper \citep{2014ApJ...791..135H} is that there is a transition in the dynamical properties of stars at a metallicity of [Fe/H]~$\sim -0.4$. Below this value stars have a large range of eccentricities, while above it stars are only on low eccentricity orbits. There is also significantly more scatter in [$\alpha$/Fe] below this [Fe/H] value, as if there were a mix of populations \citep[as in fact shown by][and reproduced here in \textbf{Figure~\ref{fig:thickd-ecc}}]{2019MNRAS.482.3426M}. 
Bearing in mind differences in metallicity scales, this is the [Fe/H] at which a clear distinction can be made between stars formed in the thick disk before/after the merger with Gaia-Enceladus \citep{2019NatAs.tmp..407G}. Only the thick disk stars below this value (i.e. more metal-poor) have been kicked-out on to more extreme orbits \citep[and there may even be some contamination from Gaia-Enceladus, see][]{2018arXiv181208232D}. Meanwhile, stars with higher metallicities formed and have stayed on (proper) thick disk-like orbits. 
This was not the original interpretation given by \citet{2014ApJ...791..135H} \citep[and follow up papers such as][]{2012A&A...541A.157S,2013A&A...555A...6S,2015A&A...576A.113Z} where the features were attributed to the presence of merger debris, whereas we now believe the latter is a minor contributor and what is seen is simply the imprint of an important transition in the history of the disk 
\citep[similar conclusions using different orbital parameters were in fact reached earlier by][]{2012MNRAS.425.2144L}.

This is an important point as there has been some propensity to attribute substructure or overdensities to accretion events. As vehemently argued by \citet{2017A&A...604A.106J}
this is not necessarily the case. A merger can induce asymmetries and substructure also in the populations formed in-situ \citep[as shown already in e.g.][for the thick disk simulations of \citealt{2008MNRAS.391.1806V}]{2010MNRAS.401.2285G,2012MNRAS.419.2163G}. On the other hand, 
asymmetries and substructures can also arise from internal dynamical processes by such as resonances with, for example, the Galactic bar and which is responsible for the Hercules stream \citep{2000AJ....119..800D}. 
Also stars in the thick disk are affected by the bar, as shown in \citet{2015ApJ...800L..32A}, and as evidenced for example in the left panel of \textbf{Figure \ref{fig:vels}}, where the $v_R$ velocity distribution of the (hot) thick disk is clearly asymmetric in the same way as the thin disk, whose asymmetry is explained as being due to the bar.

The above discussion serves to stress that care is required in the interpretation of substructure. Nonetheless, if substructures can be proven to be related to mergers, this is interesting from the archaeological point of view. As just discussed such substructures can reveal both the response of the in-situ system (and hence contain information about its properties and the nature of the encounter), as well as on the accreted population. 

\section{Discussion}
\label{sec:next}

{\it Gaia} DR2 data supplemented with that from existing large spectroscopic surveys, have made it relatively straightforward to identify what was plausibly a very important milestone in Galactic history: the merger with Gaia-Enceladus. Stars from this galaxy distinguish themselves kinematically, having slightly retrograde and very eccentric orbital motions, and more patently via their chemical abundances. In particular, these stars define a tight chemical sequence in [$\alpha$/Fe] vs [Fe/H], which is especially apparent for [Fe/H] $\gtrsim -1.3$. The sequence merges with that of the (metal-weak) thick disk for lower metallicities, and the values of [$\alpha$/Fe] here depict significantly more scatter (see \textbf{Figure~\ref{fig:thickd-ecc}}). It seems improbable that the scatter is due to the overlap of stars from just these two systems (as each follow tighter relations at higher metallicities), but more likely it is indicative of other accretion events, whose debris may have been identified already in part (and discussed in e.g. Sec.~\ref{sec:more_subs}). Because of the correlation between mass and metallicity, and between mass and star formation history, small mass objects will have lower average values of [Fe/H] (hence populate the [Fe/H]~$\lesssim -1.3$ regime), and also typically depict lower [$\alpha$/Fe] due to their less efficient star formation. Both these facts seem to be consistent with the findings reported above. 

\citet{1997A&A...326..751N} were among the first to show that $\alpha$-poor stars are interesting markers for accretion events, and as discussed in \citet{2019arXiv190809623I}, large surveys are helping to identify larger numbers of low-$\alpha$ stars which are then being followed-up with high-resolution \citep[see e.g.][]{2019NatAs...3..631X}. Similarly also r-process enhanced stars are receiving more attention \citep[see e.g.][]{2018ApJ...868..110S}, particularly because the large scatter known to be present in the field halo population at very low metallicities has been suggested to be an indication of several accreted small galaxies \citep[see e.g.][]{2018AJ....156..179R}.  Although the formation channels of Carbon-enhanced metal-poor (CEMP) stars are not fully understood, they also provide interesting insights. For example, most CEMP-no stars (CEMP stars with no over-abundance of neutron capture elements) have [Fe/H]~$\lesssim -2.5$, while the CEMP-s stars (enhanced in s-process elements) typically have [Fe/H]~$> -2.5$ \citep{2007ApJ...655..492A}. Since s-process elements would be produced on longer timescales, such stars would have their origin (when not in a binary), in systems that have sustained star formation longer, i.e. more massive hosts, and thus be preferentially found in the inner halo \citep[as the recent analysis of][seems to support]{2019ApJ...885..102L}. On the other hand, CEMP-no would form more predominantly in low mass galaxies (hence their lower average metallicity), which when accreted would remain in the outer regions of the halo, because they would not be able to sink in via dynamical friction \citep[see e.g.][]{2017MNRAS.465.2212S}, which appears to be in line with the trends observed in the Milky Way halo \citep{2014ApJ...788..180C}. 

The general picture emerging, that of a few large building blocks dominating the inner stellar halo, the largest possibly being Gaia-Enceladus, is consistent with that expected from cosmological simulations for a galaxy like the Milky Way. Zoom-in cosmological hydrodynamical simulations such as AURIGA \citep{2017MNRAS.467..179G}, FIRE \citep{2014MNRAS.445..581H}, or those based on N-body methods combined with semi-analytic models such as Aquarius \citep{2010MNRAS.406..744C}, are very useful for understanding the general properties of stellar halos and how they might relate to e.g. merger history \citep{2013MNRAS.432.3391T,2015ApJ...799..184P,2018MNRAS.480..652E}. Some of these simulations are beginning to find look-alikes to the Milky Way also in terms of their merger history, as reported for example in \citet{2019MNRAS.484.4471F}. Large fully cosmological hydrodynamical simulations such as the Illustris suite and its successor IllustrisTNG \citep{2018MNRAS.477.1206N,2019ComAC...6....2N} also contain Milky Way-like galaxies. The largest EAGLE cosmological simulation \citep{2015MNRAS.446..521S} for example, contains 100 objects that are close analogs to the Milky Way in terms of stellar and dark mass, SFR and bulge-to-disk ratio, of which a handful with similar stellar halos in terms of their dynamical properties \citep[see][]{2019arXiv190807080B}. 

\begin{figure}[h]
\includegraphics[width=\textwidth]{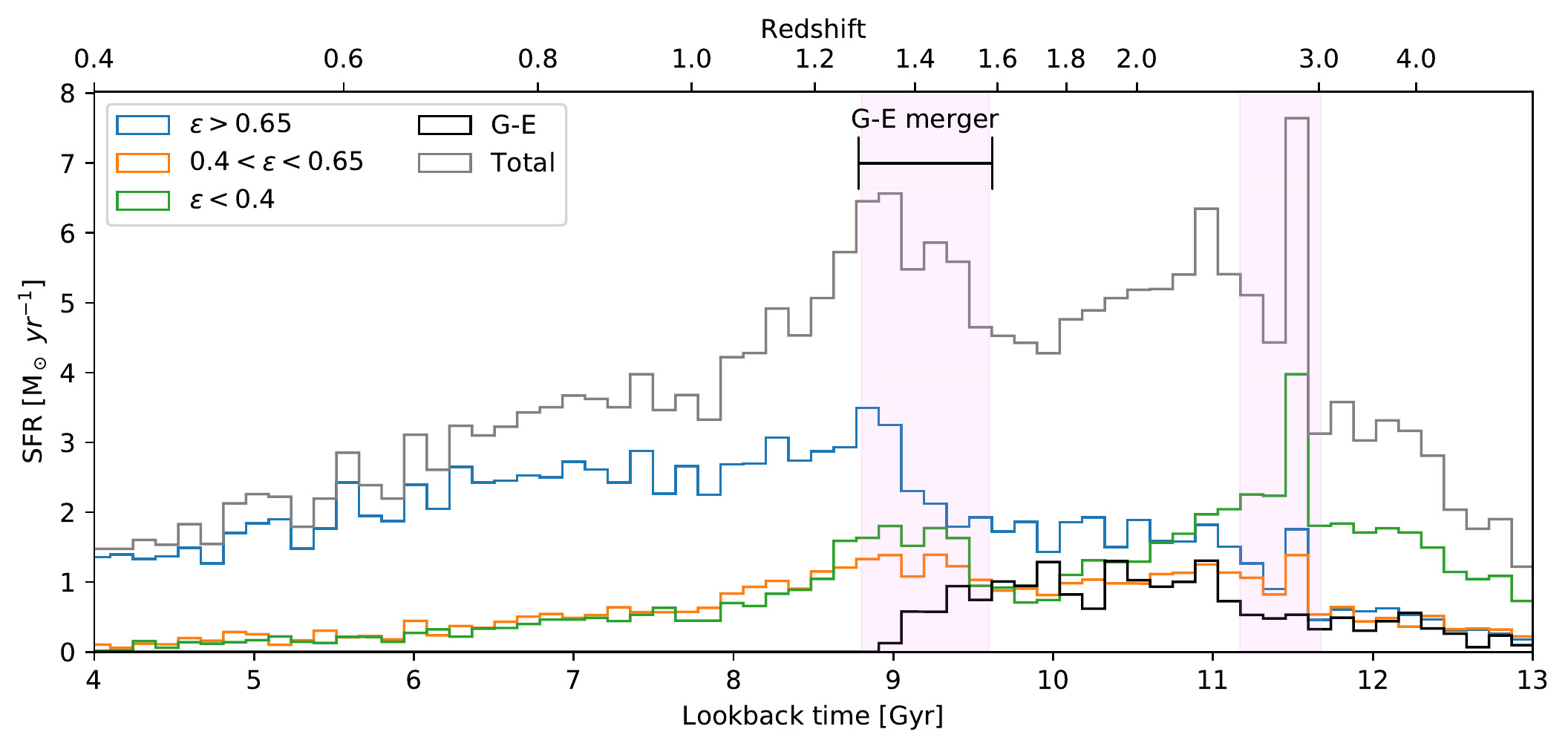}
\includegraphics[width=\textwidth]{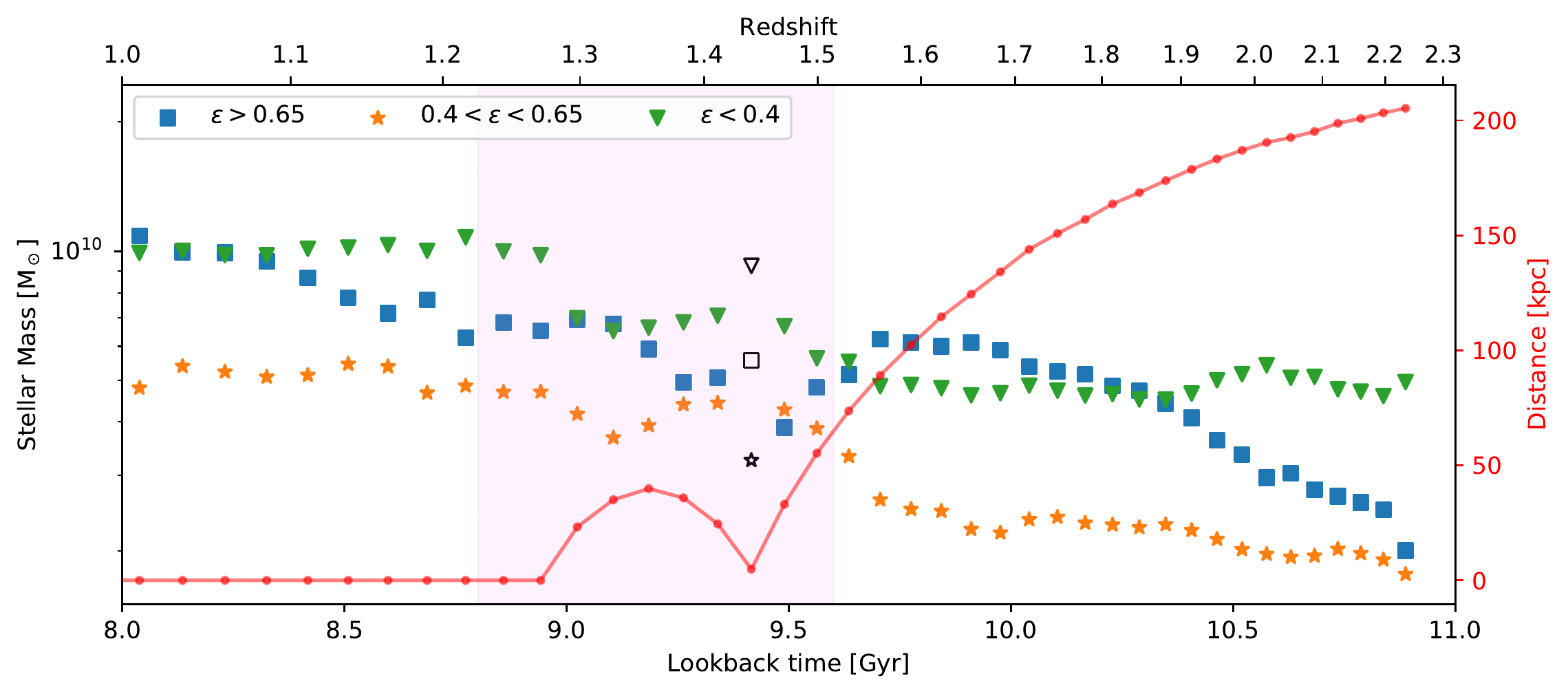}
\caption{Evolution of a Milky Way-like galaxy identified in the largest EAGLE cosmological simulation. The galaxy's spheroid component has dynamical properties similar to those of the Milky Way (a significant population of stars on very eccentric orbits, i.e. the ``Gaia-sausage") as a result of a merger with another system that was completed at $z \sim 1.2$. Stars have been associated to ``components" on the basis of their circularity (computed at each point in time), with low circularity (green) representing the spheroid,  intermediate (orange) the thick disk and high circularity (blue) the thin disk. The thick disk in this simulation originates in part from a heated disk (star particles ``transferred" from the thin disk and put on less circular orbits during the merger), but also from stars formed during the merger itself. Notice the increase in the SFR of all components during the merger. {\it Credits: Figure adapted from \citet{2019arXiv190807080B} and reproduced with permission from the AAS}.}
\label{fig:lucas}
\end{figure}

The evolution of the galaxy identified by \citet{2019arXiv190807080B} in the EAGLE suite to be a good Milky Way-alike is shown in \textbf{Figure \ref{fig:lucas}} around the time it merges with a Gaia-Enceladus analogue.  The panels of the figure show the evolution of the star formation rate (top) and stellar mass (bottom) for the different ``components" identified according to the circularity of their stars.  At the time of the merger, there is a significant increase in the SFR in the whole system, with that of the fiducial thin disk increasing dramatically towards the end of the merger (fueled in part by gas from the accreted object, which also helps its further growth). Both panels show that all components are affected by the merger, suggesting the existence of populations in the thick disk and bulge that are co-eval with the timing of the merger. 

The figure also shows that the Gaia-Enceladus-analogue barely completes two orbits before it is fully disrupted. Its mass ratio in this simulation is only 20\%, and both the host as the infalling object have more than 50\% of the baryons in cold gas. Such an event is thus quite different from those typically modeled in the context of dwarf galaxy accretion for various reasons. Firstly the debris is expected to be much more complex. On the one hand because of dynamical friction as stars lost early might have different eccentricities (and lower metallicities), and hence their orbits differ from those lost later on. On the other hand, intricate tidal tail morphologies become apparent when a disky galaxy is accreted \citep{1984ApJ...279..596Q}.  Secondly, the gas will respond strongly, and it is conceivable that some star formation might have taken place in the tidal arms or even the formation of star clusters to have been triggered. The degree of complexity of such an event evidences the need for tailored simulations including gas, star formation and chemical evolution  to fully interpret and model the observations currently available.

\subsection{Next steps: Simulations}

Because of their limited resolution, many of the cosmological simulations just mentioned have been re-sampled to produce more particles \citep{2015MNRAS.446.2274L, 2018MNRAS.481.1726G, 2018arXiv180610564S}, as typically a star particle in a simulation nowadays should be seen as representing a stellar population of approximately $10^3 - 10^4 \sm$. Nonetheless, there are still important limitations 
including the ability to trace the true phase-space distribution of stars originating in objects with a stellar mass lower than $10^6 \sm$ or thereabouts. Furthermore, not all these simulations follow chemical enrichment properly and as discussed in previous sections, this is necessary for the guidance in the interpretation of the various structures that are being (and will in the future, likely be) identified. 

The study by \citet{2019arXiv190807080B} makes clear that cosmological simulations (even with low resolution) are useful also for exploring or making links between the formation paths of the different galactic components. Now that the identification of true analogues in cosmological simulations has become possible, it will be of great 
interest to carry out new zoom-in simulations of these objects. They will allow us to address a variety of questions, including for example gas physics, star formation and chemical enrichment processes, as well as to establish the true link between different events in the assembly history of the Milky Way. 
Furthermore, such simulations will be necessary to guide dark matter detection experiments which often assume that the dark matter particles follow Maxwellian velocity distributions. We now know that the stellar halo near the Sun is complex and has multiple kinematic components \citep[see e.g.][for a discussion of the impact on direct detection experiments]{2019arXiv190904684O}.
It has stars with kinematics corresponding to the tail of the thick disk, and from a component that is mildly retrograde which is associated with Gaia-Enceladus. But we do not know how the dark matter should be distributed given the particular Galactic history just unraveled \citep[although see][]{2019ApJ...883...27N}. 
It is therefore very important to carry out such simulations now that the amount of freedom has been significantly reduced and that the boundary conditions are better known so that 
we can provide concrete constraints on the initial conditions. Such simulations can be also used to understand some peculiarities about the Milky Way, such as the possibly fairly low mass for the super-massive black hole in the center of the Galaxy, or the distribution of satellites and the origin of their configuration. 

\subsection{Next steps: Statistical analyses}

To trace the assembly history of the Milky Way as far back as possible, it will also be necessary to work in a more systematic fashion than until now. This is essential for establishing for example, the mass spectrum of the objects accreted and their internal characteristics. 
It will require the application and development of statistical methods, some of which are available in the literature (e.g. IoM, frequency space), Fourier analysis, machine learning, clustering algorithms such as (H)DBSCAN \citep[as used by][]{2019arXiv190702527B,2019arXiv191007684C,2019arXiv190707681N} or based on neural networks \citep{2019arXiv191007538Y}. An important aspect is the assessment of the statistical significance of a given feature or clump, and this can be done either through comparison to models or to suitably randomized samples. It would also be useful if there was more consistency in the different structures reported in the literature. Sometimes, a structure is reported as newly discovered, while it has been reported before \citep[see][for a recent example]{2019arXiv190904684O}. This leads to confusion in the field (in the naming and in the reality of the features) and also does not help in building up a coherent picture of Galactic history. A possible improvement would be to assign membership probabilities to the different stars, and to publish these together with the different structures and stars' IDs.

Similar issues arise for the globular cluster population of the Milky Way. Although \citet{2019arXiv190608271M} have tentatively associated many (at least 35\%) of the globular clusters of the Milky Way to what we may identify as the main building blocks of the halo (at least near the Sun), namely Gaia-Enceladus, Sagittarius, the Helmi Streams and Sequioa, for many globular clusters the assignment is not unique, particularly in Integrals of Motion space \citep[see for example][]{2018ApJ...863L..28M}. More precise ages for the clusters could potentially aid as well because although the age-metallicity relations are well-defined, they are not fully unambiguous \citep[see also][]{2019MNRAS.486.3180K}.

\begin{figure}[h]
\includegraphics[width=\textwidth]{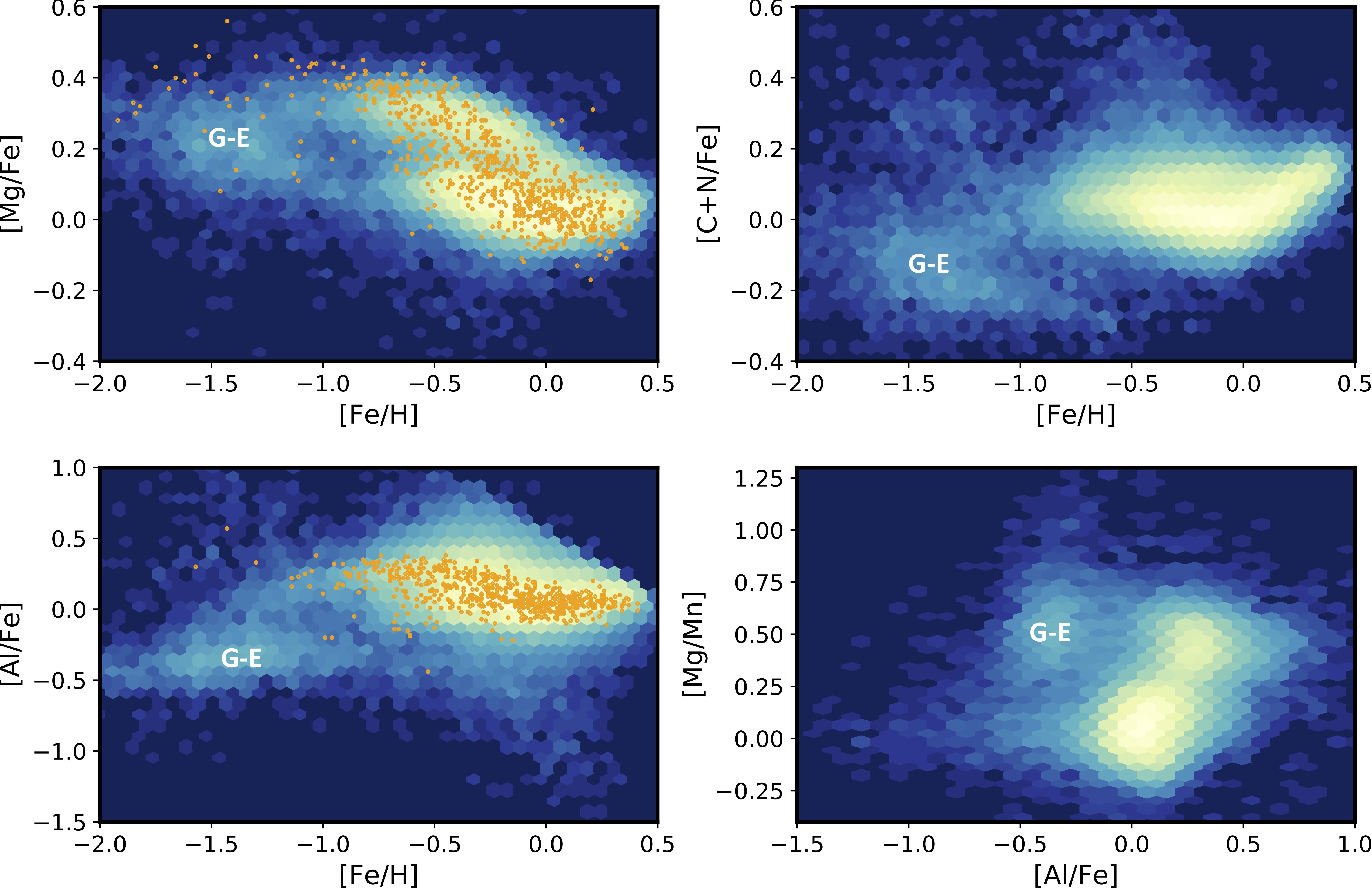}
\caption{Two-dimensional abundance distribution of APOGEE DR14 stars in different chemical abundance planes. The filled orange circles correspond to kinematically-selected disk stars from \citet{2014A&A...562A..71B}. Note the presence of Gaia-Enceladus (marked as ``G-E" in all panels), and its clear distinction from  the thick disk, not only in [Mg/Fe] vs [Fe/H] but also in lighter elements such as [Al/Fe] and [C+N/Fe]. {\it Credits: Adapted from \citet{2019arXiv190309320D}, their Fig.~3.}}
\label{fig:chem-lab}
\end{figure}

For the interpretation of the various structures, it is clear that there is an urgent need for detailed chemical abundances of the stars with full phase-space information. Such samples will also aid in disentangling the different events, in assessing their origin (particularly if substructures overlap in e.g. IoM space, as will necessarily happen in the majority of cases), and characterizing their properties and history. A very nice example of what is currently possible along these lines was given in \citet{2019arXiv190309320D}. Their analysis of APOGEE DR14 data is shown in \textbf{Figure \ref{fig:chem-lab}}, where the different main structures discussed in this review are clearly separated in chemical abundance space.

\subsection{Next steps: Surveys}

The need for spectroscopic surveys for large numbers of stars has long been recognized since the pioneering ideas of \citet{2002ARA&A..40..487F}
and discussed in the context of e.g. ESO-ESA synergies by \citet{2008Msngr.134...46T}. This interest has lead to a significant increase in the number and scope of the surveys. 
Surveys such as RAVE, SEGUE and Gaia-ESO\footnote{\tt https://www.gaia-eso.eu} have been carried out over the past decade; some like APOGEE, LAMOST and GALAH have been running over the past years. In the next few years projects such as WEAVE, 4MOST and DESI using 4m class telescopes and MOONS\footnote{\tt https://vltmoons.org} on the VLT will see the light. The Galactic surveys that will be carried out using these facilities have two complementary goals. The first is to obtain the missing radial velocity for stars for which {\it Gaia} has (at best) measured their 5D phase-space location. 
The second goal is to obtain high resolution spectra for brighter stars to do chemical labelling and characterization of the most metal-poor populations for example in the halo. 
The first goal is important for the dynamics of distant halo stars, for which tangential velocity constraints available from {\it Gaia} data releases will be available but less precise. The radial velocity measurements will allow mapping the mass distribution in our Galaxy at large radii \citep{2019Msngr.175...23H}. Also for nearby faint dwarf stars, obtaining the missing radial velocity component is useful because such stars could be used to trace the time-variations in the gravitational potential of the Milky Way for example in frequency space, as discussed in Sec.~\ref{sec:dynamics}. 

On the other hand, high resolution spectroscopic follow-up is of utmost importance as the discovery of Gaia-Enceladus has taught us. It is arguably the most powerful way to fully pin-down history and to identify debris with certainty, as well as to disentangle accretion events from one another. 
The high-resolution surveys planned by e.g. WEAVE and 4MOST will be carried out using relatively bright stars, i.e. down to $G \sim 16$, because of the
use of 4m class telescopes \citep[see e.g.][]{2019Msngr.175...26C}. Although they will be invaluable, it is already clear that a wide-field spectroscopic survey on an 8--12m class telescope would be fantastic as it would really match the capabilities of {\it Gaia}, and of e.g. LSST\footnote{\tt https://www.lsst.org} \citep{2019ApJ...873..111I} in the coming years. 
In particular for the identification and characterization of debris from low mass systems, it will be necessary to target main sequence stars, as they
are much more numerous than the few RGB stars present in ultra-faint-like galaxies.  
Such galaxies are particularly interesting because of questions related to the presence of thresholds for galaxy evolution, the impact of reionization and feedback processes, and because they may host some of the most metal-poor stars. These stars reveal the imprint of just a few supernovae and possibly of the initial mass function in the early Universe. Because of the low density of tidal debris and  of the stellar halo more generally, follow-up must be carried out using a wide-field, and an 8-10m telescope may well necessary to reach the required depth. The PFS\footnote{\tt https://pfs.ipmu.jp/intro.html} on Subaru \citep{2016SPIE.9908E..1MT} is an instrument that could potentially help with the chemical labelling, although its highest resolution mode has $R \sim 5000$ and so obtaining detailed chemical abundances for many elements will not be feasible.
The MSE\footnote{\tt https://mse.cfht.hawaii.edu}$^,$\footnote{See {\tt https://mse.cfht.hawaii.edu/misc-uploads/MSE\_Project\_Book\_20181017.pdf}} is another interesting facility being considered, but there are no (other) concrete plans at the time of writing of this review, although \citet{2018IAUS..334..242P} discuss in some detail a concept developed at ESO whose main science driver is high-resolution follow-up of {\it Gaia} targets, in a case termed ``the Milky way as a Model Galaxy Organism". 

\section{Conclusions}
\label{sec:concl}

\begin{summary}[SUMMARY POINTS]
\begin{enumerate}
\item The evidence discussed in this review supports the view that a milestone in Galactic history has been unveiled, namely the last big merger experienced by the Milky Way 10 Gyr ago. The merged galaxy, known as Gaia-Enceladus, appears to be responsible for a large fraction of the inner stellar halo.

\item A disk component was present at the time of merger with Gaia-Enceladus, which is possibly related to the metal-weak thick disk. This disk component was significantly perturbed as a consequence of the merger, and it is currently responsible for roughly 50\% of the stars with halo-like kinematics near the Sun, i.e. it seems to constitute what has been known as the ``in-situ" halo. There is also some evidence that the merger with Gaia-Enceladus triggered significant star formation in the thick disk, and it seems plausible that it was responsible for the formation of a large fraction of its stars.

\item These results are consistent with the predictions of the $\Lambda$CDM model to first order. Objects in cosmological simulations with similar merger histories have been identified and are providing new insights in the evolution of the various Galactic components, and into how their histories may be interlinked.

\item Several other kinematic substructures have been identified amongst nearby stars and are likely associated to past accretion events. The Helmi streams ($\sim 10^8 \sm$), (bonsai) Sequoia ($\sim 10^7 \sm$), Thamnos ($\sim 5 \times 10^6 \sm$),  together with Gaia-Enceladus ($\sim 10^9 \sm$) and Sagittarius ($\sim 5 \times 10^8 \sm$), appear to be the largest building blocks of the halo of our Galaxy, although it is not always clear how and if they are related to each other. For some of these (Helmi streams, Gaia-Enceladus, and Sagittarius) it has been possible
to derive (sketchy) star formation histories, and their chemical characterization has only just began via chemical labelling. The biggest current limitation is the availability of large samples of stars with detailed chemical abundance information, but when available this information turns out to be crucial.

\item When such samples become available it will be possible to do true Galactic archaeology, and establish the properties of the galaxies (star formation and chemical enrichment histories, mass, size) before they were accreted, i.e. really explore the high redshift universe from our own backyard. It should also be possible to carry out a similar exercise for the thick disk, or the proto-disk, which we know was present 10 Gyr ago (at $z \sim 1.8$), and seems to have been traced back to metallicities [Fe/H]~$\lesssim -4$ \citep{2019MNRAS.484.2166S}. Establishing the properties of this disk (structure, star formation) will allow us to make a direct link to high-redshift disks being currently observed in-situ.

\item Most of the above mentioned building blocks were identified using full phase-space information of stars in the Solar vicinity. At larger distances (beyond 20 kpc from the Galactic center, 10 kpc from the Sun), little is known about kinematics and chemistry of the stars but several large substructures have been known for a while thanks to wide-field photometric surveys. These include Hercules-Aquila and the Virgo overdensity, as well as several others. The relation between these and the so-far identified building blocks if any, has not been really pinned-down although we present here a way forward using orbital integrations. With {\it Gaia} DR3 and subsequent releases (DR4 and beyond, since the mission has been extended over the nominal lifetime for at least 2 years), and in combination with spectroscopic surveys, it should be possible to address this and many more questions. 

\item Also in the direction towards the Galactic center, little is known, yet this is where the halo reaches its highest density. Merger debris from another massive building block might well be present \citep[as suggested by the analysis of the age-metallicity relations and dynamics of globular clusters by][and which could be a ``Kraken"-like object]{2019MNRAS.486.3180K,2019arXiv190608271M}. It is not clear at this point if {\it Gaia} can answer this, but an astrometric mission in the near infrared \citep[see e.g.][]{2015IAUGA..2247720G,2019arXiv190712535H} could perhaps address this open question. 

\item There is more to be gained from the analysis of substructures in the vicinity of the Sun. For example, some of these may well be due to internal mechanisms (such as the Galactic bar or non-integrability of a generic Galactic potential), or even reveal the response of the Galaxy to an accretion event. For the analysis of the substructures as well as to address the issues mentioned in previous items, more detailed modeling is needed, also from a dynamical perspective, and preferably through zoom-in cosmological simulations which now could model systems which at least in terms of their merger history, would be much more representative of the Milky Way. 
\end{enumerate}
\end{summary}

Enormous progress has been made in recent years in our understanding of the evolution of the Milky Way from the perspective of the stellar halo and thick disk disk. This has been possible particularly thanks to {\it Gaia} DR2 and the many spectroscopic surveys currently available, and especially those surveys with detailed chemical abundance information have been proven to be crucial.  Yet a plethora of questions remain open for the next decade. Fortunately we already know we will have many of the necessary tools to answer them. There are very exciting times ahead of us in the field of Galactic archaeology. 

\begin{issues}[FUTURE ISSUES]
\begin{enumerate}
\item Although challenging, it would be extremely interesting to identify accretion events that have taken place even before the merger with Gaia-Enceladus 10 Gyr ago. 
\item The characterization (in terms of dynamical, chemical and star formation history) of the disk present at the time of the merger with Gaia-Enceladus,  should be pursued. An important link is there to be made between the disks observed in-situ in high-$z$ observational studies, and that revealed in the Milky Way. 
\item Zoom-in cosmological simulations of systems with a merger history and dynamics similar to that suggested by recent data would be particularly useful for understanding Galactic history and the link between its various components and detailed properties. Such simulations would also allow to make robust predictions for direct detection dark matter experiments. 
\item There is an ever increasing need for large high-resolution spectroscopic surveys of relatively faint stars ($G \gtrsim 16$) to supplement the dynamical information that is becoming available thanks to the {\it Gaia} mission. Determination of precise ages for such a sample of stars would be highly valuable and useful to date various events in Galactic history, particularly at early times. 
\end{enumerate}
\end{issues}

%disclosure
\section*{DISCLOSURE STATEMENT}
The author is not aware of any affiliations, memberships, funding, or financial holdings that
might be perceived as affecting the objectivity of this review. 

\section*{ACKNOWLEDGMENTS}
%Acknowledgements, general annotations, funding.

Very many thanks to all my collaborators throughout the years. I want to especially acknowledge all my  former bachelor, MSc and PhD students as well as my former postdocs, who with their work contributed directly or indirectly to this review with their ideas, enthusiasm and dedication. I am particularly indebted to Helmer Koppelman, Davide Massari, Maarten Breddels and Jovan Veljanoski for the incredible ride since {\it Gaia} DR1, in the search of truth and Milky Way's history. The {\it Gaia} consortium (DPAC) and especially Anthony Brown are particularly thanked for their fantastic work and very friendly working atmosphere. I am also grateful to my PhD advisors, Simon White and Tim de Zeeuw for their mentorship throughout my career. Several colleagues, including Helmer Koppelman, Tadafumi Matsuno, Eline Tolstoy and Tim de Zeeuw have read early drafts and contributed with comments which helped improve this review. I am also grateful to the editor, Joss Bland-Hawthorn for the open and constructive remarks, and to Eduardo Balbinot and Helmer Koppelman who helped with several of the figures included in this review. My son Manuel is especially thanked for his patience and continuous encouragement. Data from the European Space Agency
mission {\it Gaia} ({\tt http://www.cosmos.esa.int/gaia}), processed by the
{\it Gaia} DPAC (see 
{\tt http://www.cosmos.esa.int/web/gaia/dpac/consortium}) as been used in this work. Funding for
DPAC has been provided by national institutions, in particular the
institutions participating in the {\it Gaia} Multilateral Agreement. Data from the APOGEE survey, which is part of
Sloan Digital Sky Survey IV, has also been used. SDSS-IV is managed by the Astrophysical
Research Consortium for the Participating Institutions of the SDSS
Collaboration. Financial support from NOVA, and from NWO through a Vici grant and more recently the Spinoza prize, are also gratefully acknowledged. 

% References

\bibliographystyle{ar-style2} 
%\bibliography{export-bibtex-1,export-bibtex-2-1,export-bibtex-3c,export-bibtex-rv-1}
\bibliography{export-bibtex-all-1,additional-refs-1}

\end{document}